\documentclass[12pt,a4paper]{article}
\usepackage{epsfig}
\usepackage{mathtools}
\usepackage{amsfonts}
\usepackage{amssymb}
\usepackage{amsmath}
\usepackage{color}
\usepackage{slashed}
\usepackage{cite}
\usepackage{caption}
\usepackage{subcaption}

\usepackage{tikz}
\usetikzlibrary{shapes}
\usetikzlibrary{trees}
\usetikzlibrary{calc}
\usetikzlibrary{decorations.pathmorphing}
\usetikzlibrary{decorations.markings}

\def\la              {\langle}
\def\ra              {\rangle}

\newcommand{\dotalpha}{{\dot{\alpha}}}
\newcommand{\dotbeta}{{\dot{\beta}}}
\newcommand{\dotdelta}{{\dot{\delta}}}
\newcommand{\dotgamma}{{\dot{\gamma}}}

\newcommand{\sigmapart}{\tilde{\sigma}}

\newcommand{\refspina}{\xi_A}
\newcommand{\refspinb}{\xi_B}
\newcommand{\refspinal}{\xi_A}
\newcommand{\refspinbl}{\xi_B}

\newcommand{\abr}[1]{\langle #1 \rangle}
\newcommand{\sbr}[1]{\left[ #1 \right]}

\newcommand{\vll}{{\smash{\lambda}}}
\newcommand{\vlt}{{\smash{\tilde{\lambda}}}}
\newcommand{\vlet}{{\smash{\tilde{\eta}}}}
\newcommand{\vle}{{\smash{\eta}}}

\newcommand{\vllu}{\smash{\underline{\smash{\lambda}}}}
\newcommand{\vltu}{\smash{\underline{\smash{\tilde{\lambda}}}}}
\newcommand{\vleu}{\smash{\underline{\smash{\tilde{\eta}}}} }

\newcommand{\vlluu}{\smash{\underline{\underline{\smash{\lambda}}}}}
\newcommand{\vltuu}{\smash{\underline{\underline{\smash{\tilde{\lambda}}}}}}
\newcommand{\vleuu}{\smash{\underline{\underline{\smash{\tilde{\eta}}}}}}
\newcommand{\vletuu}{\smash{\underline{\underline{\smash{\eta}}}}}

\newcommand{\vlluuu}{\vlluu}
\newcommand{\vltuuu}{\vltuu}
\newcommand{\vleuuu}{\vleuu}
\newcommand{\vletuuu}{\vletuu}

\newcommand{\vmmuuu}{{\smash{\underline{\underline{\smash{\mu}}}}}}
\newcommand{\vmtuuu}{{\smash{\underline{\underline{\smash{\tilde{\mu}}}}}}}

\newcommand{\dd}{\mathrm{d}}

\newcommand{\eqndot}{\, . }
\newcommand{\eqncom}{\, , }

\newcommand{\splus}{\! + \!}
\newcommand{\sminus}{\! - \!}
\newcommand{\ssep}{\;}

\definecolor{grayn}{gray}{0.7}
\definecolor{lightgrayn}{gray}{0.8}

\def\vacuumheight{1}

\def\labelvdist{0.3}

\def\labelddist{\labelvdist*0.70710678118}

\newlength{\vacuumradius}
\newlength{\onshellradius}
\tikzstyle{db}=[circle, black, fill=black, minimum width=\onshellradius, draw, inner sep=0pt]
\tikzstyle{dw}=[circle, black, fill=white, minimum width=\onshellradius, draw, inner sep=0pt]
\tikzstyle{dvac}=[circle, black, fill=lightgrayn, minimum width=\vacuumradius, inner sep=0pt]
\tikzstyle{dl}=[circle, black, fill=white, inner sep=2pt]

\tikzset{
	gluon/.style={decorate, decoration={coil, amplitude=2pt, segment length=3.5pt, aspect=1}, draw=black}
}

\begin{document}

\begin{center}

\vspace{1cm}

{\bf \Large Wilson lines, Grassmannians and Gauge Invariant \\ \vspace{0.2cm}  Off-shell Amplitudes in $\mathcal{N}=4$ SYM.} \vspace{1cm}

{\large L.V. Bork$^{1,2}$ A.I. Onishchenko$^{3,4,5}$}\vspace{0.5cm}

{\it $^1$Institute for Theoretical and Experimental Physics, Moscow,
	Russia,\\
	$^2$The Center for Fundamental and Applied Research, All-Russia
	Research Institute of Automatics, Moscow, Russia, \\
	$^3$Bogoliubov Laboratory of Theoretical Physics, Joint
	Institute for Nuclear Research, Dubna, Russia, \\
	$^4$Moscow Institute of Physics and Technology (State University), Dolgoprudny, Russia, \\
	$^5$Skobeltsyn Institute of Nuclear Physics, Moscow State University, Moscow, Russia}\vspace{1cm}

\abstract{In this paper we consider tree-level gauge invariant off-shell amplitudes (Wilson line form factors) in $\mathcal{N}=4$ SYM. For the off-shell amplitudes with one leg off-shell we present a conjecture for their Grassmannian integral representation in spinor helicity, twistor and momentum twistor parameterizations. The presented conjecture is successfully checked against BCFW results for MHV$_n$, NMHV$_4$ and NMHV$_5$ off-shell amplitudes. We have also  verified that our Grassmannian integral representation correctly reproduces soft (on-shell) limit for the off-shell gluon momentum. It is shown that the (deformed) off-shell amplitude expressions could be also obtained using quantum inverse scattering method for auxiliary $\mathfrak{gl}(4|4)$ super spin chain.}
\end{center}

\begin{center}
\end{center}

\newpage

\tableofcontents{}\vspace{0.5cm}

\renewcommand{\theequation}{\thesection.\arabic{equation}}

\section{Introduction}

$\mathcal{N}=4$ SYM theory is an excellent playground for testing new computational methods for $D=4$ dimensional gauge theories. In the last decade we have witnessed a serious progress in understanding the structure of amplitudes (S-matrix) in $\mathcal{N}=4$ SYM  as well as in other gauge theories (see for a review \cite{Reviews_Ampl_General,Henrietta_Amplitudes}). The main role
in these achievements was played by a number of so called unitarity methods based on the exploration of the amplitude analytical structure\cite{Reviews_Ampl_General,Henrietta_Amplitudes}. The latter are given for example by BCFW recursion \cite{BCFW1,BCFW2} for tree amplitudes and generalized unitarity (see \cite{Reviews_Ampl_General} and references therein) for loop amplitudes. We should especially mention the use of new variables, such as helicity spinors (see appendix \ref{aA} for details) and momentum twistors \cite{MomentumTwistors}, together with superspace methods \cite{Nair,DualSuperConformalSymmetry} (see also recent review \cite{IvanovReview}). The application of these new methods provided us with the explicit answers for amplitudes both at high orders of perturbation theory and/or with large number of external legs (see  \cite{Reviews_Ampl_General,Henrietta_Amplitudes} for a review and reference therein), which in their turn lead to several important all-loop results as well as to the discovery of underlying integrable structure behind amplitudes of $\mathcal{N}=4$ SYM
\cite{YangianSymmetryTreeAmplitudes,BeisertYangianRev,Beisert_SpectralReg_New,AmplitudesSpectralParameter1,AmplitudesSpectralParameter2,Chicherin_YangBaxterScatteringAmplitudes,Frassek_BetheAnsatzYangianInvariants,Kanning_ShortcutAmplitudesIntegrability,Broedel_DictionaryRoperatorsOnshellGraphsYangianAlgebras,Broedel_DeformedOneLoopAmplitudes}.

Another novel way of studying scattering amplitudes is based on their
Grassmannian integral representation
\cite{DualitySMatrix,AmplitudesPositiveGrassmannian,AllLoopIntegrandN4SYM,GrassmanianOriginDualConformalInvariance,UnificationResiduesGrassmannianDualities,DualSupercondormalInvarianceMomentumTwistorsGrassmannians}. It naturally unifies different BCFW representations for
tree level amplitudes and loop level integrands  \cite{DualitySMatrix,AmplitudesPositiveGrassmannian}. Moreover, it shed light on the integrable structure behind $\mathcal{N}=4$ SYM amplitudes (S-matrix) \cite{Drummond_Grassmannians_Tduality,Drummond_Yangian_origin_Grassmannian_integral,Chicherin_YangBaxterScatteringAmplitudes}. This representation also naturally relates perturbative $\mathcal{N}=4$ SYM and twistor string theories amplitudes\cite{UnificationResiduesGrassmannianDualities}.
In addition  a possible geometrical interpretation of $\mathcal{N}=4$ SYM (so called Amplituhedron) was discovered within Grassmannian picture
\cite{MomentumTwistors,Arcani_Hamed_Polytopes,Amplituhdron_1,Amplituhdron_2,Amplituhdron_3,Amplituhdron_4,Amplituhdron_5,Amplituhdron_6}.

The unitary based methods were applied not only to study on-shell amplitudes, but also for partially off-shell objects, such as form factors in $\mathcal{N}=4$ SYM theory  \cite{FormFactorMHV_component_Brandhuber,HarmonyofFF_Brandhuber,BORK_NMHV_FF,FF_MHV_3_2loop,Roiban_FormFactorsOfMultipleOperators,FormFactorMHV_half_BPS_Brandhuber,FormFactorMHV_Remainder_half_BPS_Brandhuber,BORK_POLY,Wilhelm_Integrability_1,Wilhelm_Integrability_2,Wilhelm_Integrability_3,Wilhelm_Integrability_4,Wilhelm_Twisors_1,Wilhelm_Twisors_2,LHC_1,LHC_2,LHC_3,BoFeng_BoundaryContributions,Henn_Different_Reg_FF,3loopSudakovN4SYM,FF_Colour_Kinematic,masters4loopSudakovN4SYM,Oluf_Tang_Engelund_Lagrangian_Insertion,Brandhuber_DynamicFormFactors}. The form factors are
the matrix elements of the form\footnote{The on-shell amplitudes in ''all ingoing" notation may be viewed as a particular case of  form factors of unity operator $\langle p_1^{\lambda_1}, \ldots, p_n^{\lambda_n}|0 \rangle$. }
\begin{equation}
\langle p_1^{\lambda_1}, \ldots,
p_n^{\lambda_n}|\mathcal{O}|0\rangle,
\end{equation}
where $\mathcal{O}$ is some gauge invariant operator, which when acting on the vacuum state of the theory produces multi-particle state $\langle p_1^{\lambda_1}, \ldots, p_n^{\lambda_n}|$ with momenta $p_1, \ldots, p_n$ and helicities $\lambda_1, \ldots, \lambda_n$. One can
view form factors as amplitudes of the processes where classical field coupled through gauge invariant operator $\mathcal{O}$ produces quantum state. Grassmannian representation is no exception and can be applied to form factors as well \cite{SoftTheoremsFormFactors,FormFactorsGrassmanians,WilhelmThesis,q2zeroFormFactors}.

Another interesting off-shell objects are gauge invariant off-shell amplitudes  \cite{vanHamerenBCFW,LipatovEL1,LipatovEL2,KirschnerEL1,KirschnerEL2,KotkoWilsonLines,vanHamerenWL1,vanHamerenWL2} (reggeon amplitudes within the context of Lipatov's effective lagrangian), which typically appear
in  $k_T$ - or high-energy factorization approach
\cite{GribovLevinRyskin,CataniCiafaloniHautmann,CollinsEllis,CataniHautmann} as well as in the study of processes at multi-regge kinematics. We should mention, that there are also other studies of off-shell currents and amplitudes \cite{BerendsGiele,recursion_solution,offsellBCFWcurrents,Kallosh_1,Kallosh_2}. (see also \cite{DixonReview} for a review and references to original papers). However, the off-shell objects studied there are either gauge dependent \cite{BerendsGiele,recursion_solution,offsellBCFWcurrents} or lack Lorentz invariance \cite{Kallosh_1,Kallosh_2}. Usually within the context of application of unitarity based methods to form factors  $\mathcal{O}$ is local
gauge invariant color singlet operator, for example operators from stress-tensor operator supermultiplet  \cite{FormFactorMHV_component_Brandhuber,HarmonyofFF_Brandhuber,BORK_NMHV_FF,FF_MHV_3_2loop,Roiban_FormFactorsOfMultipleOperators,BKV_Form_Factors_N=4SYM,BKV_SuperForm,Zhiboedov_Strong_coupling_FF,Strong_coupling_FF_Yang_Gao}. However in general one can consider also gauge invariant non-local operators, such as Wilson loops (lines). This brings us to the following important observation.  We can formulate \emph{gauge invariant off-shell amplitudes} in Yang-Mills theory in terms of form factors of
Wilson line operators \cite{vanHamerenBCFW,LipatovEL1,LipatovEL2,KirschnerEL1,KirschnerEL2,
KotkoWilsonLines,vanHamerenWL1,vanHamerenWL2}. See also \cite{vanHameren:2015bba} for the discussion of amplitudes with off-shell fermions. An insertion of Wilson line operator plays the role of off-shell or reggeized gluon in such formulation. Keeping in mind an extremely important role played by Wilson loops in $\mathcal{N}=4$ SYM theory it is very interesting to study the possibility of Grassmannian integral representation  for gauge invariant off-shell amplitudes (Wilson line form factors). The aim of this article is to investigate this question in detail.

This paper is organized as follows. In section 2 we discuss the definition and kinematics of gauge invariant off-shell amplitudes in terms of form factors of Wilson line operators corresponding to off-shell gluons in $\mathcal{N}=4$ SYM theory. Here we also derive an expression for the ``minimal'' 2+1 point tree level off-shell amplitude, such that two on-shell particles are treated in manifestly supersymmetric manner (using on-shell momentum superspace), while the off-shell gluon is left unsupersymmetrized. In section 3 we use generalization
of BCFW recursion for off-shell amplitudes \cite{vanHamerenBCFW} to derive some explicit answers for gauge invariant amplitudes with one leg off-shell in $\mbox{MHV}$ and $\mbox{NMHV}$ sectors. In section 4 after reminding the reader the basic facts about Grassmannian integral representation for $\mathcal{N}=4$ SYM on-shell amplitudes (on-shell diagram formalism, et.c.) we present our conjecture for Grassmannian integral representation for all tree-level gauge invariant  amplitudes with one leg off-shell (Wilson-line form factors). We present our conjecture using spinor helicity, twistor and momentum twistor variables. In section 5 we verify our conjecture for the Grassmannian integral representation by reproducing the results obtained
in section 2 together with the appropriate soft (on-shell) limit for off-shell momentum. We also derive a conjecture for
$\mbox{NMHV}_{n+1}$ off-shell tree level amplitudes in terms of $[abcde]$ momentum twistor invariants. In section 6 we discuss integrability properties of tree level gauge invariant off-shell amplitudes
in $\mathcal{N}=4$ SYM theory. We show that off-shell amplitudes with one leg off-shell (Wilson-line form factor) are eigenvectors of the transfer matrix of $\mathfrak{gl}(4|4)$ super spin chain.
In conclusion we summarize all the obtained results and discuss open questions. The appendixes contain details about spinor helicity formalism, $\mathcal{N}=4$ SYM theory and evaluation of Grassmannian integrals by residues.

\section{Wilson lines and off-shell amplitudes in  $\mathcal{N}=4$ SYM}
\label{WilsonLinesSection}

Before discussing gauge invariant off-shell amplitudes in the context of $\mathcal{N}=4$ SYM let us recall the corresponding description in the context of pure Yang-Mills theory.

It is convenient to define off-shell  amplitudes we are interested in to be color ordered. For example, any tree diagram for $n$-gluon scattering could be reduced to a sum of single trace terms. Indeed, if we eliminate structure constants $f^{abc}$ in favor of $SU(N_c)$ - generators $t^a$, using\footnote{We use normalization $\text{tr}~t^at^b = \delta^{ab}$, so that $[t^a,t^b] = i\sqrt{2} f^{abc}t^c$}
\begin{eqnarray}
f^{abc} = -\frac{i}{\sqrt{2}} \left\{
\text{tr}~(t^at^bt^c) - \text{tr}~(t^bt^at^c),
\right\} \label{adjointTofundamental}
\end{eqnarray}
together with Fierz rearrangement
\begin{eqnarray}
(t^a)_{i_1}^{~j_1}(t^a)_{i_2}^{~j_2} = \delta_{i_1}^{~j_2}\delta_{i_2}^{~j_1}
- \frac{1}{N_c}\delta_{i_1}^{~j_1}\delta_{i_2}^{~j_2}
\end{eqnarray}
to reduce number of traces in each diagram, it is easy to rewrite $n$-gluon tree amplitude in a color decomposed form:
\begin{eqnarray}\label{ColourOrderedAmplitudeDefinition}
\mathcal{A}_n^{tree}({p_i,h_i,a_i}) = g^{n-2}\sum_{\sigma \in S_n/Z_n} \text{tr}
~(t^{a_{\sigma (1)}}\cdots t^{a_{\sigma (n)}}) A_n^{tree} (\sigma(1^{h_1}),\ldots ,\sigma (n^{h_n})).
\end{eqnarray}
Here $g$ is the gauge coupling $\frac{g^2}{4\pi} = \alpha_s$, $p_i$, $h_i$ are the gluon momenta and helicities. $A_n^{tree} (1^{h_1},\ldots , n^{h_n})$ are the partial amplitudes, which contain all the kinematic information. $S_n$ is the set of all permutations of $n$ objects, while $Z_n$ is the subset of cyclic permutations.

Next, the kinematics of scattering amplitudes involving off-shell gluons could be conveniently described using off-shell momentum decomposition typically employed
within high-energy factorization or $k_T$ - factorization approach
\cite{GribovLevinRyskin,CataniCiafaloniHautmann,CollinsEllis,CataniHautmann}. That is, the off-shell gluon momentum is written as
\begin{eqnarray}\label{kT}
k^{\mu} = x p^{\mu} + k_T^{\mu},
\end{eqnarray}
where $p$ is the gluon direction (also known as the off-shell gluon polarization vector), such that $p^2=0$, $p\cdot k = 0$ and $x \in [0,1]$. It is clear, that there is a freedom in such decomposition, which is typically parametrized by an auxiliary light-like four-vector $q^{\mu}$, so that
\begin{eqnarray}
k_T^{\mu} (q) = k^{\mu} - x(q) p^{\mu}\quad \text{with}\quad x(q) = \frac{q\cdot k}{q\cdot p} \;\; \text{and} \;\; q^2 = 0.
\end{eqnarray}
Using the fact, that now $k_T^{\mu}$ is transverse both with respect to $p^\mu$ and $q^\mu$ we can write down off-shell gluon transverse momentum $k_T^{\mu}$ in the basis of two ``polarization'' vectors as \cite{vanHamerenBCFW}:
\begin{eqnarray}
k_T^{\mu} (q) = -\frac{\kappa}{2}\frac{\la p|\gamma^{\mu}|q]}{[pq]}
- \frac{\kappa^{*}}{2}\frac{\la q|\gamma^{\mu}|p]}{\la qp\ra}\quad
\text{with} \quad \kappa = \frac{\la q|\slashed{k}|p]}{\la qp\ra},\;
\kappa^{*} = \frac{\la p|\slashed{k}|q]}{[pq]}.
\end{eqnarray}
It is easy to see, that $k^2 = -\kappa\kappa^{*}$. Moreover, using Schouten identities it could be shown, that both $\kappa$ and $\kappa^{*}$ are independent of auxiliary four-vector $q^{\mu}$ \cite{vanHamerenBCFW}. The on-shell states are described by their on-shell momenta and  polarization vectors as usual. Having said that, let us continue with the gauge invariance of the off-shell scattering amplitudes.

It is well known, that off-shell scattering amplitudes are gauge dependent in general. To insure gauge invariance one typically needs to add additional non-standard contributions, that is those, which are not calculable from standard QCD Feynman rules. The first rigorous consideration of the gauge-invariant off-shell scattering amplitudes to our knowledge was performed within the context of Lipatov's effective lagrangian \cite{LipatovEL1,LipatovEL2} used to describe the QCD high-energy scattering in multi-regge kinematics\footnote{See also \cite{KirschnerEL1,KirschnerEL2} for earlier studies of effective action for high-energy QCD scattering.}. Within Lipatov's effective lagrangian approach an off-shell gluon with additional contributions is interpreted as an effective reggeized gluon. Recently, a new manifestly gauge invariant definition of scattering amplitudes with an arbitrary number of off-shell external gluons appeared in  \cite{KotkoWilsonLines}, where off-shell gluons are described in terms of infinite Wilson lines\footnote{Earlier studies  preceding this construction employed eikonal quark lines and Slavnov-Taylor identities \cite{vanHamerenWL1,vanHamerenWL2}.}. For example, the gauge-invariant scattering amplitude with one leg off-shell is given by the following matrix element\footnote{The asterisk denotes an off-shell gluon} \cite{KotkoWilsonLines}
\begin{eqnarray}
\mathcal{A}_{n+1} \left(1,\ldots ,n,(n+1)^*\right) = \la k_1, \epsilon_1, c_1;\ldots
;k_n,\epsilon_n,c_n|\mathcal{W}_p^{c_{n+1}}(k)|0\ra
\end{eqnarray}
of Wilson line operator\footnote{The color generators are normalized as $\mathrm{Tr} (t^a t^b) = \delta^{a b}$}
\begin{eqnarray}\label{WilsonLineOperDef}
\mathcal{W}_p^c(k) = \int d^4 x e^{ix\cdot k} \mathrm{Tr} \left\{
\frac{1}{\pi g} t^c \; \mathcal{P} \exp\left[\frac{ig}{\sqrt{2}}\int_{-\infty}^{\infty}
ds \; p\cdot A_b (x+ sp) t^b\right]
\right\} .
\end{eqnarray}
Here $p$ is the direction of the off-shell gluon, $k$ is its off-shell momentum and $c$ - color index. Similarly $|k_i,\varepsilon_i, c_i\ra$ describes on-shell gluon state with momentum $k_i$, polarization vector $\varepsilon_i$ and color index $c_i$. Colour ordered version of this
object can be obtained via (\ref{ColourOrderedAmplitudeDefinition}). The helicities of on-shell
gluons are not shown.

In the case of  $\mathcal{N}=4$ super Yang-Mills theory we have both scalars and gluons in the adjoint representation of gauge group in addition to the gluons of pure Yang-Mills theory\footnote{see appendix \ref{aA} for $\mathcal{N}=4$ SYM lagrangian and field content.}. To keep track of the components of the on-shell states it is highly convenient to introduce an on-shell superspace \cite{Nair}. For each external on-shell leg we introduce four Grassmann variables $\tilde{\eta}^i_A$ labeled by the $SU(4)$ index $A = 1,2,3,4$ and leg index $i$. This allows us to collect the 16 states (creation/annihilation operators) into $\mathcal{N}=4$ on-shell chiral superfield
\begin{eqnarray}
\Omega_i= g_i^+ + \vlet^i_A\psi_i^A + \frac{1}{2!}\vlet^i_{A}\vlet^i_{B}\phi_i^{AB}
+ \frac{1}{3!}\vlet^i_A\vlet^i_B\vlet^i_C\epsilon^{ABCD}\bar{\psi}_{i,D}
+ \frac{1}{4!}\vlet^i_A\vlet^i_B\vlet^i_C\vlet^i_D\epsilon^{ABCD}g_i^{-},
\end{eqnarray}
where $g^{\pm}$ denote two physical polarizations of gluon, n-particle on-shell state is then given by $|\Omega_{1\ldots n}\rangle=\prod_{i=1}^n \Omega_i|0\rangle$.

The Wilson line operator above could be also supersymmetrized. First of all, there is both chiral \cite{WilsonSmoothChiral1,WilsonSmoothChiral2,WilsonSmoothNonChiral} and non-chiral \cite{WilsonSmoothNonChiral} versions of smooth supersymmetric Maldacena-Wilson loops \cite{MaldacenaWilsonLoops,ReyYeeWilsonLoops}. The  supersymmetric chiral  version of Maldacena-Wilson loop is given by \cite{WilsonSmoothNonChiral}:
\begin{eqnarray}
\mathcal{W}_{MW} = \frac{1}{N_c} \mathrm{Tr}\mathcal{P}\exp
\left(-\oint_Z d\tau (p^{\mu} A_{\mu} + \dot{\theta}^{\alpha} A_{\alpha} + q^i \Phi_i)\right) ,
\end{eqnarray}
where the superpath $Z = \{x^{\mu}, \theta^{\alpha} , q^i \}$ and $q^i$ is constrained by $q^2 = q^iq^i = p^{\mu}p_{\mu}, \; (\mu = 1,\ldots , 4 ,\; i=1,\ldots 6)$, so that $q^i = \sqrt{p^{\mu}p_{\mu}}n^i$ with $n^i$ being a unit vector. In our case $p^{\mu}p_{\mu} = 0$ and the contribution of scalars could be neglected. In this case Maldacena-Wilson loop turns into a light-like supersymmetric Wilson loop \cite{WilsonLoopTwistorSpace,NotesAmplitudesWilsonLoopsDuality,NullWilsonLoopsFullSuperspace}. Its chiral version is written as follows \cite{WilsonLoopTwistorSpace,NotesAmplitudesWilsonLoopsDuality}:
\begin{eqnarray}
\mathcal{W}_{LL} = \frac{1}{N_c} \mathrm{Tr}\mathcal{P}\exp \left(
	-\int \mathcal{A}\right) ,
\end{eqnarray}
where
\begin{align}
\mathcal{A} = & \frac{d x}{2} \left(
A + \bar{\psi}_A\theta^A + \frac{1}{2} D\phi_{A B}\theta^A\theta^B
- \frac{1}{3!}\varepsilon_{ABCD} D\psi^A\theta^B\theta^C\theta^D
+ \frac{1}{4!}\varepsilon_{ABCD} DF\theta^A\theta^B\theta^C\theta^D + \ldots
\right) \nonumber \\
& + d\theta^A \left(-\frac{1}{2}\phi_{AB}\theta^B +
\frac{1}{3}\varepsilon_{ABCD}\psi^B\theta^C\theta^D +
\frac{1}{6}\varepsilon_{ABCD}F\theta^B\theta^C\theta^D + \ldots \right) .
\end{align}
Here the summation over dotted and undotted Weyl indexes is assumed. The factor $1/2$ comes from the substitution $dx^{\mu}A_{\mu} \to \frac{d x_{\alpha\dotalpha}}{2}A^{\alpha\dotalpha}$.
\begin{figure}[h]
	\centering
\begin{eqnarray}
 \begin{tikzpicture}[baseline={($(n1.base) - (0,0.05)$)},transform shape, scale=1]
  \node[left] (n1) at (-0.6,0) {$\mu$};
  \node[right] (n2) at (0.6,0) {$\nu$};
  \draw[gluon] (n1) -- (n2);
 \end{tikzpicture}
 = \frac{-g^{\mu\nu}}{k^2}, \quad
  \begin{tikzpicture}[baseline={($(n1.base) - (0,0.1)$)},transform shape, scale=1]
  \node[left] (n1) at (-0.6,0) {};
  \node[right] (n2) at (0.6,0) {};
  \draw[gluon] (n1) -- (n2);
  \draw[black,very thick] (n1) -- (n2);
  \end{tikzpicture}
  = \frac{1}{2 p\cdot k}, \quad
\begin{tikzpicture}[baseline={($(n1.base) - (0,0.1)$)},transform shape, scale=1]
\node[left] (n1) at (-0.6,0) {};
\node[right] (n2) at (0.6,0) {};
\draw[black,very thick] (n1) -- (n2);
\end{tikzpicture} = \frac{1}{\slashed{k}}\quad
\begin{tikzpicture}[baseline={($(n1.base) - (0,0.1)$)},transform shape, scale=1]
\node[left] (n1) at (-0.6,0) {};
\node[right] (n2) at (0.6,0) {};
\draw[dashed,very thick] (n1) -- (n2);
\end{tikzpicture} = \frac{1}{k^2} ,
\nonumber
\end{eqnarray}
\begin{eqnarray}
\begin{tikzpicture}[baseline={($(c1.base) - (0,0.5)$)},transform shape, scale=1]
\coordinate (c1) at (-1,0);
\coordinate (c2) at (0,0);
\coordinate (c3) at (1,0);
\node[below] (n1) at (0,-1) {$\mu$};
\draw[gluon] (c1) -- (c2);
\draw[black,very thick] (c1) -- (c2);
\draw[fill] (c2) circle [radius=0.05];
\draw[gluon] (c2) -- (c3);
\draw[black,very thick] (c2) -- (c3);
\draw[gluon] (c2) -- (n1);
\end{tikzpicture}
= \sqrt{2} p^{\mu}, \quad\quad
\begin{tikzpicture}[baseline={($(n1.base) - (0,0.8)$)},transform shape, scale=1]
\node[above] (n1) at (0,0) {$\mu$};
\coordinate (n2) at (0,-0.8);
\draw[fill] (n2) circle [radius=0.05];
\node[right] (n3) at ($(n2)+(-45:0.8)$) {};
\node[left] (n4) at ($(n2)+(225:0.8)$) {};
\draw[gluon] (n1) -- (n2);
\draw[black,very thick] (n2) -- (n3);
\draw[black,very thick] (n2) -- (n4);
\end{tikzpicture}
= \frac{1}{\sqrt{2}}\sigma^{\mu}, \quad\quad
\begin{tikzpicture}[baseline={($(n1.base) - (0,0.8)$)},transform shape, scale=1]
\node[above] (n1) at (0,0) {$\mu$};
\coordinate (n2) at (0,-0.8);
\draw[fill] (n2) circle [radius=0.05];
\node[right] (n3) at ($(n2)+(-45:0.8)$) {2};
\node[left] (n4) at ($(n2)+(225:0.8)$) {3};
\draw[gluon] (n1) -- (n2);
\draw[dashed,very thick] (n2) -- (n3);
\draw[dashed,very thick] (n2) -- (n4);
\end{tikzpicture}
= \frac{1}{\sqrt{2}}(p_2-p_3)^{\mu} ,
\nonumber
\end{eqnarray}
\begin{eqnarray}
\begin{tikzpicture}[baseline={($(n1.base) - (0,0.8)$)},transform shape, scale=1]
\node[above] (n1) at (0,0) {1};
\coordinate (n2) at (0,-0.8);
\draw[fill] (n2) circle [radius=0.05];
\node[right] (n3) at ($(n2)+(-45:0.8)$) {2};
\node[left] (n4) at ($(n2)+(225:0.8)$) {3};
\draw[gluon] (n1) -- (n2);
\draw[gluon] (n2) -- (n3);
\draw[gluon] (n2) -- (n4);
\end{tikzpicture}
= \frac{1}{\sqrt{2}}\left[
(k_1-k_2)^{\mu_3}g^{\mu_1\mu_2} + (k_2-k_3)^{\mu_1}g^{\mu_2\mu_3}
+ (k_3-k_1)^{\mu_2}g^{\mu_3\mu_1}
\right] . \nonumber
\end{eqnarray}
\caption{Color ordered Feynman rules from Wilson lines expansion together with $\mathcal{N}=4$ SYM propagators and 3-point vertexes. A coil denotes ordinary gluon, a coil crossed by line stands for off-shell gluon coming from Wilson line, solid line denotes fermion and dashed line stands for scalar.}
\label{fig:feynman_rules}
\end{figure}

In the present paper we however decided to keep off-shell gluon and corresponding light-like Wilson line non-supersymmetric, while treating the on-shell states in a supersymmetric fashion. The explicit calculations of different 3-point off-shell component amplitudes (interactions of the off-shell gluon with on-shell gluons, fermions and scalars) with the use of Feynman rules in Fig. \ref{fig:feynman_rules}  showed that the 3-point amplitude with one off-shell gluon and two other supersymmetric on-shell states could be written in the following form:
\begin{eqnarray}
A_{2,2+1}^*(1^*,2,3) = \frac{1}{\kappa_1^*}\prod_{A=1}^4\frac{\partial}{\partial\vlet_p^A} \left[\frac{
	\delta^4(k+\vll_2\vlt_2+\vll_3\vlt_3)
	\delta^8(\vll_p\vlet_p+\vll_2\vlet_2+\vll_3\vlet_3)
}{\abr{p2}\abr{23}\abr{3p}} \right].
\end{eqnarray}
The momenta of on-shell states are given by $k_i = \vll_i\vlt_i$, while for the momentum of the off-shell gluon denoted by  $k$ we use  decomposition (\ref{kT}). The off-shell gluon direction $p$ and axillary ``momentum'' (four-vector) $q$  are given by $p=\lambda_p\tilde{\lambda}_p$ and $q=\lambda_q\tilde{\lambda}_q$ light like four-vectors. We will label corresponding spinors as $\lambda_i\equiv|i\ra$, $\tilde{\lambda}_i\equiv[i|$, $\lambda_p\equiv|p\ra$, $\tilde{\lambda}_p\equiv[p|$ and $\lambda_q\equiv|q\ra$, $\tilde{\lambda}_q\equiv[q|$. Also to denote helicities of the amplitudes the following notation will be used. We will call amplitude with $n$ on-shell particles
with overall helicity $\lambda_{\Sigma}=n+2-2k$ as $\mbox{N}^{k-2}\mbox{MHV}$ amplitude and refer to it as $A_{k,n+1}^*$. For on-shell amplitudes the standard notation will be used, that is
$A_{k,n}$ will denote $\mbox{N}^{k-2}\mbox{MHV}$ amplitude.
In what follows to simplify notation we will however drop the derivatives over the Grassmann variables corresponding to the off-shell gluon and just write
\begin{eqnarray}
A_{2,2+1}^*(1^*,2,3) = \frac{1}{\kappa_1^*}\frac{
	\delta^4(k+\vll_2\vlt_2+\vll_3\vlt_3)
	\delta^8(\vll_p\vlet_p+\vll_2\vlet_2+\vll_3\vlet_3)
}{\abr{p2}\abr{23}\abr{3p}}.
\end{eqnarray}
Below we will also always drop the indexes of $\kappa^*$ and $\kappa$, as the amplitudes we are going to consider in this paper will have only one off-shell gluon.

\section{Off-shell BCFW recursion in $\mathcal{N}=4$ SYM}
\label{offshellBCFW}

The off-shell BCFW recursion for gauge invariant gluon off-shell scattering amplitudes with an arbitrary number of off-shell gluons was worked out in \cite{vanHamerenBCFW}. Here, we will remind the reader the main results of \cite{vanHamerenBCFW}, comment on supersymmetric extension of the off-shell BCFW and perform some example calculations, which will be later compared with the results obtained from our Grassmannian representation.

The BCFW recursion \cite{BCFW1,BCFW2} is based on the observation, that a contour integral of an analytical function $f$ vanishing at infinity equals to zero, that is
\begin{eqnarray}
\oint\frac{d z}{2\pi i}\frac{f (z)}{z} = 0 .
\end{eqnarray}
and the integration contour expands to infinity. If $f$ is a rational function of a complex variable we have
\begin{eqnarray}
f (0) = - \sum_i \frac{\text{res}_i f(z)}{z_i} , \label{fpolessum}
\end{eqnarray}
where the sum is over all poles of $f$ and $\text{res}_i f(z)$ is a residue of $f$ at pole $z_i$. In the original on-shell BCFW recursion the $z$-dependence of scattering amplitude is obtained by a $z$-dependent shift of particle's momenta. Similarly, the off-shell gluon BCFW recursion of \cite{vanHamerenBCFW} is formulated using a shift of momenta for two external gluons $i$ and $j$ with a vector
\begin{eqnarray}
e^{\mu} = \frac{1}{2}\la i|\gamma^{\mu}|j] ,\qquad p_i\cdot e = p_j\cdot e = e\cdot e = 0 ,
\end{eqnarray}
so that
\begin{align}
\hat{k}_i^{\mu}(z) &\equiv  k_i^{\mu} + z e^{\mu} = x_i (p_j) p_i^{\mu}
- \frac{\kappa_i - [i j] z}{2}\frac{\la i|\gamma^{\mu} |j]}{[i j]}
- \frac{\kappa_i^{*}}{2}\frac{\la j|\gamma^{\mu}|i]}{\la j i\ra}\; , \\
\hat{k}_j^{\mu}(z) &\equiv k_j^{\mu} - z e^{\mu} = x_j (p_i) p_j^{\mu}
- \frac{\kappa_j}{2}\frac{\la j|\gamma^{\mu} |i]}{[j i]}
- \frac{\kappa_j^{*} + \la i j\ra z}{2}\frac{\la i|\gamma^{\mu} |j]}{\la i j\ra} .
\end{align}
This shift does not violate momentum conservation and we still have $p_i\cdot \hat{k}_i (z) = 0$ and $p_j\cdot\hat{k}_j (z) = 0$. We would like to note, that the overall effect of shifting momenta is that the values of $\kappa_i$ and $\kappa_j^{*}$ shift while $\kappa_i^{*}$ and $\kappa_j$ are not effected. In the on-shell limit this shift corresponds to the usual $[i j\ra$ BCFW shift. Note also, that we could also have chosen another shift vector $e^{\mu} = \frac{1}{2}\la j|\gamma^{\mu}| i]$ and shift $\kappa_i^{*}$ and $\kappa_j$ instead. The off-shell amplitudes we consider in this paper do also have a correct large $z$ ($z\to\infty$) behavior \cite{vanHamerenBCFW}, so that we should not worry about boundary terms at infinity.

The sum over the poles (\ref{fpolessum}) for $z$-dependent off-shell gluon  scattering amplitude is given by the following graphical representation\footnote{We are considering color ordered scattering amplitudes and without loss of generality may use shift of two adjacent legs $1$ and $n$.} \cite{vanHamerenBCFW}:
\begin{eqnarray}
\begin{tikzpicture}[baseline={($(nc.base) - (0,0)$)},transform shape, scale=0.6]
\node (nc) at (0,0) {};
\node[below] (n1) at (-1.5,-1) {1};
\node[below] (nn) at (1.5,-1) {$n$};
\node[left] (n2) at (-1.5,0.2) {2};
\node[right] (nnm1) at (1.5,0.2) {$n-1$};
\node[right] (d1) at (-1,1) {};
\node[above] (d2) at (0,1.2) {};
\node[left] (d3) at (1,1) {};
\draw[double, thick] (nc) -- (n1);
\draw[double, thick] (nc) -- (nn);
\draw[double, thick] (nc) -- (n2);
\draw[double, thick] (nc) -- (nnm1);
\draw[very thick] (nc) circle [radius=0.6];
\draw[fill,grayn] (nc) circle [radius=0.6];
\draw[fill] (d1) circle [radius=0.05];
\draw[fill] (d2) circle [radius=0.05];
\draw[fill] (d3) circle [radius=0.05];
\end{tikzpicture}
= \sum_{i=2}^{n-2}\sum_{h} \mathbb{A}_{i,h} + \sum_{i=2}^{n-1}\mathbb{B}_i + \mathbb{C} + \mathbb{D},
\end{eqnarray}
where
\begin{gather}
\mathbb{A}_{i,h} =
\begin{tikzpicture}[baseline={($(nc.base) - (0,0.1)$)},transform shape, scale=0.6]
\node (nc) at (0,0) {};
\coordinate (ch) at (1,0);
\node[below] (n1) at (-0.2,-1.5) {$\hat{1}$};
\node[above] (ni) at (-0.2,1.5) {$i$};
\node[above] (nh) at (1,0) {$h$};
\node[left] (d1) at (-0.8,-0.5) {};
\node[left] (d2) at (-1,0) {};
\node[left] (d3) at (-0.8,0.5) {};
\draw[double, thick] (nc) -- (n1);
\draw[double, thick] (nc) -- (ni);
\draw[very thick] (nc) -- (ch);
\draw[very thick] (nc) circle [radius=0.6];
\draw[fill,grayn] (nc) circle [radius=0.6];
\draw[fill] (d1) circle [radius=0.05];
\draw[fill] (d2) circle [radius=0.05];
\draw[fill] (d3) circle [radius=0.05];
\end{tikzpicture}
\; \frac{1}{k_{1,i}^2} \;
\begin{tikzpicture}[baseline={($(nc.base) - (0,0.1)$)},transform shape, scale=0.6]
\node (nc) at (0,0) {};
\coordinate (ch) at (-1,0);
\node[below] (nn) at (0.2,-1.5) {$\hat{n}$};
\node[above] (nip1) at (0.2,1.5) {$i+1$};
\node[above] (nh) at (-1,0) {$-h$};
\node[right] (d1) at (0.8,-0.5) {};
\node[right] (d2) at (1,0) {};
\node[right] (d3) at (0.8,0.5) {};
\draw[double, thick] (nc) -- (nn);
\draw[double, thick] (nc) -- (nip1);
\draw[very thick] (nc) -- (ch);
\draw[very thick] (nc) circle [radius=0.6];
\draw[fill,grayn] (nc) circle [radius=0.6];
\draw[fill] (d1) circle [radius=0.05];
\draw[fill] (d2) circle [radius=0.05];
\draw[fill] (d3) circle [radius=0.05];
\end{tikzpicture}
\qquad
\mathbb{B}_i =
 \begin{tikzpicture}[baseline={($(nc.base) - (0,0.1)$)},transform shape, scale=0.6]
 \node (nc) at (0,0) {};
 \node[below] (n1) at (-0.2,-1.5) {$\hat{1}$};
 \node[above] (ni) at (-0.2,1.5) {$i$};
 \node[left] (d1) at (-0.8,-0.5) {};
 \node[left] (d2) at (-1,0) {};
 \node[left] (d3) at (-0.8,0.5) {};
 \draw[double, thick] (nc) -- (n1);
 \draw[double, thick] (nc) -- (ni);
 \draw[very thick] (nc) circle [radius=0.6];
 \draw[fill,grayn] (nc) circle [radius=0.6];
 \draw[fill] (d1) circle [radius=0.05];
 \draw[fill] (d2) circle [radius=0.05];
 \draw[fill] (d3) circle [radius=0.05];
 \end{tikzpicture}
 \;\; \frac{1}{2 p_i\cdot k_{i,n}} \;\;
 \begin{tikzpicture}[baseline={($(nc.base) - (0,0.1)$)},transform shape, scale=0.6]
 \node (nc) at (0,0) {};
 \node[below] (nn) at (0.2,-1.5) {$\hat{n}$};
 \node[above] (ni) at (0.2,1.5) {$i$};
 \node[right] (d1) at (0.8,-0.5) {};
 \node[right] (d2) at (1,0) {};
 \node[right] (d3) at (0.8,0.5) {};
 \draw[double, thick] (nc) -- (nn);
 \draw[double, thick] (nc) -- (ni);
 \draw[very thick] (nc) circle [radius=0.6];
 \draw[fill,grayn] (nc) circle [radius=0.6];
 \draw[fill] (d1) circle [radius=0.05];
 \draw[fill] (d2) circle [radius=0.05];
 \draw[fill] (d3) circle [radius=0.05];
 \end{tikzpicture} \nonumber \\ \nonumber \\
 \mathbb{C} = \frac{1}{\kappa_1} \;
 \begin{tikzpicture}[baseline={($(nc.base) - (0,0)$)},transform shape, scale=0.6]
 \node (nc) at (0,0) {};
 \node[below] (n1) at (-1.5,-1) {$\hat{1}$};
 \node[below] (nn) at (1.5,-1) {$\hat{n}$};
 \node[left] (n2) at (-1.5,0.2) {2};
 \node[right] (nnm1) at (1.5,0.2) {$n-1$};
 \node[right] (d1) at (-1,1) {};
 \node[above] (d2) at (0,1.2) {};
 \node[left] (d3) at (1,1) {};
 \draw[very thick] (nc) -- (n1);
 \draw[double, thick] (nc) -- (nn);
 \draw[double, thick] (nc) -- (n2);
 \draw[double, thick] (nc) -- (nnm1);
 \draw[very thick] (nc) circle [radius=0.6];
 \draw[fill,grayn] (nc) circle [radius=0.6];
 \draw[fill] (d1) circle [radius=0.05];
 \draw[fill] (d2) circle [radius=0.05];
 \draw[fill] (d3) circle [radius=0.05];
 \end{tikzpicture}
 \qquad
  \mathbb{D} = \frac{1}{\kappa_n^{*}} \;
  \begin{tikzpicture}[baseline={($(nc.base) - (0,0)$)},transform shape, scale=0.6]
  \node (nc) at (0,0) {};
  \node[below] (n1) at (-1.5,-1) {$\hat{1}$};
  \node[below] (nn) at (1.5,-1) {$\hat{n}$};
  \node[left] (n2) at (-1.5,0.2) {2};
  \node[right] (nnm1) at (1.5,0.2) {$n-1$};
  \node[right] (d1) at (-1,1) {};
  \node[above] (d2) at (0,1.2) {};
  \node[left] (d3) at (1,1) {};
  \draw[double, thick] (nc) -- (n1);
  \draw[very thick] (nc) -- (nn);
  \draw[double, thick] (nc) -- (n2);
  \draw[double, thick] (nc) -- (nnm1);
  \draw[very thick] (nc) circle [radius=0.6];
  \draw[fill,grayn] (nc) circle [radius=0.6];
  \draw[fill] (d1) circle [radius=0.05];
  \draw[fill] (d2) circle [radius=0.05];
  \draw[fill] (d3) circle [radius=0.05];
  \end{tikzpicture}
\end{gather}
$k_{i,j}^{\mu}\equiv k_i^{\mu} + k_{i+1}^{\mu} + \cdots + k_j^{\mu}$ and $h$ is an internal on-shell gluon helicity or a summation index over all on-shell states in the Nair on-shell supermultiplet in the supersymmetric case discussed later. The $\mathbb{A}_{i,h}$ terms are the usual on-shell BCFW terms, which correspond to the $z$ - poles at which internal gluon propagator $\hat{k}_{1,i}^2 (z)$ vanishes. The $\mathbb{B}_i$ terms refer
to the situation when the denominators of eikonal propagators coming from Wilson line expansion vanish, that is $p_i\cdot\hat{k}_{i,n} (z) = 0$ and $p_i^{\mu}$ is the direction of Wilson line associated with the off-shell gluon. We want to stress, that this term is present only if $i$ labels an off-shell external gluon. The $\mathbb{C}$ term is only present if the gluon number $1$ is off-shell. It appears due to vanishing of the external momentum square $\hat{k}_1^2 (z)$. Similarly, the $\mathbb{D}$ term is due to vanishing of the external momentum square $\hat{k}_n^2 (z)$. It turns our that both these contributions could be calculated in terms of the same scattering diagrams with the off-shell gluons $1$ or $n$ exchanged for the on-shell ones. The helicity of the on-shell gluons depends on type
of the term ($\mathbb{C}$ or $\mathbb{D}$) and the shift vector $e^{\mu}$ ($\frac{1}{2}\la i|\gamma^{\mu}|j]$ or $\frac{1}{2}\la j|\gamma^{\mu}|i]$) used. We refer the reader to \cite{vanHamerenBCFW} for further details. In what follows we will not see $\mathbb{C}$ or $\mathbb{D}$ contributions as
the only shifts we are going to use involve only on-shell legs.

The use of shifts involving only on-shell legs also allows us easily perform the supersymmetrization of the off-shell BCFW recursion introduced in \cite{vanHamerenBCFW}. Indeed, it is easy to see, that the supersymmetric shifts of momenta and corresponding Grassmann variables are given by the on-shell BCFW $[i j\ra$ super-shifts\footnote{These shifts respect both momentum and supermomentum conservation.}:
\begin{eqnarray}
|\hat{i}] = |1] + z|j], \qquad |\hat{j}\ra = |j\ra - z|i\ra, \qquad \hat{\eta}_A^i = \eta_A^i + z\eta_A^j.
\end{eqnarray}
No other spinors or Grassmann variables shift.

Now let us consider the solution of the described BCFW recursion for some
particular cases of the scattering amplitudes with one leg off-shell. In the case of 4-point MHV off-shell amplitude $A^{*}_{2, 3+1}(1^+2^+3^-4^*)$ (4-th leg is off-shell) the  BCFW contributions relevant to $[1 2\ra$ - shift are depicted in Fig. \ref{fig:BCFW_MHV_4}. The $[1 2\ra$ - shift itself is given by the following expressions
\begin{eqnarray}
|\widehat{1} ] = |1] + z |2], \quad |\widehat{2}] = |2], \quad
|\widehat{1}\ra = |1\ra , \quad |\widehat{2}\ra = |2\ra - z |1\ra ,
\end{eqnarray}
that is\footnote{For on-shell states in our convention $k_i = p_i$.
} $\widehat{p}_1 = p_1 + z e$ and $\widehat{p}_2 = p_2 - z e$, where $e^{\mu} = \frac{1}{2}\la 1|\gamma^{\mu} |2]$ . The $A^{* (a)}_{2, 4}$ contribution is given by
\begin{eqnarray}
A^{* (a)}_{2, 4} =  \frac{1}{\kappa^{*}}
\frac{\la\widehat{P} 4\ra^3}{\la p\widehat{1}\ra\la\widehat{1}\widehat{P}\ra}\times\frac{1}{(p_1 + k)^2}
\times - \frac{[\widehat{P}\widehat{2}]^4}{[\widehat{2} 3][3 \widehat{P}][\widehat{P} \widehat{2}]} ,
\end{eqnarray}
where $\widehat P = \widehat{p}_1 + k$. Now noticing, that $\la\widehat{P} 4\ra [\widehat P 2] = \la p 3\ra [3 2] $ and $\la 1\widehat{P}\ra [3 \widehat{P}] = \la 1 2\ra [2 3]$ we get
$
A^{* (a)}_{2,4} = \frac{1}{\kappa^{*}}\frac{\la 3 p\ra^4}{\la 1 2\ra\la 2 3\ra\la 3 p\ra\la p 1\ra} .
$
The $A^{* (b)}_{2,4}$ contribution on the other hand is given by
\begin{eqnarray}
A^{* (b)}_{2,4} = \frac{\la q|\gamma_{\mu}|\widehat{1}]}{\sqrt{2}\la q\widehat{1}\ra} \sqrt{2} p^{\mu} \frac{-1}{2p\cdot p_1} \;\; \begin{tikzpicture}[baseline={($(nc.base) - (0,0.1)$)},transform shape, scale=0.6]
\node (nc) at (0,0) {};
\node[below] (nn) at (0.5,-1.5) {$\bf 4^{*}$};
\node[above] (ni) at (0.5,1.5) {$\bf\widehat{2}^{+}$};
\node[right] (nr) at (1.5,0) {$\bf 3^{-}$};
\draw[very thick] (nc) -- (nn);
\draw[very thick] (nc) -- (ni);
\draw[very thick] (nc) -- (nr);
\draw[very thick] (nc) circle [radius=0.4];
\draw[fill,grayn] (nc) circle [radius=0.4];
\end{tikzpicture}
\;\; = - \frac{\la q p\ra [p \widehat{1}]}{\la q 1\ra\la p 1\ra [1 p]} \;\;
\begin{tikzpicture}[baseline={($(nc.base) - (0,0.1)$)},transform shape, scale=0.6]
\node (nc) at (0,0) {};
\node[below] (nn) at (0.5,-1.5) {$\bf 4^{*}$};
\node[above] (ni) at (0.5,1.5) {$\bf\widehat{2}^{+}$};
\node[right] (nr) at (1.5,0) {$\bf 3^{-}$};
\draw[very thick] (nc) -- (nn);
\draw[very thick] (nc) -- (ni);
\draw[very thick] (nc) -- (nr);
\draw[very thick] (nc) circle [radius=0.4];
\draw[fill,grayn] (nc) circle [radius=0.4];
\end{tikzpicture}
\end{eqnarray}
where $q$ is the polarization reference momentum for the first gluon. This contribution however turns out to be zero as ($z = -\frac{[1 p]}{[2 p]}$ from $p\cdot \widehat{p}_1 = 0$):
\begin{eqnarray}
|\widehat{1} ] = |1] - \frac{[1 p]}{[2 p]}|2] = \frac{[2 1]}{[2 p]}
\left\{\frac{|1][2|}{[2 1]} - \frac{|2] [1|}{[2 1]} \right\} | p] = \frac{[2 1]}{[2 p]} | p] .
\end{eqnarray}
In the last step we have used Schouten identity. So, finally (hereafter in this section we will drop the total momentum
conservation $\delta$ - function $\delta^4\left(\lambda_1\tilde{\lambda}_1+\ldots+\lambda_n\tilde{\lambda}_n+k\right)$)
\begin{eqnarray}
A^{*}_{2,3+1}(1^+2^+3^-4^*) = \frac{1}{\kappa^{*}}\frac{\la 3 p\ra^4}{\la 1 2\ra\la 2 3\ra\la 3 p\ra\la p 1\ra} .
\end{eqnarray}
Using this result we can also immediately write down the answer for anti-$\mbox{MHV}$ four
point amplitude
\begin{eqnarray}\label{NMHV4component}
A^{*}_{2,3+1}(1^-2^-3^+4^*) = \frac{1}{\kappa}\frac{[ 3 p]^4}{[1 2][ 2 3][ 3 p][p 1]} .
\end{eqnarray}
\begin{figure}[h]
\centering
\begin{gather}
A^{* (a)}_{2,4} =
\begin{tikzpicture}[baseline={($(nc.base) - (0,0.1)$)},transform shape, scale=0.6]
\node (nc) at (0,0) {};
\coordinate (ch) at (1,0);
\node[below] (n1) at (-0.5,-1.5) {$\bf 4^{*}$};
\node[above] (ni) at (-0.5,1.5) {$\bf\widehat{1}^{+}$};
\node[above] (nh) at (1,0) {$-$};
\draw[double, thick] (nc) -- (n1);
\draw[very thick] (nc) -- (ni);
\draw[very thick] (nc) -- (ch);
\draw[very thick] (nc) circle [radius=0.4];
\draw[fill,grayn] (nc) circle [radius=0.4];
\end{tikzpicture}
\; \frac{1}{k_{2,3}^2} \;
\begin{tikzpicture}[baseline={($(nc.base) - (0,0.1)$)},transform shape, scale=0.6]
\node (nc) at (0,0) {};
\coordinate (ch) at (-1,0);
\node[below] (nn) at (0.5,-1.5) {$\bf 3^{-}$};
\node[above] (nip1) at (0.5,1.5) {$\bf\widehat{2}^{+}$};
\node[above] (nh) at (-1,0) {$+$};
\draw[very thick] (nc) -- (nn);
\draw[very thick] (nc) -- (nip1);
\draw[very thick] (nc) -- (ch);
\draw[very thick] (nc) circle [radius=0.4];
\draw[fill,grayn] (nc) circle [radius=0.4];
\end{tikzpicture}
\qquad
A_{2,4}^{* (b)} =
\begin{tikzpicture}[baseline={($(nc.base) - (0,0.1)$)},transform shape, scale=0.6]
\node (nc) at (0,0) {};
\node[below] (n1) at (-0.5,-1.5) {$\bf 4^{*}$};
\node[above] (ni) at (-0.5,1.5) {$\bf\widehat{1}^{+}$};
\draw[double, thick] (nc) -- (n1);
\draw[very thick] (nc) -- (ni);
\draw[very thick] (nc) circle [radius=0.4];
\draw[fill,grayn] (nc) circle [radius=0.4];
\end{tikzpicture}
\;\; \frac{-1}{2 p\cdot k_{2,3}} \;\;
\begin{tikzpicture}[baseline={($(nc.base) - (0,0.1)$)},transform shape, scale=0.6]
\node (nc) at (0,0) {};
\node[below] (nn) at (0.5,-1.5) {$\bf 4^{*}$};
\node[above] (ni) at (0.5,1.5) {$\bf\widehat{2}^{+}$};
\node[right] (nr) at (1.5,0) {$\bf 3^{-}$};
\draw[very thick] (nc) -- (nn);
\draw[very thick] (nc) -- (ni);
\draw[very thick] (nc) -- (nr);
\draw[very thick] (nc) circle [radius=0.4];
\draw[fill,grayn] (nc) circle [radius=0.4];
\end{tikzpicture}
\nonumber
\end{gather}
\caption{BCFW contributions for $A^{*}_{2,3+1}(1^+2^+3^-4^*)$ for the case of $[1,2\ra$ shift}
\label{fig:BCFW_MHV_4}
\end{figure}
In the case of NMHV off-shell amplitude with $5$ legs $A^{*}_{3,4+1}(1^+2^+3^-4^-5^*)$ corresponding BCFW contributions could be found in Fig. \ref{fig:BCFW_NMHV_5}. The contribution $A^{* (c)}_{3,5}$ is zero for the same reason the similar contribution turns out to be zero in the on-shell BCFW recursion.  On the other hand, contribution $A^{* (d)}_{3, 5}$ is zero for the same reason as $A^{* (b)}_{2, 4}$ contribution in previous example.
\begin{figure}[h]
	\centering
	\begin{gather}
	A^{* (a)}_{3,5} =
	\begin{tikzpicture}[baseline={($(nc.base) - (0,0.1)$)},transform shape, scale=0.6]
	\node (nc) at (0,0) {};
	\coordinate (ch) at (1,0);
	\node[below] (n1) at (-0.5,-1.5) {$\bf 5^{*}$};
	\node[above] (ni) at (-0.5,1.5) {$\bf\widehat{1}^{+}$};
	\node[above] (nh) at (1,0) {$-$};
	\draw[double, thick] (nc) -- (n1);
	\draw[very thick] (nc) -- (ni);
	\draw[very thick] (nc) -- (ch);
	\draw[very thick] (nc) circle [radius=0.4];
	\draw[fill,grayn] (nc) circle [radius=0.4];
	\end{tikzpicture}
	\; \frac{1}{k_{2,4}^2} \;
	\begin{tikzpicture}[baseline={($(nc.base) - (0,0.1)$)},transform shape, scale=0.6]
	\node (nc) at (0,0) {};
	\coordinate (ch) at (-1,0);
	\node[below] (nn) at (0.5,-1.5) {$\bf 4^{-}$};
	\node[above] (nip1) at (0.5,1.5) {$\bf\widehat{2}^{+}$};
	\node[above] (nh) at (-1,0) {$+$};
    \node[right] (nr) at (1.5,0) {$\bf 3^{-}$};
	\draw[very thick] (nc) -- (nn);
	\draw[very thick] (nc) -- (nip1);
	\draw[very thick] (nc) -- (nr);
	\draw[very thick] (nc) -- (ch);
	\draw[very thick] (nc) circle [radius=0.4];
	\draw[fill,grayn] (nc) circle [radius=0.4];
	\end{tikzpicture}
	\qquad
	A_{3,5}^{* (b)} =
	\begin{tikzpicture}[baseline={($(nc.base) - (0,0.1)$)},transform shape, scale=0.6]
	\node (nc) at (0,0) {};
	\coordinate (ch) at (1,0);
	\node[below] (n1) at (-0.5,-1.5) {$\bf 4^{-}$};
	\node[above] (ni) at (-0.5,1.5) {$\bf\widehat{1}^{+}$};
	\node[above] (nh) at (1,0) {$-$};
	\node[left] (nl) at (-1.5,0) {$\bf 5^{*}$};
	\draw[very thick] (nc) -- (n1);
	\draw[double,thick] (nc) -- (nl);
	\draw[very thick] (nc) -- (ni);
	\draw[very thick] (nc) -- (ch);
	\draw[very thick] (nc) circle [radius=0.4];
	\draw[fill,grayn] (nc) circle [radius=0.4];
	\end{tikzpicture}
	\;\; \frac{1}{k_{2,3}^2} \;\;
    \begin{tikzpicture}[baseline={($(nc.base) - (0,0.1)$)},transform shape, scale=0.6]
    \node (nc) at (0,0) {};
    \coordinate (ch) at (-1,0);
    \node[below] (nn) at (0.5,-1.5) {$\bf 3^{-}$};
    \node[above] (nip1) at (0.5,1.5) {$\bf\widehat{2}^{+}$};
    \node[above] (nh) at (-1,0) {$+$};
    \draw[very thick] (nc) -- (nn);
    \draw[very thick] (nc) -- (nip1);
    \draw[very thick] (nc) -- (ch);
    \draw[very thick] (nc) circle [radius=0.4];
    \draw[fill,grayn] (nc) circle [radius=0.4];
    \end{tikzpicture} \nonumber \\
    A_{3,5}^{* (c)} =
    \begin{tikzpicture}[baseline={($(nc.base) - (0,0.1)$)},transform shape, scale=0.6]
    \node (nc) at (0,0) {};
    \coordinate (ch) at (1,0);
    \node[below] (n1) at (-0.5,-1.5) {$\bf 4^{-}$};
    \node[above] (ni) at (-0.5,1.5) {$\bf\widehat{1}^{+}$};
    \node[above] (nh) at (1,0) {$+$};
    \node[left] (nl) at (-1.5,0) {$\bf 5^{*}$};
    \draw[very thick] (nc) -- (n1);
    \draw[double,thick] (nc) -- (nl);
    \draw[very thick] (nc) -- (ni);
    \draw[very thick] (nc) -- (ch);
    \draw[very thick] (nc) circle [radius=0.4];
    \draw[fill,grayn] (nc) circle [radius=0.4];
    \end{tikzpicture}
    \;\; \frac{1}{k_{2,3}^2} \;\;
    \begin{tikzpicture}[baseline={($(nc.base) - (0,0.1)$)},transform shape, scale=0.6]
    \node (nc) at (0,0) {};
    \coordinate (ch) at (-1,0);
    \node[below] (nn) at (0.5,-1.5) {$\bf 3^{-}$};
    \node[above] (nip1) at (0.5,1.5) {$\bf\widehat{2}^{+}$};
    \node[above] (nh) at (-1,0) {$-$};
    \draw[very thick] (nc) -- (nn);
    \draw[very thick] (nc) -- (nip1);
    \draw[very thick] (nc) -- (ch);
    \draw[very thick] (nc) circle [radius=0.4];
    \draw[fill,grayn] (nc) circle [radius=0.4];
    \end{tikzpicture} \qquad
    A_{3,5}^{* (d)} =
    \begin{tikzpicture}[baseline={($(nc.base) - (0,0.1)$)},transform shape, scale=0.6]
    \node (nc) at (0,0) {};
    \node[below] (n1) at (-0.5,-1.5) {$\bf 5^{*}$};
    \node[above] (ni) at (-0.5,1.5) {$\bf\widehat{1}^{+}$};
    \draw[double, thick] (nc) -- (n1);
    \draw[very thick] (nc) -- (ni);
    \draw[very thick] (nc) circle [radius=0.4];
    \draw[fill,grayn] (nc) circle [radius=0.4];
    \end{tikzpicture}
    \;\; \frac{-1}{2 p\cdot k_{2,4}} \;\;
    \begin{tikzpicture}[baseline={($(nc.base) - (0,0.1)$)},transform shape, scale=0.6]
    \node (nc) at (0,0) {};
    \node[below] (nn) at (0.5,-1.5) {$\bf 5^{*}$};
    \node[above] (ni) at (0.5,1.5) {$\bf\widehat{2}^{+}$};
    \node[right] (nr1) at (1.5,0.5) {$\bf 3^{-}$};
    \node[right] (nr2) at (1.5,-0.5) {$\bf 4^{-}$};
    \draw[very thick] (nc) -- (nn);
    \draw[very thick] (nc) -- (ni);
    \draw[very thick] (nc) -- (nr1);
    \draw[very thick] (nc) -- (nr2);
    \draw[very thick] (nc) circle [radius=0.4];
    \draw[fill,grayn] (nc) circle [radius=0.4];
    \end{tikzpicture}
    \nonumber
	\end{gather}
	\caption{BCFW contributions for $A^{*}_{3,4+1}(1^+2^+3^-4^-5^*)$ for the case of $[1,2\ra$ shift}
	\label{fig:BCFW_NMHV_5}
\end{figure}
The $A^{* (a)}_{3,4+1}$ contribution is given by
\begin{eqnarray}
A^{* (a)}_{3,4+1} = \frac{1}{\kappa^{*}}
\frac{\la\widehat{P} p\ra^3}{\la p\widehat{1}\ra\la\widehat{1}\widehat{P}\ra}\times
\frac{1}{(p_2 + p_3 + p_4)^2}\times \frac{\la 3 4\ra^4}{\la\widehat{2} 3\ra\la 3 4\ra\la 4\widehat{P}\ra\la\widehat{P}\widehat{2}\ra} ,
\end{eqnarray}
where $\widehat{P} = \widehat{p}_1 + k$ and $k$ is off-shell momentum of 5-th leg. To simplify this expression it is convenient to multiply both its numerator and denominator by $[\widehat{P} 2]^3$. Now using easily derived relations ($z = \frac{p_{2,4}^2}{\la 1|3+4|2]}$ is determined from the condition $(\widehat{p}_1 + k)^2 = 0$)
\begin{gather}
\la\widehat{1}\widehat{P}\ra [\widehat{P} 2] = - \la 1|3+4|2],
\quad \la 4\widehat{P}\ra [\widehat{P} 2] = \la 3 4\ra [3 2], \quad
\la\widehat{P} \widehat{2}\ra [\widehat{P} 2] = \la 3 4\ra [3 4], \nonumber \\
\la\widehat{P} p\ra [\widehat{P} 2] = \la p| 3+4|2] ,\quad \la 5 \widehat{1}\ra = \la p 1\ra,
\quad \la\widehat{2} 3\ra = \la 2 3\ra - z\la 1 3\ra = \la 2 3\ra - \frac{p_{2,4}^2}{\la 1|3+4|2]}\la 1 3\ra \nonumber \\
\end{gather}
the expression for $A^{* (a)}_{3,4+1}$ contribution could be written in the following form
\begin{eqnarray}
A^{* (a)}_{3,4+1} = \frac{1}{\kappa^{*}}
 \frac{\la p| 3+4|2]^3 \la 3 4\ra}{\la p 1\ra [2 3] [3 4] p_{2,4}^2 (\la 1|3+4|2]\la 2 3\ra - p_{2,4}^2 \la 1 3\ra)}.
\end{eqnarray}
This expression can be further simplified to the form
\begin{eqnarray}
A^{* (a)}_{3,4+1} = \frac{1}{\kappa^{*}}
 \frac{\la p| 3+4|2]^3}{\la p 1\ra [2 3] [3 4] p_{2,4}^2 \la 1|2+3|4]}.
\end{eqnarray}
Analogously for the $A^{* (b)}_{3, 4+1}$ contribution we have
\begin{eqnarray}
A^{* (b)}_{3,4+1} = \frac{1}{\kappa} \frac{[p\widehat{1}]^4}{[p 4] [4\widehat{P}][\widehat{P}\widehat{1}][\widehat{1} p]} \times
\frac{1}{(p_2+p_3)^2} \times \frac{[\widehat{2}\widehat{P}]^4}{[\widehat{P} 3][3\widehat{2}][\widehat{2}\widehat{P}]} ,
\end{eqnarray}
where $\widehat{P} = \widehat{p}_1 + p_4 + k$. Again, to simplify this expression, it is convenient to multiply both its numerator and denominator by $\la\widehat{P} 1\ra^3$. Using the relations ($z = \frac{\la 2 3\ra}{\la 1 3\ra}$ is determined from the condition $(\widehat{p}_2 + p_3)^2 = 0$)
\begin{gather}
[2\widehat{P}]\la\widehat{P} 1\ra = - [2 3]\la 3 1\ra, \quad
[4\widehat{P}]\la\widehat{P} 1\ra = \la 1|2+3|4], \quad
[\widehat{P} 3]\la\widehat{P} 1\ra = -\la 1 2\ra [2 3], \nonumber \\
[\widehat{1} p] = [1 p] + z [2 p] =-\frac{\la 3|1+2|p]}{\la 1 3\ra}, \quad
[3 \widehat{2}] = [3 2] , \nonumber \\
\quad [\widehat{P}\widehat{1}]\la\widehat{P} 1\ra = \la 1|2+3|\widehat{1} ] = \la 1|2+3|1] + p_{2,3}^2=p_{1,3}^2,
\end{gather}
we get
\begin{eqnarray}
A^{* (b)}_{3,5} = \frac{1}{\kappa}\frac{\la 3|1+2|p]^3}{[4 p]\la 1|2+3|4]\la 1 2\ra\la 2 3\ra (\la 1|2+3|1]+p_{2,3}^2)},
\end{eqnarray}
which after additional simplifications can be written as
\begin{eqnarray}
A^{* (b)}_{3,5} = \frac{1}{\kappa}\frac{\la 3|1+2|p]^3}{\la 1 2\ra\la 2 3\ra[4 p] p^2_{1,3} \la 1|2+3|4] }.
\end{eqnarray}
Combining all terms together we finally obtain:
\begin{eqnarray}
A^{*}_{3,4+1}(1^+2^+3^-4^-5^*)= \frac{1}{\kappa^{*}}
 \frac{\la p| 3+4|2]^3}{\la p 1\ra [2 3] [3 4] p_{2,4}^2 \la 1|2+3|4]}+\frac{1}{\kappa}\frac{\la 3|1+2|p]^3}{\la 1 2\ra\la 2 3\ra[4 p] p^2_{1,3} \la 1|2+3|4] }.\nonumber\\
\end{eqnarray}
It is interesting to compare this result with $[1,2\ra$ BCFW shift representation of 6 - point on-shell amplitude $A_{3,6}(1^+2^+3^-4^-5^-6^+)$:
\begin{eqnarray}
A_{3,6}(1^+2^+3^-4^-5^-6^+)=
 \frac{\la 5| 3+4|2]^3}{\la 5 6\ra\la 6 1\ra [2 3] [3 4] p_{2,4}^2 \la 1|2+3|4]}+\frac{\la 3|1+2|6]^3}{\la 1 2\ra\la 2 3\ra[4 5][5 6] p^2_{1,3} \la 1|2+3|4] }.\nonumber\\
\end{eqnarray}
This expression could be further supersymmetrized and written as
\begin{eqnarray}\label{NMHV6super}
A_{3,6}= A_{2,6}R_{142}+A_{2,6}R_{153}+A_{2,6}R_{152} ,
\end{eqnarray}
where
\begin{eqnarray}\label{NMHV6}
A_{2,6}R_{152}&=&\frac{\delta^8(q_{1...6})\hat{\delta}^4(126)}
{\langle34\rangle\langle45\rangle[12][16]\langle5|3+4|2]\langle3|4+5|6]p_{3,5}^2},\\
A_{2,6}R_{142}&=&\frac{\delta^8(q_{1...6})\hat{\delta}^4(234)}
{\langle56\rangle\langle16\rangle[43][23]\langle1|5+6|4]\langle5|4+3|2]p_{2,4}^2},\\
A_{2,6}R_{153}&=&\frac{\delta^8(q_{1...6})\hat{\delta}^4(456)}
{\langle12\rangle\langle23\rangle[45][65]\langle1|3+2|4]\langle3|5+4|6]p^2_{1,3}}.
\end{eqnarray}
Here to simplify notation we used abbreviation $\hat{\delta}^4(ijk)\equiv \hat{\delta}^4(\eta_i[jk]+\mbox{perm.})$
Component expression obtained before can be extracted from supersymmetric version as coefficient in front of $\tilde{\eta}_3^4\tilde{\eta}_4^4\tilde{\eta}_5^4$. The third term vanishes for this particular component. Next, each term in (\ref{NMHV6super}) could be associated with a particular residue in the Grassmannian integral considered in next section and will be discussed in more detail later on.

\section{Grassmannian representation for off-shell amplitudes in $\mathcal{N}=4$ SYM}

The Grassmannian representation for the off-shell  amplitudes with one leg off-shell could be obtained in the same way as the Grassmannian representation for the form factors \cite{FormFactorsGrassmanians,SoftTheoremsFormFactors,q2zeroFormFactors}, see also \cite{WilhelmThesis}. It was noticed in \cite{FormFactorsGrassmanians}, that the top-cell diagrams\footnote{In general, there will be several top-cell diagrams for a particular form factor compared to single top-cell diagram in the case of off-shell amplitudes.} for the form factors could be obtained from the top-cell diagrams for amplitudes by applying to them square and merge/unmerge moves until a box appears on the boundary of the corresponding on-shell diagram, which should be then replaced with the corresponding minimal form factor. Graphically, this relation reads\footnote{We have borrowed this nice picture from \cite{FormFactorsGrassmanians}}
\begin{equation}
	\scalebox{0.9}{\(
		\begin{aligned}
		\begin{tikzpicture}[scale=0.8, baseline=-0.7cm]
		\draw (1,1) -- (1,2) -- (2,2) -- (2,1) -- (1,1);
		\draw (1,0) -- (1,1);
		\draw (2,0) -- (2,1);
		\draw (2.5,2.5) -- (2,2);
		\draw (0.5,2.5) -- (1,2);
		\draw (-0.5,-0.75) -- (-0.5,-2.5);
		\draw (1.5,-0.75) -- (1.5,-2.5);
		\draw (2.5,-0.75) -- (2.5,-2.5);
		\draw (3.5,-0.75) -- (3.5,-2.5);
		\node at (-0.5,-2.5-\labelvdist) {$n$};
		\node at (0.5,-2.5-\labelvdist) {$\cdots$};
		\node at (1.5,-2.5-\labelvdist) {$3$};
		\node at (2.5,-2.5-\labelvdist) {$2$};
		\node at (3.5,-2.5-\labelvdist) {$1$};
		\node at (2.5+\labelddist,2.5+\labelddist) {$n+2$};
		\node at (0.5+\labelddist,2.5+\labelddist) {$n+1$};
		\node[dw] at (2,1) {};
		\node[dw] at (1,2) {};
		\node[db] at (1,1) {};
		\node[db] at (2,2) {};
		\node[ellipse, black, fill=grayn, minimum width=4 cm, minimum height=2 cm, draw, inner sep=0pt] at (1.5,-0.75) {};
		\end{tikzpicture}
		\quad
		\longrightarrow
		\quad
		\begin{tikzpicture}[scale=0.8, baseline=-0.7cm]
		\draw[thick,double] (1.5,-0+1.5) -- (1.5,-0.5+1.5); 
		\draw (1.5,-0.5+1.5) -- (2,-\vacuumheight+1.4);
		\draw (1.5,-0.5+1.5) -- (1,-\vacuumheight+1.4);
		\draw (-0.5,-0.75) -- (-0.5,-2.5);
		\draw (1.5,-0.75) -- (1.5,-2.5);
		\draw (2.5,-0.75) -- (2.5,-2.5);
		\draw (3.5,-0.75) -- (3.5,-2.5);
		\node at (-0.5,-2.5-\labelvdist) {$n$};
		\node at (0.5,-2.5-\labelvdist) {$\cdots$};
		\node at (1.5,-2.5-\labelvdist) {$3$};
		\node at (2.5,-2.5-\labelvdist) {$2$};
		\node at (3.5,-2.5-\labelvdist) {$1$};
		\node[ellipse, black, fill=grayn, minimum width=4 cm, minimum height=2 cm, draw, inner sep=0pt] at (1.5,-0.75) {};
		\fill[black!60!white] (1.43,1.95) rectangle (1.56,0.9);
		\end{tikzpicture}
		\quad
		\eqncom
		\end{aligned}
		\)} \label{amplitudeToformfactor}
\end{equation}
where the box at the legs $n+1$ and $n+2$ was replaced for the sake of concreteness. A similar relation of form factor on-shell diagrams to the amplitude on-shell diagrams was obtained in \cite{SoftTheoremsFormFactors,q2zeroFormFactors} based on soft limit procedure. There the corresponding box diagram is deformed by extra soft factor, what makes the box equivalent to the corresponding minimal form factor.  One should check, that the obtained this way top-cell diagrams for off-shell amplitudes do have correct color ordering compared to the case of form factors of color singlet operators. In the latter case we always have correct color ordering. In what follows, after a brief remainder of the on-shell diagrams and Grassmannians, we will present the details of the  derivation of the Grassmannian representation for N$^k$MHV amplitudes with one leg off-shell.

\subsection{On-shell diagrams and Grassmannians}

The on-shell N$^{k-2}$MHV tree level scattering amplitudes or leading singularities of their loop counterparts in the planar $\mathcal{N}=4$ SYM could be written in terms of contour integrals over the Grassmannians $G(k,n)$ \cite{DualitySMatrix}. The Grassmannian $G(k,n)$ is defined as a space of $k$-dimensional planes in $\mathbb{C}^n$, passing through origin, so that its points are given by $k\times n$ matrices $C$ modulo $GL (k)$ transformation related to the basis choice. Thus, for N$^{k-2}$MHV on-shell amplitude we have
\begin{eqnarray}
A_{k,n} = \int_{\Gamma_{k,n}} \frac{d^{k\times n} C}{\text{Vol}[GL(k)]}
\frac{\delta^{k\times 2} (C\cdot\vlt)\delta^{k\times 4} (C\cdot\vlet)\delta^{(n-k)\times 2} (C^{\perp}\cdot\vll)}{(1\cdots k)(2\cdots k+1)\cdots (n\cdots k-1)} , \label{GrassmannianIntegralAmplitudes}
\end{eqnarray}
where $\Gamma_{k,n}$ is the integration contour\footnote{The integration goes not over the all points of the Grassmannian, but only those, which belong to the so called positive Grassmannian $G_{+}(k,n)$ \cite{AmplitudesPositiveGrassmannian}. The latter is a submanifold of $G (k, n)$, such that all consecutive minors for its points are positive. In what follows we will always understand positive Grassmannian $G_{+} (k,n)$ when referring to Grassmannian $G (k,n)$}, that is the prescription describing which particular combinations of consecutive minors of matrix $C$ should vanish when computing residues. Here, $(i_1, \ldots i_k)$ denotes minor corresponding to columns $i_1,\ldots ,i_k$ and $C^{\perp}$ is the orthogonal complement of $C$ fulfilling $C (C^{\perp})^T = 0$ . The $\delta$ - functions in the formula above encode momentum and supermomentum conservation, that is, for example
\begin{equation}
(C\cdot\vlt)^{\dot\alpha}_I=\sum_{i=1}^nC_{Ii}\vlt_i^{\dot\alpha}=0 \quad\text{and}\quad (C^\perp\cdot\vll)^{\alpha}_J=\sum_{i=1}^nC^\perp_{Ji}\vll_i^\alpha=0 \quad\text{imply}\quad
\vll\cdot\vlt=0 .
\end{equation}
and similarly for supermomentum. The appearance of the Grassmannians in the description of on-shell scattering amplitudes was fully understood with the introduction of {\it on-shell diagrams} \cite{AmplitudesPositiveGrassmannian}. These diagrams (graphs) are built though gluing two basic trivalent vertices - "black" and "white", corresponding to three-point MHV $A_{2,3}$ and anti-MHV $A_{1,3}$ amplitudes:
\begin{equation}
\label{eq: amplitude building blocks for on-shell diagrams}
\begin{aligned}
\begin{aligned}
\begin{tikzpicture}[scale=0.8]
\draw (1,1) -- (1,1+0.65);
\draw (1,1) -- (1-0.5,1-0.5);
\draw (1,1) -- (1+0.5,1-0.5);
\node[db] at (1,1) {};
\end{tikzpicture}
\end{aligned}
&=
A_{2,3}=\frac{
	\delta^4(\vll_1\vlt_1+\vll_2\vlt_2+\vll_3\vlt_3)
	\delta^8(\vll_1\vlet_1+\vll_2\vlet_2+\vll_3\vlet_3)
}{\abr{12}\abr{23}\abr{31}} \eqncom\\
\begin{aligned}
\begin{tikzpicture}[scale=0.8]
\draw (1,1) -- (1,1+0.65);
\draw (1,1) -- (1-0.5,1-0.5);
\draw (1,1) -- (1+0.5,1-0.5);
\node[dw] at (1,1) {};
\end{tikzpicture}
\end{aligned}
&=
A_{1,3}
=\frac{
	\delta^4(\vll_1\vlt_1+\vll_2\vlt_2+\vll_3\vlt_3)
	\delta^4(\sbr{12}\vlet_3+\sbr{23}\vlet_1+\sbr{31}\vlet_2)
}{\sbr{12}\sbr{23}\sbr{31}} \eqndot
\end{aligned}
\end{equation}
For a research preceding on-shell diagram formalism see \cite{DualitySMatrix,AllLoopIntegrandN4SYM,GrassmanianOriginDualConformalInvariance,UnificationResiduesGrassmannianDualities}. The on-shell diagrams are used to describe different BCFW terms (Yangian invariants) in the BCFW decomposition of the tree level scattering amplitudes or  integrands in the case of loop amplitudes. The MHV level $k$ and the number of legs (multiplicity) of the amplitude the on-shell graph corresponds to could be related to the numbers of white $n_w$ and black $n_b$ vertexes together with the number of internal lines $n_I$ as
\begin{eqnarray}
n = 3 (n_w + n_b) - 2 n_I , \qquad k = n_w + 2 n_b - n_I .
\end{eqnarray}
In the Grassmannian representation $A_{2,3}$ and $A_{1,3}$ amplitudes are given by an integral over Grassmannians $G(2,3)$ and $G(1,3)$ correspondingly. The gluing procedure then give rise to a larger Grassmannian $G(k,n)$. The number of degrees of freedom $d$ of a general on-shell diagram is  given by the number of its edges minus number of its internal nodes (we subtract $GL(1)$ gauge redundancy associated to every internal node)
\begin{eqnarray}
d = n_I - (n_b + n_w) .
\end{eqnarray}
For a planar\footnote{For nonplanar on-shell diagrams see \cite{NonplanarOnshell} and references therein.} on-shell diagram with $F$ - faces this is equal to $d = F - 1$. The corresponding Grassmannian integral written in terms of graph's degrees of freedom takes so called "$d\log$" form \cite{AmplitudesPositiveGrassmannian}:
\begin{eqnarray}
\oint\frac{f_1}{f_1}\oint \frac{f_2}{f_2}\cdots \oint \frac{f_d}{f_d} \;
\delta^{k\times 2} (C (f_i)\cdot\vlt)\delta^{k\times 4} (C (f_i)\cdot\vlet)\delta^{(n-k)\times 2} (C^{\perp} (f_i)\cdot\vll) ,
\end{eqnarray}
where we used face variables $f_i$\footnote{The face variables are given by products of all edge variables around the faces and are subject to constraint $\Pi_{i=1}^{F} f_i = 1$}, however similar form could be also obtained using $d$ independent edge variables. The $k\times n$ matrix $C$ is expressed in terms of faces or edge variables using so called {\it boundary measurement} operation \cite{TotalPositivityGrassmanniansNetworks}. To do so, one first introduces a {\it perfect matching} $P$, which is a subset of edges in the graph, such that every internal node is the endpoint of exactly one edge in $P$ and external nodes belong to one or no edge in $P$.   In one-to-one correspondence to perfect matching is a {\it perfect orientation}. A perfect orientation is an assignment of special orientation to edges, such that each white vertex has a single incoming arrow and each black vertex has a single outgoing arrow. The edge with a special orientation at each internal node (directed from black to white vertex in our case) is precisely the edge belonging to the perfect matching subset \cite{TotalPositivityGrassmanniansNetworks,BipartiteFieldTheories}. Given a perfect orientation all external nodes are divided into sources and sinks. Then entries of the matrix $C$ are given by \cite{TotalPositivityGrassmanniansNetworks}:
\begin{eqnarray}
C_{ij} (\alpha) = \sum_{\Gamma \; \in \; \{i\to j\}} (-1)^{s_{\Gamma}}
\prod_{e \; \in \; \Gamma}\alpha_e^{\{-1,1\}} ,
\end{eqnarray}
where  $i$ runs over sources, $j$ runs over all external nodes and $\Gamma$ is an oriented path from $i$  to $j$ consistent with perfect orientation. If the edge is traversed in the direction from white to black node\footnote{It is just a convention for assigning edge variables, which could have been chosen differently.}, then the power of edge variable is $1$, and $-1$ when traversing in opposite direction. The $s_{\Gamma}$ in the formula above is the number of sources strictly between nodes $i$ and $j$.

\begin{figure}[h]
	\centering
\begin{gather*}
\begin{tikzpicture}[baseline={($(n1.base) + (0,1)$)},scale=0.8,rotate=0]
\draw (1,1) -- (1,2) -- (2,2) -- (2,1) -- (1,1);
\draw (0.5,0.5) -- (1,1);
\draw (2.5,0.5) -- (2,1);
\draw (2.5,2.5) -- (2,2);
\draw (0.5,2.5) -- (1,2);
\node[dw] at (2,1) {};
\node[dw] at (1,2) {};
\node[db] at (1,1) {};
\node[db] at (2,2) {};
\node (n1) at (2.5+\labelddist,0.5-\labelddist) {1};
\node at (0.5-\labelddist,2.5+\labelddist) {3};
\node at (0.5-\labelddist,0.5-\labelddist) {2};
\node at (2.5+\labelddist,2.5+\labelddist) {4};
\end{tikzpicture}
=
\begin{tikzpicture}[baseline={($(n1.base) + (0,1)$)},scale=0.8,rotate=0]
\draw (1,1) -- (1,2) -- (2,2) -- (2,1) -- (1,1);
\draw (0.5,0.5) -- (1,1);
\draw (2.5,0.5) -- (2,1);
\draw (2.5,2.5) -- (2,2);
\draw (0.5,2.5) -- (1,2);
\node[db] at (2,1) {};
\node[db] at (1,2) {};
\node[dw] at (1,1) {};
\node[dw] at (2,2) {};
\node (n1) at (2.5+\labelddist,0.5-\labelddist) {1};
\node at (0.5-\labelddist,2.5+\labelddist) {3};
\node at (0.5-\labelddist,0.5-\labelddist) {2};
\node at (2.5+\labelddist,2.5+\labelddist) {4};
\end{tikzpicture} \qquad\quad
\begin{tikzpicture}[baseline={($(n1.base) + (0,1)$)},scale=0.8,rotate=0]
\draw (1,1) -- (1,2);
\draw (0.5,0.5) -- (1,1);
\draw (1.5,0.5) -- (1,1);
\draw (1.5,2.5) -- (1,2);
\draw (0.5,2.5) -- (1,2);
\node[db] at (1,1) {};
\node[db] at (1,2) {};
\node (n1) at (1.5+\labelddist,0.5-\labelddist) {1};
\node at (0.5-\labelddist,2.5+\labelddist) {3};
\node at (0.5-\labelddist,0.5-\labelddist) {2};
\node at (1.5+\labelddist,2.5+\labelddist) {4};
\end{tikzpicture}
=
\begin{aligned}
\begin{tikzpicture}[baseline={($(n1.base) + (0,1)$)},scale=0.8,rotate=90]
\draw (0.5,0.5) -- (1,1);
\draw (1.5,0.5) -- (1,1);
\draw (1.5,1.5) -- (1,1);
\draw (0.5,1.5) -- (1,1);
\node[db] at (1,1) {};
\node at (1.5+\labelddist,0.5-\labelddist) {4};
\node at (0.5-\labelddist,1.5+\labelddist) {2};
\node (n1) at (0.5-\labelddist,0.5-\labelddist) {1};
\node at (1.5+\labelddist,1.5+\labelddist) {3};
\end{tikzpicture}
\end{aligned}
=
\begin{tikzpicture}[baseline={($(n1.base) + (0,0.6)$)},scale=0.8,rotate=90]
\draw (1,1) -- (1,2);
\draw (0.5,0.5) -- (1,1);
\draw (1.5,0.5) -- (1,1);
\draw (1.5,2.5) -- (1,2);
\draw (0.5,2.5) -- (1,2);
\node[db] at (1,1) {};
\node[db] at (1,2) {};
\node at (1.5+\labelddist,0.5-\labelddist) {4};
\node at (0.5-\labelddist,2.5+\labelddist) {2};
\node (n1) at (0.5-\labelddist,0.5-\labelddist) {1};
\node at (1.5+\labelddist,2.5+\labelddist) {3};
\end{tikzpicture}
\\
a) \hspace*{6cm} b) \\
\begin{tikzpicture}[baseline={($(n1.base) + (0,0.6)$)},scale=0.8,rotate=90]
\draw (0.5,0.5) -- (1,1);
\draw (1.5,0.5) -- (1,1);
\draw (1.5,2.5) -- (1,2);
\draw (0.5,2.5) -- (1,2);
\draw plot [smooth,tension=2] coordinates {(1,1) (1.2,1.5) (1,2)};
\draw plot [smooth,tension=2] coordinates {(1,1) (0.8,1.5) (1,2)};
\node[db] at (1,1) {};
\node[dw] at (1,2) {};
\node at (1.5+\labelddist,0.5-\labelddist) {4};
\node at (0.5-\labelddist,2.5+\labelddist) {2};
\node (n1) at (0.5-\labelddist,0.5-\labelddist) {1};
\node at (1.5+\labelddist,2.5+\labelddist) {3};
\end{tikzpicture}
=
\begin{tikzpicture}[baseline={($(n1.base) + (0,0.6)$)},scale=0.8,rotate=90]
\draw (1,1) -- (1,2);
\draw (0.5,0.5) -- (1,1);
\draw (1.5,0.5) -- (1,1);
\draw (1.5,2.5) -- (1,2);
\draw (0.5,2.5) -- (1,2);
\node[db] at (1,1) {};
\node[dw] at (1,2) {};
\node at (1.5+\labelddist,0.5-\labelddist) {4};
\node at (0.5-\labelddist,2.5+\labelddist) {2};
\node (n1) at (0.5-\labelddist,0.5-\labelddist) {1};
\node at (1.5+\labelddist,2.5+\labelddist) {3};
\end{tikzpicture} \\
c)
\end{gather*}
\caption{Equivalence moves for on-shell diagrams: a) square move, b) merge/unmerge move for black nodes (similar for white nodes), c) bubble reduction.}
\label{EquivalenceMoves}
\end{figure}
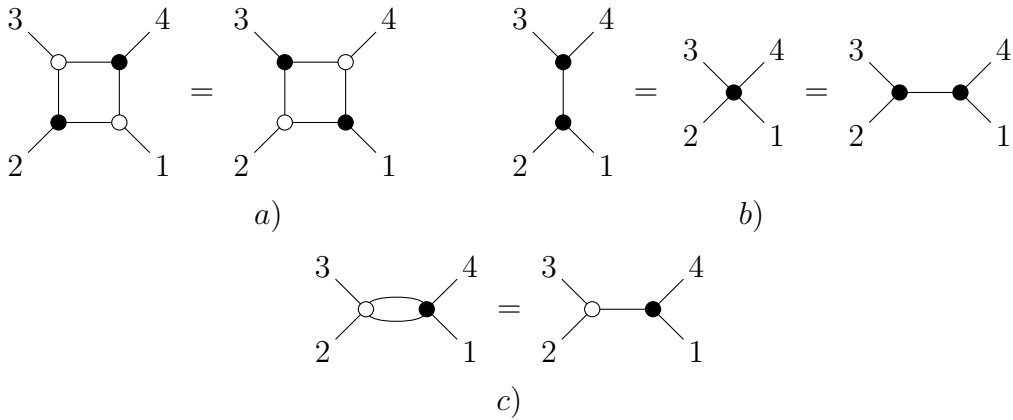

The encodings of the scattering amplitudes in terms of on-shell graphs is not unique \cite{AmplitudesPositiveGrassmannian}. On-shell  diagrams form equivalence classes. Equivalent diagrams are related by a sequence of equivalence moves, such as square move, merge/unmerge move and bubble reduction, see Fig. \ref{EquivalenceMoves}. It should be noted, that while the bubble reduction decreases the number of degrees of freedom in the diagram by one, the region of Grassmannian parametrized by the diagram (cell) stays the same.
It turns out that there is a one-to-one correspondence between the reduced modulo equivalence transformations on-shell diagrams and decorated permutations \cite{AmplitudesPositiveGrassmannian}. A decorated permutation is an injective map
\begin{eqnarray}
\sigma : \{ 1, \ldots , n\} \to \{1, \ldots , 2n\} ,
\end{eqnarray}
such that $1 \leq \sigma (i) \leq i+n$ and $\sigma$ modulo $n$ is an ordinary permutation. The permutation is constructed from the on-shell diagram as follows: starting from $i$-th leg one follows internal edges of the graph turning right at each black vertex and left at each white vertex, the external leg $j$ where this path ends is given by the image $\sigma (i)$.  There is also a correspondence between submanifolds of $G (k, n)$ ({\it positroid cells}) and on-shell diagrams labeled by decorated permutations \cite{AmplitudesPositiveGrassmannian}. The permutation in this case encodes a linear dependence between columns of $C$-matrix describing  points of  the Grassmannian: $\sigma (i) \geq i$ labels the first column $c_{\sigma (i)}$ such that $c_{i} \in \text{span} \{c_{i+1}, \ldots , c_{\sigma (i)} \}$. It is possible also to construct in a systematic way an on-shell diagram starting from a corresponding permutation \cite{AmplitudesPositiveGrassmannian} . The procedure is known as a BCFW bridge addition construction. First, the permutation is decomposed into a chain of consequent transpositions. Then each transposition $(i,j)$ is interpreted as a BCFW bridge. And finally these BCFW bridges are applied to a corresponding empty vacuum diagram with the prescribed values of $k$ and $n$\footnote{See \cite{AmplitudesPositiveGrassmannian} for more details.}.
The BCFW bridge addition operation is given by
\begin{eqnarray}\label{bridgeadditionoperation}
R_{i j} f(\vll_i , \vlt_i , \vlet_i , \vll_j , \vlt_j , \vlet_j  )
= \int \frac{d\alpha}{\alpha} f (\vll_i - \alpha\vll_j , \vlt_i, \vlet_i,
\vll_j , \vlt_j + \alpha\vlt_i , \vlet_j + \alpha\vlet_i)
\end{eqnarray}

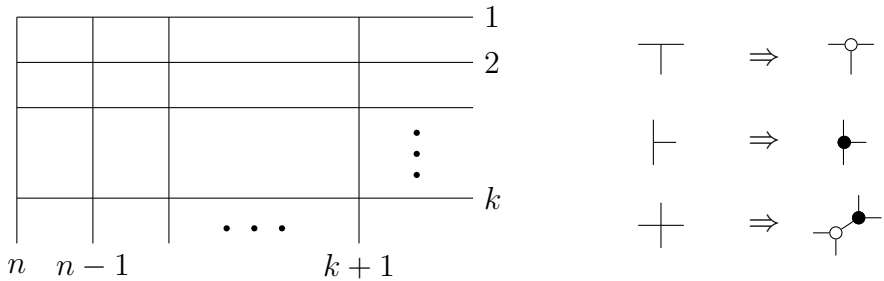
\begin{figure}
\centering	
\begin{equation}
\begin{tikzpicture}[baseline={($(n1.base) - (0,0)$)},transform shape, scale=1]
\node[right] (n1) at (6,0) {$1$};
\node[right] (n2) at (6,-0.6) {$2$};
\node[right] (nk) at (6,-2.4) {$k$};
\node[right] (nvdots) at (5,-1.6) {$\scalebox{2}{\vdots}$};
\node[below] (nn) at (0,-3.07) {$n$};
\node[below] (nnm1) at (1,-3) {$n-1$};
\node[below] (nkp1) at (4.5,-3) {$k+1$};
\node[right] (ncdots) at (2.5,-2.8) {$\scalebox{2}{\ldots}$};
\draw (0,0) -- (6,0);
\draw (0,-0.6) -- (6,-0.6);
\draw (0,-1.2) -- (6,-1.2);
\draw (0,-2.4) -- (6,-2.4);
\draw (0,0) -- (0,-3);
\draw (1,0) -- (1,-3);
\draw (2,0) -- (2,-3);
\draw (4.5,0) -- (4.5,-3);
\end{tikzpicture}
\qquad\qquad
\begin{tikzpicture}[baseline={($(n1.base) + (0,0.5)$)},transform shape, scale=1]
\node[right] (n1) at (1.5,-0.7) {$\Rightarrow$};
\node[right] (n2) at (1.5,-1.8) {$\Rightarrow$};
\node[right] (n3) at (1.5,-2.9) {$\Rightarrow$};
\coordinate (nc1) at (3,-0.505) {};
\coordinate (nc2) at (2.91,-1.8) {};
\coordinate (nc3) at (2.8,-3);
\coordinate (nc4) at (3.1,-2.8);
\draw (0.2,-0.5) -- (0.8,-0.5);
\draw (0.5,-0.5) -- (0.5,-0.9);
\draw (2.7,-0.5) -- (3.3,-0.5);
\draw (3,-0.5) -- (3,-0.9);
\draw[fill,white] (nc1) circle [radius=0.08];
\draw[black] (nc1) circle [radius=0.08];
\draw (0.4,-1.5) -- (0.4,-2.1);
\draw (0.4,-1.8) -- (0.7,-1.8);
\draw (2.9,-1.5) -- (2.9,-2.1);
\draw (2.9,-1.8) -- (3.2,-1.8);
\draw[fill,black] (nc2) circle [radius=0.08];
\draw[black] (nc2) circle [radius=0.08];
\draw (0.2,-2.9) -- (0.8,-2.9);
\draw (0.5,-2.6) -- (0.5,-3.2);
\draw (nc3) -- (nc4);
\draw (nc3) -- (2.5,-3);
\draw (nc3) -- (2.8,-3.3);
\draw (nc4) -- (3.4,-2.8);
\draw (nc4) -- (3.1,-2.5);
\draw[fill,white] (nc3) circle [radius=0.08];
\draw[black] (nc3) circle [radius=0.08];
\draw[fill,black] (nc4) circle [radius=0.08];
\draw[black] (nc4) circle [radius=0.08];
\end{tikzpicture}
\nonumber
\end{equation}
\caption{Top-cell on-shell diagram for $A_{k,n}$ on-shell amplitude.}
\label{TopCellAkn}
\end{figure}

All BCFW terms in the BCFW decomposition of the on-shell scattering amplitude $A_{k,n}$ could be obtained starting from a single on-shell diagram (the {\it top-cell} diagram) corresponding to a permutation which is a cyclic shift by $k$
\begin{eqnarray}
A_{k,n} : \quad\sigma = (k+1, \ldots n,1,\ldots k).
\end{eqnarray}
A representative on-shell top-cell diagram could be easily constructed as follows\footnote{See also \cite{AmplitudesSpectralParameter1},\cite{Broedel_DictionaryRoperatorsOnshellGraphsYangianAlgebras} for review.} \cite{TotalPositivityGrassmanniansNetworks}: draw  $k$ horizontal lines, $(n-k)$ vertical lines so that the left most and topmost are boundaries and substitute three and four-crossings according to the rules in Fig. \ref{TopCellAkn}. The on-shell diagrams corresponding to BCFW channels are then obtained by removing $(k-2)(n-k-2)$ edges from top cell diagram. It should be noted that not all edges are removable, but only those which removal lowers the dimension of the on-shell diagram  by exactly one. The corresponding positroid cells are given by submanifolds with extra linear dependencies between $k$ consecutive columns of the points $C$ of the Grassmannian.  The positroid cells with larger number of linear dependent columns are boundaries of positroid cells with smaller number of linear dependent columns. In the case of top-cell there are no $k$ linear dependent consecutive columns.

The on-shell diagrams for scattering amplitudes with one leg off-shell are given by the corresponding on-shell diagrams for on-shell scattering amplitudes with one of the vertexes exchanged for the off-shell vertex introduced in Section \ref{WilsonLinesSection}. The cutting off off-shell vertex (for certainty we will assume that the number of the off-shell leg is $n+1$) from a diagram with $n+1$ legs results in the on-shell diagram with $n+2$ legs containing only black and white on-shell vertexes. The same on-shell diagram could also be obtained starting from the on-shell diagram for on-shell scattering amplitude with $n+2$ legs and cutting off the box\footnote{One generally needs to perform a series of square and merge/unmerge moves to get a box at a prescribed position.} at the boundary of the diagram containing legs $n+1$ and $n+2$. We will need the latter on-shell diagram as a building block later in this section in the gluing procedure \cite{FormFactorsGrassmanians} used to derive a Grassmannian representation for the scattering amplitudes with one leg off-shell. It is not hard to see, that the permutation for this diagram is given by (the exchanged legs $n+1$ and $n+2$ in the top-cell permutation for $A_{k,n+2}$) \cite{FormFactorsGrassmanians} :
\begin{eqnarray}
\tilde{\sigma} = (k+1,\ldots n,n+2,n+1,1,2,\ldots k,k-1) .
\end{eqnarray}

\subsection{Grassmannian representation for amplitudes with one leg off-shell}\label{GrassmannianMinimalOffshellVertex}

Now we are ready to proceed with the derivation of Grassmannian representation for scattering amplitudes with one leg off-shell. First, let us derive the Grassmannian representation for 3-point off-shell vertex presented in section \ref{WilsonLinesSection}. The easiest way\footnote{Here we should note, that this is not a valid spin chain description for 3-point off-shell amplitude, but just a technical trick.} to get it - is through the action of $R$ - matrix operators (bridge addition operators) on the deformed three particle vacuum state analogues to the case of form factors\cite{FormFactorsGrassmanians} ($\lambda_1 = \lambda_p$):
\begin{eqnarray}
\frac{1}{\kappa^{*}}\delta^2 (\vll_1)\delta^2 (\vltu_2)\delta^4 (\vlet_2)
\delta^2 (\vltu_3)\delta^4 (\vlet_3),
\end{eqnarray}
where ($k$ is the off-shell gluon momentum and $p$ is its direction)
\begin{eqnarray}
\vltu_2 = \vlt_2 + \frac{\la 3|q}{\la 32\ra},\quad
\vltu_3 = \vlt_3 + \frac{\la 2|q}{\la 23\ra}\quad \text{and} \quad
q = k - p
\end{eqnarray}
so that the sum of particle momenta are
\begin{eqnarray}
\vll_1^{\alpha}\vlt_1^{\dotalpha} + \vll_2^{\alpha}\vltu_2^{\dotalpha}
+ \vll_3^{\alpha}\vltu_3^{\dotalpha} &=&
\vll_1^{\alpha}\vlt_1^{\dotalpha} + \vll_2^{\alpha}\vlt_2^{\dotalpha}
+ \vll_3^{\alpha}\vlt_3^{\dotalpha}
- \frac{\varepsilon_{cb}}{\la 23\ra} \left(
\vll_2^{\alpha}\vll_3^{c} - \vll_2^c\vll_3^{\alpha}
\right) q^{b\dotalpha} \nonumber \\ &=& \vll_1^{\alpha}\vlt_1^{\dotalpha} + \vll_2^{\alpha}\vlt_2^{\dotalpha}
+ \vll_3^{\alpha}\vlt_3^{\dotalpha} + q^{\alpha\dotalpha} .
\end{eqnarray}
The deformed (the case of non-zero spectral parameters $u_i$) off-shell three-point amplitude is then given by
\begin{eqnarray}
A_{2,2+1}^*(1^*,2,3) = \frac{1}{\kappa^{*}} R_{23} (u_{32}) R_{12} (u_{31})
\delta^2 (\lambda_1)\delta^2 (\vltu_2)\delta^4 (\vleu_2)
\delta^2 (\vltu_3)\delta^4 (\vleu_3), \label{3pointRoperators}
\end{eqnarray}
where as before ($u$ is the spectral parameter)
\begin{eqnarray}
R_{ij} (u) f(\vll_i,\vlt_i,\vlet_i,\vll_j,\vlt_j,\vlet_j) =
\int\frac{d\alpha}{\alpha^{1+u}} f (\vll_i - \alpha\vll_j,
\vlt_i,\vlet_i, \vll_j,\vlt_j + \alpha\vlt_i,\vlet_j+\alpha\vlet_i) . \nonumber \\
\end{eqnarray}
and $u_{ij} = u_i - u_j$.  So, we have:
\begin{align}
A_{2,2+1}^*(1^{*},2,3) =& \frac{1}{\kappa^{*}} R_{23} (u_{23}) \int
\frac{d\alpha_1}{\alpha_1^{1+u_{31}}} \delta^2 (\vll_1 - \alpha_1\vll_2)
\delta^2 (\vltu_2 + \alpha_1\vlt_1)\delta^4 (\vlet_2 + \alpha_1\vlet_1)
\delta^2 (\vltu_3)\delta^4 (\vlet_3) \nonumber \\ =&
\frac{1}{\kappa^{*}}\int\frac{d\alpha_2}{\alpha_2^{1+u_{32}}}
\int\frac{d\alpha_1}{\alpha_1^{1+u_{31}}}\delta^2 (\vll_1 - \alpha_1\vll_2 + \alpha_1\alpha_2\vll_3)\times \nonumber \\ & \times
\delta^2 (\vltu_2 + \alpha_1\vlt_1)\delta^4 (\vlet_2 + \alpha_1 \vlet_1)
\delta^2 (\vltu_3 + \alpha_2\vltu_2)\delta^4 (\vlet_3 + \alpha_2\vlet_2)
\nonumber \\ = &
\frac{1}{\kappa^{*}}\int\frac{d\alpha_2}{\alpha_2^{1+u_{32}}}
\int\frac{d\alpha_1}{\alpha_1^{1+u_{31}}}
\delta^4 (C (\alpha_1,\alpha_2)\cdot \vltu)
\delta^8 (C (\alpha_1,\alpha_2)\cdot \vlet)
\delta^2 (C^{\perp} (\alpha_1, \alpha_2)\cdot \vll) , \label{deformed-offshell-vertex}
\end{align}
where
\begin{eqnarray}
C (\alpha_1,\alpha_2) = \begin{pmatrix}
\alpha_1 & 1 & 0 \\ 0 & \alpha_2 & 1
\end{pmatrix} \quad \text{and} \quad
C^{\perp} = \begin{pmatrix}
1 & -\alpha_1 & \alpha_1\alpha_2
\end{pmatrix}
\end{eqnarray}
The same formula could be written as an integral over $G(2,3)$ Grassmanian as
\begin{align}
A_{2,2+1}^*(1^*,2,3) = \frac{1}{\kappa^{*}}\int
\frac{d^{2\times 3} C}{\text{Vol} [GL(2)]}
\frac{1}{(12)^{1-u_{23}}(23)^{1-u_{31}}(31)^{1-u_{12}}}
\delta^4 (C\cdot \vltu)\delta^8 (C\cdot \vlet) \delta^4 (C^{\perp}\cdot \vll ) , \label{3pointGrassmanianDeformed}
\end{align}
where $(i i+1)$ denote consecutive minors of $C$-matrix. In the following when deriving Grassmannian integral representation for off-shell amplitudes we will however need only undeformed 3-point off-shell amplitude at zero values of spectral parameters, which is given by
\begin{align}
A_{2,2+1}^*(1^*,2,3) = \frac{1}{\kappa^{*}}\int
\frac{d^{2\times 3} C}{\text{Vol} [GL(2)]}
\frac{1}{(12)(23)(31)}
\delta^4 (C\cdot \vltu)\delta^8 (C\cdot \vlet) \delta^4 (C^{\perp}\cdot \vll ) , \label{3pointGrassmanian}
\end{align}

\subsubsection{spinor helicity representation}

As we already mentioned in order to write down N$^k$MHV amplitudes as the integrals over the Grassmanians we will use  the gluing procedure as in \cite{FormFactorsGrassmanians}, that is we break the corresponding top cell on-shell diagram for amplitude into two pieces: the mentioned before box, which we replace with minimal 3-point off-shell amplitude (\ref{3pointGrassmanian}) and the remaining purely on-shell piece with $n+2$ legs, for which a Grassmannian integral representation is known. After that we glue these two pieces together, i.e. we perform the on-shell phase space integration. The on-shell piece is given by Grassmannian integral representation of corresponding on-shell amplitude and can be written as\footnote{We hope that here and later from the context it will be clear, that $k$ here is N$^{k-2}$MHV degree and not the off-shell gluon momentum} \cite{AmplitudesPositiveGrassmannian}:
\begin{equation}
I_{n+2,k}=
\int \frac{\dd\alpha_1}{\alpha_1}\cdots\frac{\dd\alpha_m}{\alpha_m} \;
\delta^{k\times 2}(C\cdot\vlt) \,
\delta^{k\times 4}(C\cdot\eta) \,
\delta^{(n+2-k)\times 2}(C^\perp\cdot\vll)
\eqncom
\end{equation}
where the matrix $C$ depends on the Grassmannian coordinates $\alpha_i$'s, $C=C(\alpha_i)\in G(k,n+2)$ and $m$ is the dimension of the corresponding cell in the Grassmannian. The concrete parametrization of matrix $C$ corresponding to on-shell diagram labeled by permutation $\sigma$ could be obtained with the help of Mathematica package positroid.m \cite{MathematicaPositroid}. So, gluing the minimal 3-point off-shell amplitude (\ref{3pointGrassmanian}) to the legs $n+1$ and $n+2$ and accounting for all helicity configurations in the gluing channel we get\footnote{Without loss of generality we may choose off-shell leg to lie between legs $1$ and $n$.}
\begin{eqnarray}
A^{*}_{k,n+1} = \int\prod_{i=n+1}^{n+2}\frac{d^2\vll_i d^2\vlt_i}{\text{Vol}[GL(1)]} d^4\vlet_i A_{2,2+1}^*(p^{*},n+1,n+2)\Big|_{\vll\to -\vll} I_{n+2,k}
\end{eqnarray}
That is
\begin{align}
A^{*}_{k,n+1} &= \frac{1}{\kappa^{*}}\int\frac{d\beta_2}{\beta_2}\int\frac{d\beta_1}{\beta_1}
\delta^2 (\vll_p + \beta_1\vll_{n+1}-\beta_1\beta_2\vll_{n+2})
\delta^2 (\vltu_{n+1} + \beta_1\vlt_p) \nonumber \\
&\times \delta^4 (\vlet_{n+1} + \beta_1\vlet_p) \delta^2 (\vltu_{n+2} + \beta_2 \vltu_{n+1})\delta^4 (\vlet_{n+2} + \beta_2\vlet_{n+1}) \nonumber \\
&\times \int\prod_{i=n+1}^{n+2}\frac{d^2\vll_i d^2\vlt_i}{\text{Vol}[GL(1)]} d^4\vlet_i \int \frac{\dd\alpha_1}{\alpha_1}\cdots\frac{\dd\alpha_m}{\alpha_m} \;
\delta^{k\times 2}(C\cdot\vlt) \,
\delta^{k\times 4}(C\cdot\eta) \,
\delta^{(n+2-k)\times 2}(C^\perp\cdot\vll) ,
\end{align}
were $\vltu_{n+1} = \vlt_{n+1} + \frac{\la n+2|q}{\la n+2 | n+1\ra}$
and $\vltu_{n+2} = \vlt_{n+2} + \frac{\la n+1|q}{\la n+1 | n+2\ra}$. Performing integrations over $\vlt_{n+1}$, $\vlt_{n+2}$, $\vlet_{n+1}$ and $\vlet_{n+2}$ we get
\begin{align}
\vlt_{n+1} &= \frac{\la n+2|q}{\la n+2|n+1\ra} - \beta_1\vlt_p,\quad
&\vlt_{n+2} &= \frac{\la n+1|q}{\la n+1|n+2\ra} + \beta_1\beta_2\vlt_p, \\ \quad  \vlet_{n+1} &= -\beta_1\vlet_p , \quad &\vlet_{n+2} &= \beta_1\beta_2\vlet_p.
\end{align}
To remove the $\text{Vol} [GL(1)]^2$ redundancy in the remaining integrations over  $\vll$, we use their parametrization as in \cite{FormFactorsGrassmanians}:
\begin{equation}
\vll_{n+1}=\refspina-\beta_3 \refspinb
\eqncom\qquad
\vll_{n+2}=\refspinb-\beta_4 \refspina
\eqndot
\end{equation}
Where $\refspina$ and $\refspinb$ are two arbitrary but linearly independent reference spinors. Then $\abr{n \splus 1 \ssep n \splus 2 }=(\beta_3\beta_4-1)\abr{\refspinbl\refspinal}$,
\begin{eqnarray}
\int\frac{d^2\vll_{n+1}}{\text{Vol}[GL(1)]}\frac{d^2\vll_{n+2}}{\text{Vol}[GL(1)]} = - \la\refspina\refspinb\ra^2 \int d\beta_3 d\beta_4
\end{eqnarray}
and\footnote{In deriving this formula it was convenient to put $\vlt_p$ to zero, so that $q = k$. Here $k= q + p$ and $q$ should not be confused with the off-shell gluon polarization vector.}
\begin{align}
A^{*}_{k,n+1} =& \frac{1}{\kappa^{*}}\la\refspina\refspinb\ra^2
\int\frac{d\alpha_1}{\alpha_1}\ldots \frac{d\alpha_m}{\alpha_m}
\frac{d\beta_1}{\beta_1}\frac{d\beta_2}{\beta_2}
\frac{d\beta_3 d\beta_4}{(1-\beta_3\beta_4)^2} \nonumber \\
&\times \delta^2 (\lambda_p + \beta_1 (1+\beta_2\beta_4)\refspina
- \beta_1 (\beta_2 + \beta_3)\refspinb ) \nonumber \\
&\times \delta^{k\times 2} \left( C' (\alpha_i,\beta_i)
\cdot \vltuu \right)
\delta^{k\times 4} \left(C' (\alpha_i, \beta_i)\cdot \vleuu \right)
\delta^{(n+2-k)\times 2} \left(C'^{\perp} (\alpha_i,\beta_i)\cdot \vlluu \right). \label{GrassmannianBeforeb3bIntegration}
\end{align}
Here, the following notation was introduced
\begin{align}
C'_{n+1} &= \frac{1}{1-\beta_3\beta_4}C_{n+1} + \frac{\beta_3}{1-\beta_3\beta_4}C_{n+2},
&C'^{\perp}_{n+1} &= C^{\perp}_{n+1} - \beta_4 C^{\perp}_{n+2}, \nonumber \\
C'_{n+2} &= \frac{1}{1-\beta_3\beta_4}C_{n+2} + \frac{\beta_4}{1-\beta_3\beta_4} C_{n+1},
&C'^{\perp}_{n+2} &= C^{\perp}_{n+2} - \beta_3 C^{\perp}_{n+1},
\end{align}
and
\begin{align}
&\vlluu_i = \vll_i, & i = 1,\ldots  n& , &\vlluu_{n+1} &= \refspina ,
&\vlluu_{n+2} &= \refspinb \nonumber \\
&\vltuu_i = \vlt_i, & i = 1,\ldots  n& ,
&\vltuu_{n+1} &= \frac{\la\refspinb |k}{\la\refspinb\refspina\ra},
&\vltuu_{n+2} &= - \frac{\la\refspina |k}{\la\refspinb\refspina\ra} , \nonumber \\
&\vleuu_i = \vlet_i,  & i = 1,\ldots  n& , &\vleuu_{n+1} &= \vlet_p ,
 &\vleuu_{n+2} &= 0 . \nonumber \\
\end{align}
The factor of $1/(1-\beta_3\beta_4)^2$ in  (\ref{GrassmannianBeforeb3bIntegration}) is a Jacobian from reorganizing the $C^\perp\cdot\vll$ $\delta$ - functions (see \cite{FormFactorsGrassmanians} for further details). Now, rewriting first $\delta$ - function in (\ref{GrassmannianBeforeb3bIntegration}) as
\begin{multline}
\delta^2 (\lambda_p + \beta_1 (1+\beta_2\beta_4)\refspina - \beta_1 (\beta_2 + \beta_3)\refspinb) \\ = \frac{1}{\beta_1^2\beta_2 \la\refspina\refspinb\ra}\delta (\beta_3 - \frac{\la\refspina p\ra}{\beta_1\la\refspina\refspinb\ra} + \beta_2)\cdot
\delta (\beta_4 - \frac{\la\refspinb p\ra}{\beta_1\beta_2 \la\refspina\refspinb\ra} + \frac{1}{\beta_2})
\end{multline}
choosing $\refspina = \vll_p$, $\refspinb = \xi$ and performing integrations over $\beta_3$, $\beta_4$ we get
\begin{multline}
A^{*}_{k,n+1} = \frac{1}{\kappa^{*}}\la\xi p\ra \int
\frac{d\alpha_1}{\alpha_1}\ldots \frac{\alpha_m}{\alpha_m}
\frac{d\beta_1 d\beta_2}{\beta_1\beta_2^2}\times \\
\times\delta^{k\times 2} \left( C' (\alpha_i,\beta_i)
\cdot \vltuu \right)
\delta^{k\times 4} \left(C' (\alpha_i, \beta_i)\cdot \vleuu \right)
\delta^{(n+2-k)\times 2} \left(C'^{\perp} (\alpha_i,\beta_i)\cdot \vlluu \right) , \label{GrassmannianEdgeIntegral}
\end{multline}
where now
\begin{align}
C'_{n+1} &= -\beta_1 C_{n+1} + \beta_1\beta_2 C_{n+2},
&C'^{\perp}_{n+1} &= C^{\perp}_{n+1} + \frac{1+\beta_1}{\beta_1\beta_2} C^{\perp}_{n+2}, \nonumber \\
C'_{n+2} &= -\beta_1 C_{n+2} + \frac{1+\beta_1}{\beta_2} C_{n+1},
&C'^{\perp}_{n+2} &= C^{\perp}_{n+2} + \beta_2 C^{\perp}_{n+1},
\end{align}
and
\begin{align}\label{SpinorsInDeformadGrassmannian}
&\vlluu_i = \vll_i, & i = 1,\ldots  n& , &\vlluu_{n+1} &= \lambda_p ,
&\vlluu_{n+2} &= \xi \nonumber \\
&\vltuu_i = \vlt_i, & i = 1,\ldots  n& ,
&\vltuu_{n+1} &= \frac{\la\xi |k}{\la\xi p\ra},
&\vltuu_{n+2} &= - \frac{\la p |k}{\la\xi p\ra} , \nonumber \\
&\vleuu_i = \vlet_i,  & i = 1,\ldots  n& , &\vleuu_{n+1} &= \vlet_p ,
&\vleuu_{n+2} &= 0 . \nonumber \\
\end{align}
To write down (\ref{GrassmannianEdgeIntegral}) in terms of the integral over the points of the $G(k,n+2)$ Grassmannian parametrized by the elements of $C'$ matrix we have considered a number of particular examples. For example, in the case of $A^{*}_{2,3+1}$  scattering amplitude the corresponding permutation and $C$-matrix are shown in Fig.
\ref{IMHV4}. The $C'$ matrix in this case is given by
\begin{eqnarray}
C' = \begin{pmatrix}
1 & 0 & -\alpha_2 & \beta_1 (\alpha_2\alpha_4 - \alpha_1\beta_2) &
\alpha_1\beta_1 - \frac{\alpha_2\alpha_4 (1+\beta_1)}{\beta_2} \\
0 & 1 & \alpha_3 & -\alpha_3\alpha_4\beta_1  & \frac{\alpha_3\alpha_4 (1+\beta_1)}{\beta_2}
\end{pmatrix}
\end{eqnarray}
Taking into account the Jacobian of transformation from the Grassmannian $C_{ij}$ matrix elements to edge variables $\alpha_i ,\beta_j$, that is $J = \frac{\alpha_1\alpha_3^2\alpha_4\beta_1^2}{\beta_2}$, one can easily see that the result for $A^{*}_{2,3+1}$ off-shell MHV$_{3+1}$ scattering amplitude could be written in the following form
\begin{eqnarray}
A^{*}_{2,3+1} = \frac{ \la\xi p\ra}{\kappa^{*}}
\int\frac{d^{2\times 5} C'}{\text{Vol}[GL(2)]} \frac{\delta^{2\times 2} \left( C'
\cdot \vltuu \right)
\delta^{2\times 4} \left(C' \cdot \vleuu \right)
\delta^{3\times 2} \left(C'^{\perp} \cdot \vlluu \right)}{(12)(23)(34)(45)(41)}.
\end{eqnarray}
\begin{figure}[htbp]
	\(
	I=
	\begin{aligned}
	\begin{tikzpicture}[scale=0.8]
	\draw (1,-1) -- (0,-0.7);
	\draw (1,-1) -- (2,-0.7);
	\draw (1,-1) -- (1,-1-0.65);
	\draw (0,-0.7) -- (0-0.3,-0.7-0.6);
	\draw (2,-0.7) -- (2+0.3,-0.7-0.6);
	\draw (0,-0.7) -- (0+0.3,-0.7+0.6);
	\draw (2,-0.7) -- (2-0.3,-0.7+0.6);
	\node[db] at (1,-1) {};
	\node[dw] at (0,-0.7) {};
	\node[dw] at (2,-0.7) {};
	\node[] at (-0.45,-1.6) {3};
	\node[] at (2.45,-1.6) {1};
	\node[] at (0.4,0.2) {4};
	\node[] at (1.6,0.2) {5};
	\node[] at (1,-2) {2};
	\end{tikzpicture}
	\end{aligned}
	\eqncom
	\qquad
	\sigmapart=(3,5,4,2,1)
	\eqncom
	\qquad
	C=\begin{pmatrix}
	1&0&-\alpha_2&-\alpha_2\alpha_4&-\alpha_1\\
	0&1&\alpha_3&\alpha_3\alpha_4&0
	\end{pmatrix}
	\eqndot
	\)
	\centering
	\caption{On-shell sub-diagram $I$ obtained by removing off-shell vertex from $A^{*}_{2,3+1}$ diagram, corresponding permutation $\sigmapart$ and $C$ matrix.}
	\label{IMHV4}
\end{figure}
Similar consideration of the other examples allows us to write down a general conjecture for the integral over Grassmannian for a general set of values for $n$ and $k$. Lets consider
the following integral over Grassmannian:
\begin{equation}
\Omega_{n+2}^k[\Gamma]=\frac{\la\xi p\ra}{\kappa^{*}}
\int_{\Gamma}\frac{d^{k\times (n+2)}C'}{\text{Vol}[GL (k)]}
\frac{\delta^{k\times 2} \left( C'
\cdot \vltuu \right)
\delta^{k\times 4} \left(C' \cdot \vleuu \right)
\delta^{(n+2-k)\times 2} \left(C'^{\perp} \cdot \vlluu \right)}{(1 \cdots k)\cdots (n+1 \cdots k-2) (n+1 \; 1\cdots k-1)}. \label{GrassmannianIntegralFinal}
\end{equation}
We claim that for appropriate choice of the integration contour $\Gamma=\Gamma_{k,n+2}$ the
following identity holds
\begin{equation}
A^{*}_{k,n+1}=\Omega_{n+2}^k[\Gamma_{k,n+2}].\label{GrassmannianIntegralFinal+treeContour}
\end{equation}

\subsubsection{twistor representation}

The transformation of the obtained result to twistor space could be performed in full analogy to the case of on-shell amplitudes \cite{DualitySMatrix} and form factors \cite{FormFactorsGrassmanians}. Super twistors are defined as
\begin{eqnarray}
\mathcal{W}_i = (\vmtuuu_i,\vltuu_i,\vleuu_i) ,
\end{eqnarray}
where $\vmtuuu_i$ is related to $\vlluu_i$ via Witten's half Fourier transformation \cite{GaugeTheoryAsTwistorString}
\begin{eqnarray}
\int d^2\vlluu_j \exp (-i \vmtuuu
_j^{\alpha}\vlluu_{j\alpha}) . \label{WittenHalfFourier}
\end{eqnarray}
Now recalling, that $\vlluu_{n+1} = \lambda_p$, $\vlluu_{n+2} = \xi$ and using (\ref{WittenHalfFourier})  $\la\xi p\ra$ could be written as
\begin{eqnarray}
\la\xi p\ra = \la\frac{\partial}{\partial\vmtuuu_{n+1}}\frac{\partial}{\partial\vmtuuu_{n+2}}\ra .
\end{eqnarray}
Rewriting $\delta$ - function $\delta^{2\times(n+2-k)}(C'^{\perp}\cdot\vlluuu)$ as \cite{FormFactorsGrassmanians}:
\begin{eqnarray}
\prod_{K=1}^{k}\int d^2\rho_K \delta^{2 \times (n+2)} (\vlluu_i - \sum_{L=1}^k\rho_L C'_{L i}) \; , \label{deltafuncrep}
\end{eqnarray}
applying half Fourier transformation (\ref{WittenHalfFourier}) to this representation and performing the integrals over $\vlluuu_i$ via $\delta$ - functions, we find
\begin{equation}
\prod_{K=1}^k \int d^2\rho_K \exp(-i\sum_{j=1}^{n+2}\sum_{L=1}^k\rho^\alpha_{L}C'_{Lj}\vmtuuu_{\alpha j})=\delta^{2k}(C'\cdot\vmtuuu)\eqndot
\end{equation}
So, we can rewrite (\ref{GrassmannianIntegralFinal}) and
(\ref{GrassmannianIntegralFinal+treeContour})
as
\begin{eqnarray}
\Omega^k_{n+2}[\Gamma] &=& \frac{1}{\kappa^{*}} \la\frac{\partial}{\partial\vmtuuu_{n+1}}\frac{\partial}{\partial\vmtuuu_{n+2}}\ra
\int_{\Gamma}\frac{d^{k\times (n+2)}C'}{\text{Vol}[GL(k)]}
\frac{\delta^{4 k|4 k} (C'\cdot \mathcal{W})}{(1\cdots k)\cdots (n+1\cdots k-2)(n+1 \; 1\cdots k-1)} . \nonumber \\
A^{*}_{k,n+1} &=& \Omega^k_{n+2}[\Gamma_{k,n+2}]
\end{eqnarray}

\subsubsection{momentum twistor representation}

\begin{figure}[htbp]
	\begin{equation*}
	\begin{tikzpicture}[scale=0.8]
	\draw[<-] (-0.5,0) -- (1.7,3);
	\draw[<-] (1.7,3) -- (3,1.9);
	\draw[<-] (3,1.9) -- (4.5,3.3);
	\draw[<-] (4.5,3.3) -- (6.5,0);
	\draw[dashed,<-] (2.2,-1.3) -- (-0.5,0);
	\draw[dashed,<-] (6.5,0) -- (2.2,-1.3);
	\node[label={[label distance=-5pt]135:$p_1$}] at (0.85,1.5) {};
	\node[label={[label distance=-7pt]45:$p_2$}] at (2.15,2.45) {};
	\node[label={[label distance=0pt]340:$p_3$}] at (3.25,2.6) {};
	\node[label={[label distance=+8pt]360:$p_4$}] at (4.95,1.65) {};
	\node[label={[label distance=-2pt]354:$p_5$}] at (4.1,-0.65) {};
	\node[label={[label distance=-2pt]270:$p_6$}] at (1.1,-0.65) {};
	\node[label={[label distance=-6pt]220:$y_1$}] at (0,0) {};
	\node[label={[label distance=-2pt]90:$y_2$}] at (1.7,3) {};
	\node[label={[label distance=-2pt]270:$y_3$}] at (3,1.9) {};
	\node[label={[label distance=-2pt]90:$y_4$}] at (4.5,3.3) {};
	\node[label={[label distance=-4pt]330:$y_5$}] at (6,0) {};
	\node[label={[label distance=-2pt]270:$y_6$}] at (2.2,-1.3) {};
	\end{tikzpicture}
	\end{equation*}
	\caption{Momenta and dual coordinates in the case of amplitude with one off-shell and $n=4$ on-shell legs. In contrast to the case of off-shell amplitudes, the $n$ on-shell momenta do not add up to zero but to the off-shell gluon momentum $k$.}
	\label{fig: dual contour}
\end{figure}
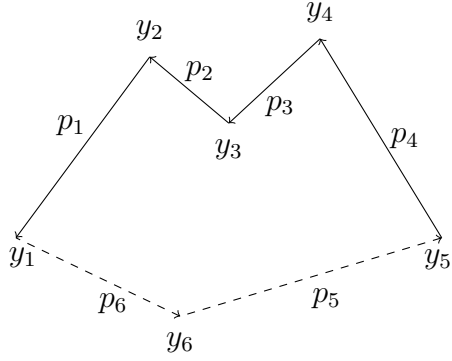

The transformation to momentum twistor space could be performed using the strategy of \cite{GrassmanianOriginDualConformalInvariance,Grassmanians-N4SYM-ABJM,FormFactorsGrassmanians}. The momentum super twistor variables $\mathcal{Z}_i = (\vlluu_i ,\vmmuuu_i , \vletuu_i)$ \cite{MomentumTwistors} are defined through introduction of dual super coordinates
\begin{eqnarray}
\vlluu_i \vltuu_i = y_i - y_{i+1} , \quad \vlluu_i \vleuu_i = \vartheta_i - \vartheta_{i+1} .
\end{eqnarray}
In Fig. \ref{fig: dual contour} we have shown a contour in the dual space formed by on-shell particles momenta together with momenta of two auxiliary states $n+1$ and $n+2$.
The components of momentum twistors are defined then through the following incidence relations
\begin{eqnarray}
\vmmuuu_i = \vlluu_i y_i = \vlluu_i y_{i+1},\quad \vletuu_i = \vlluu_i\vartheta_i = \vlluu_i\vartheta_{i+1} .
\end{eqnarray}
Inverting these relations we get
\begin{equation}
\begin{aligned}
\vltuuu_i&=\frac{\abr{i\splus1\,i}\vmmuuu_{i-1}+\abr{i\,i\sminus1}\vmmuuu_{i+1}+\abr{i\sminus1\, i\splus1}\vmmuuu_{i}}{\abr{i\sminus1\,i}\abr{i\,i\splus1}} , \\
\vleuuu_i&=\frac{\abr{i\splus1\,i}\vletuuu_{i-1}+\abr{i\,i\sminus1}\vletuuu_{i+1}+\abr{i\sminus1\, i\splus1}\vletuuu_{i}}{\abr{i\sminus1\,i}\abr{i\,i\splus1}} .
\end{aligned}
\label{MomentumToMomentumTwistors}
\end{equation}
Now following \cite{FormFactorsGrassmanians} we use representation of $\delta^{2\times (n+2-k)} (C'^{\perp}\cdot\vlluu)$ as in (\ref{deltafuncrep}).  Fixing part of $GL (k)$ redundancy as
\begin{eqnarray}
\rho = \begin{pmatrix}
0 & \cdots & 0 & 1 & 0 \\
0 & \cdots & 0 & 0 & 1
\end{pmatrix} .
\end{eqnarray}
we get
\begin{eqnarray}
C'_{k-1 \; i} = \vlluu_i^1 , \quad  C'_{k i} = \vlluu_i^2 .
\end{eqnarray}
and
\begin{multline}
\Omega^k_{n+2}[\Gamma] = \frac{ \la\xi p\ra}{\kappa^{*}}
\delta^4 (\vlluu\cdot\vltuu)\delta^8 (\vlluu\cdot\vleuu)
\int_{\Gamma}\frac{d^{(k-2)\times (n+2)}C'}{\text{Vol}[GL(k-2)\times T_{k-2}]}\times \\
\times \frac{\delta^{2\times (k-2)} (C'\cdot\vltuu )\delta^{4\times (k-2)} (C'\cdot\vltuu )}{(1\cdots k)\cdots (n+1\cdots k-2)(n+1\; 1\cdots k-1)} ,
\end{multline}
where the integral and $\delta$ - function contains only the first $k-2$ rows of $C'$. The action of the shift symmetry $T_{k-2}$ on these rows is given by
\begin{eqnarray}
C'_{I i}\to C'_{I i} + r_{1 I}\vlluu_i^1 + r_{2 I}\vlluu_i^2 ,\qquad  I = 1,\ldots , k-2 ,
\end{eqnarray}
with arbitrary $r_{1 I}$ and $r_{2 I}$. According to \cite{Grassmanians-N4SYM-ABJM}, (\ref{MomentumToMomentumTwistors}) leads to
\begin{equation}
\sum_{i=1}^{n+2}C'_{Ii}\vltuuu_i=-\sum_{i=1}^{n+2}D_{Ii}\vmmuuu_i \eqncom \qquad
\sum_{i=1}^{n+2}C'_{Ii}\vleuuu_i=-\sum_{i=1}^{n+2}D_{Ii}\vletuuu_i \eqncom \qquad I=1,\dots, k-2\eqncom
\end{equation}
where the matrix $D$ is given by
\begin{equation}
D_{Ii}=\frac{\abr{i\, i\splus1}C'_{I\,i\sminus1}+\abr{i\sminus1\,i}C'_{I\,i\splus1}+\abr{i\splus1\,i\sminus1}C'_{I\,i}}{\abr{i\sminus1\,i}\abr{i\,i\splus1}}\eqndot
\end{equation}
Next, we rewrite the minors of $C'$ in terms of minors of $D$. The consecutive minors could be rewritten as \cite{GrassmanianOriginDualConformalInvariance,Grassmanians-N4SYM-ABJM}:
\begin{equation}
(C'_1\dots C'_{k})=-\abr{1\,2}\cdots\abr{k\sminus1\,k}(D_2\dots D_{k\sminus1}) ,
\label{MinorsElvang}
\end{equation}
while the only non-consecutive minor we need is given by \cite{FormFactorsGrassmanians}:
\begin{multline}
(C'_{n+1} C_{1} \ldots C'_{k-1}) = - \la n+1 \; 1\ra\la 1 2\ra \ldots \la k-2 \; k-1\ra (D_1 \ldots D_{k-2})  \\ - \la n+1 \; n+2\ra\la 1 2\ra \ldots \la k-2 \; k-1\ra (D_{n+2} D_2 \ldots D_{k-2}) .
\label{MinorsWilhelm}
\end{multline}
Using (\ref{MinorsElvang}) and (\ref{MinorsWilhelm}) we have
\begin{equation}
(1\cdots k)_{C'}\cdots(n \splus 2\cdots k \sminus 1)_{C'}=(-1)^{n+2}(\abr{1\,2}\cdots\abr{n\splus2\,1})^{k-1}(1\cdots k)_{D}\cdots(n \splus 2\cdots k \sminus 1)_{D}\eqndot
\end{equation}
and
\begin{equation}
\frac{(n+2 \; 1 \ldots k-1)_{C'}}{(n+1 \; 1 \ldots k-1)_{C'}} =
\frac{\la n+2 \; 1\ra (1 \ldots k-2)_D}{\la n+1 \; 1\ra (1 \ldots k-2)_D + \la n+1 \; n+2\ra (n+2 \; 2\ldots k-2)_D}
\end{equation}
Next. we use the $T_{k-2}$ shift symmetry to set $C'_{I 1} = C'_{I 2} = 0$ \cite{Grassmanians-N4SYM-ABJM,FormFactorsGrassmanians}, so that
\begin{equation}
\frac{d^{(k-2)\times(n+2)}C'}{\text{Vol}[GL(k-2)\ltimes T_{k-2}]} = \abr{12}^{k-2}\frac{d^{(k-2)\times(n)}C'}{\text{Vol}[GL(k-2)]} .
\end{equation}
Changing the integration variables from $C'$ to $D$
\begin{equation}
\frac{d^{(k-2)\times(n)}C'}{\text{Vol}[GL(k-2)]}=\left(\frac{\abr{12} \cdots \abr{n+2 \;1}}{\abr{12}^2}\right)^{k-2}\frac{d^{(k-2)\times(n)}D}{\text{Vol}[GL(k-2)]} ,
\end{equation}
and undoing gauge fixing of first two colums of $C'$-matrix yields factors of
\begin{equation}
\abr{12}\delta^2(D_{Ii}\vlluuu_i)
\end{equation}
for $I=1,\dots,k-2$. The details of these steps could be found in \cite{Grassmanians-N4SYM-ABJM}. The final expression for scattering amplitude with one leg off-shell for appropriate choice of
 integration contour $\Gamma_{k,n+2}$ can be represented as
\begin{eqnarray}
\frac{A^{*}_{k,n+1}}{A^{*}_{2,n+1}}&=&\omega^k_{n+2}[\Gamma_{k,n+2}], \nonumber \\
\omega^k_{n+2}[\Gamma] &=& \int_{\Gamma}
\frac{d^{(k-2)\times (n+2)}D}{\text{Vol}[GL(k-2)]}\frac{1}{1+\frac{\la p\xi\ra}{\la p 1\ra}\frac{(n+2 \; 2 \; \ldots \; k-2)}{(1 \; \ldots \; k-2)}}
\frac{\delta^{ 4 (k-2) | 4 (k-2)} (D\cdot \mathcal{Z})}{(1 \; \ldots \; k-2) \;\ldots \; (n+2 \; \ldots \; k-3)}. \nonumber\\
 \label{GrassmannianMomentumTwistors}
\end{eqnarray}

\subsection{MHV$_{n+1}$, NMHV$_{4,5}$ and NMHV$_{n+1}$ amplitudes}

Let us now perform some checks of the formula derived in previous subsections. Fist we start with off-shell MHV$_n$ case. Using (\ref{GrassmannianIntegralFinal}) for $k=2$ and setting $\xi = \lambda_1$ the off-shell MHV$_n$ amplitude could be written as
\begin{equation}
A^{*}_{k,n+1} = \frac{\la\xi p\ra}{\kappa^{*}}
\int\frac{d^{2\times (n+2)}C'}{\text{Vol}[GL (2)]}
\frac{\delta^{4} \left( C'
\cdot \vltuu \right)
\delta^{8} \left(C' \cdot \vleuu \right)
\delta^{2 n} \left(C'^{\perp} \cdot \vlluu \right)}{(1 2) (2 3)\cdots (n+1 \; n+2) (n+1 \; 1)},
\end{equation}
where
\begin{align}
&\vlluu_i = \vll_i, & i = 1,\ldots  n& , &\vlluu_{n+1} &= \lambda_p ,
&\vlluu_{n+2} &= \vll_1 \nonumber \\
&\vltuu_i = \vlt_i, & i = 1,\ldots  n& ,
&\vltuu_{n+1} &= \frac{\la 1 |k}{\la 1 p\ra},
&\vltuu_{n+2} &= - \frac{\la p |k}{\la 1 p\ra} , \nonumber \\
&\vleuu_i = \vlet_i,  & i = 1,\ldots  n& , &\vleuu_{n+1} &= \vlet_p ,
&\vleuu_{n+2} &= 0 . \nonumber \\ \label{kindataSpinorHelicity}
\end{align}
Using the standard gauge fixing for $C'$ - matrix
\begin{eqnarray}
C' = \begin{pmatrix}
1 & 0  & c'_{13} & \cdots & c'_{1 n+2}  \\
0 & 1 & c'_{23} & \cdots & c'_{2 n+2}
\end{pmatrix} , \quad
C'^{\perp} = \begin{pmatrix}
-c'_{13} & -c'_{23} & 1 & 0 &\cdots & 0 \\
-c'_{14} & -c'_{24} & 0 & 1 &\cdots & 0 \\
\vdots & \vdots  & \vdots  & \vdots   & \vdots  & \vdots \\
-c'_{1 n+2} & -c'_{2 n+2} & 0 & 0 & \cdots & 1
\end{pmatrix}
\end{eqnarray}
and integrating out the last $2 n$ $\delta $ - functions we get
\begin{eqnarray}
C' = \begin{pmatrix}
1 & 0 & -\frac{\la 2 3\ra}{\la 1 2\ra} & \cdots &
-\frac{\la 2 n\ra}{\la 1 2\ra} & -\frac{\la 2 p\ra}{\la 1 2\ra} & 1 \\
0 & 1 & \frac{\la 1 3\ra}{\la 1 2\ra} & \cdots &
\frac{\la 1 n\ra}{\la 1 2\ra} & \frac{\la 1 p\ra}{\la 1 2\ra} & 0
\end{pmatrix}
\end{eqnarray}
and
\begin{eqnarray}
A^{*}_{k,n+1} = \frac{1}{\kappa^{*}}\frac{1}{\la 1 2\ra\la 2 3\ra \cdots
	\la n p\ra\la p 1\ra} \delta^4 (\sum_{i=1}^{n} p_i + k)
\delta^8 (\sum_{i=1}^{n}\vll_i \vlet_i + \vll_p \vlet_p) .
\end{eqnarray}

In the case of NMHV$_4$ amplitude with one leg off-shell the corresponding Grassmannian integral representation is given by (here we will not choose any specific value of axillary spinor $\xi$)
\begin{equation}
A^{*}_{3,3+1} = \frac{\la \xi p\ra}{\kappa^{*}}
\int\frac{d^{3\times 5} C'}{\text{Vol}[GL (3)]}
\frac{(512)}{(412)}\frac{\delta^6 (C'\cdot \vltuu)
\delta^8 (C\cdot \vleuu)\delta^4 (C'^{\perp} \cdot \vlluu)}{(123)(234)(345)(451)(512)},
\end{equation}
where $\vltuu_i$,$\vlluu_i$ and $\vleuu_i$ are given by (\ref{kindataSpinorHelicity}) with $n = 3$. Again using the standard gauge fixing for $C'$ - matrix
\begin{eqnarray}
C' = \begin{pmatrix}
1 & 0 & 0 & c'_{14} & c'_{15} \\
0 & 1 & 0 & c'_{24} & c'_{25} \\
0 & 0 & 1 & c'_{34} & c'_{35}
\end{pmatrix} , \qquad C'^{\perp} =
\begin{pmatrix}
-c'_{14} & -c'_{24} & -c'_{34} & 1 & 0 \\\
-c'_{15} & -c'_{25} & -c'_{35} & 0 & 1
\end{pmatrix}
\end{eqnarray}
and integrating over the first 6 $\delta$ - functions we get
%
%
%
%
%
%
%
\begin{eqnarray}\label{CmatrixNMHV3+1}
C' = \begin{pmatrix}
1 & 0 & 0 & c'_{14}=\frac{[15]}{[45]} & c'_{15}=\frac{[14]}{[45]} \\
0 & 1 & 0 & c'_{24}=\frac{[25]}{[45]} & c'_{25}=\frac{[24]}{[45]} \\
0 & 0 & 1 & c'_{34}=\frac{[35]}{[45]} & c'_{35}=\frac{[34]}{[45]}
\end{pmatrix},
\end{eqnarray}
so that
\begin{eqnarray}
A^{*}_{3,3+1} &=&\delta^4 (p_1+p_2+p_3+k)\prod_{i=1}^3\delta^4\left(\tilde{\eta}_i+\tilde{\eta}_4\frac{[i5]}{[45]}+\tilde{\eta}_5\frac{[i4]}{[45]}\right)\times\nonumber\\
&\times&
\frac{\la \xi p\ra}{\kappa^{*}}\frac{(512)}{(412)}\times[45]^2\times
\frac{1}{[45]^3}\times\frac{[45]^5}{[12][23][34][45][51]}
,
\end{eqnarray}
where using (\ref{CmatrixNMHV3+1}) we have
\begin{eqnarray}
\frac{(512)}{(412)}=\frac{[43]}{[53]}.
\end{eqnarray}
In the expression above factor $[45]^2$ comes from  $\delta^4 (C'^{\perp} \cdot \vlluu)$, which will give rise to the total momentum conservation $\delta$ - function, while $1/[45]^3$ is the contribution from integrating over $\delta^6 (C'\cdot \vltuu)$.
Combining all terms together we see that
\begin{eqnarray}
A^{*}_{3,3+1} =\frac{\la \xi p\ra}{\kappa^{*}}\frac{[45]^3}{[12][23][35][51]}
\delta^4 (p_1+p_2+p_3+k)
\prod_{i=1}^3\delta^4\left(\tilde{\eta}_i+\tilde{\eta}_4\frac{[i5]}{[45]}+\tilde{\eta}_5\frac{[i4]}{[45]}\right).\nonumber\\
\end{eqnarray}
Now one can use decomposition (\ref{kT}) of  off-shell momentum $k$ in terms of a pair on-shell momenta\footnote{Here we choose $q$ equal to $q=\xi\tilde{\xi}$, so that
$\kappa=\langle\xi|k|p]/\langle\xi p\rangle$ and $\kappa^*=\langle p|k|\xi]/[p\xi]$.} $p$ and $q=\xi\tilde{\xi}$ together with the definition (\ref{SpinorsInDeformadGrassmannian}) of spinors with numbers 4 and 5 to obtain the following relations
\begin{eqnarray}
[i5]&=&\frac{[i|k|p\rangle }{\langle \xi p \rangle},~~
[i4]=\frac{[i|k|\xi\rangle}{\langle \xi p \rangle},~~[45]=\frac{k^2}{\langle \xi p \rangle},\nonumber\\
k^2&=&\kappa^{*}\kappa,~~k|p\ra=|p]\kappa^{*}.
\end{eqnarray}
Using these relations $A^{*}_{3,3+1}$ could be rewritten in the following way
\begin{eqnarray}
A^{*}_{3,3+1} = \delta^4 (p_1+p_2+p_3+k)
\frac{\kappa^3}{[1 p][p 3][3 2][21]}
\prod_{i=1}^3\delta^4 \left(\vlet_i + \frac{[p i]}{\kappa}\vlet_p\right) , \label{A34spinorhelicity}
\end{eqnarray}
which is simplified further to the form
\begin{eqnarray}
A^{*}_{3,3+1} = \frac{\delta^4 (p_1+p_2+p_3+k)}{[1 2][2 3][3 p][p 1]}\frac{1}{\kappa}\frac{\delta^8(q_{1\ldots3}+\lambda_p\eta_p) \delta^4 ([p 2]\vlet_1 - [p 1]\vlet_2)}{\la 3 p\ra^4}.
\label{A34spinorhelicityFinal}
\end{eqnarray}
This result is in complete agreement with the result obtained previously from BCFW recursion. Indeed, taking for example component proportional to $\tilde{\eta}_1^4\tilde{\eta}_2^4\tilde{\eta}_4^4$ we reproduce (\ref{NMHV4component}).

Similar checks could be performed using momentum twistors. In the case of $k = 2$ the matrix $D$ is zero-dimensional and all consecutive minors of $D$ equal to one, while the nonconsecutive minor is zero. Thus, the integral in  (\ref{GrassmannianMomentumTwistors}) is zero-dimensional while the integrand is $1$, so that the result is given by prefactor $A^{*}_{2,n+1}$.

For $k = 3$ we have
\begin{eqnarray}
D = (d_1 \; d_2 \; \cdots \; d_{n+2}) .
\end{eqnarray}
and
\begin{equation}
\frac{A^{*}_{3,n+1}}{A^{*}_{2,n+1}} =\omega^3_{n+2}[\Gamma_{3,n+2}]= \int_{\Gamma_{3,n+2}}\frac{d^{1\times (n+2)} D}{\text{Vol}[GL(1)]}\frac{1}{1+\frac{\la p\xi\ra}{\la p 1\ra}\frac{d_{n+2}}{d_1}}
\frac{1}{d_1 d_2 \ldots d_{n+2}}\delta^{4|4} (D\cdot \mathcal{Z}) .
\end{equation}
Let us first check this formula for $n=3$. Choosing $\xi = \lambda_2$ (The choice $\xi = \lambda_1$ leads to divergences in denominator) we get
\begin{eqnarray}
\frac{A^{*}_{3,3+1}}{A^{*}_{2,3+1}} = \int\frac{d^{1\times 5} D}{\text{Vol}[GL(1)]}\frac{1}{1+\frac{\la p 2\ra}{\la p 1\ra}\frac{d_{5}}{d_1}}
\frac{1}{d_1 d_2 d_3 d_4 d_5}\delta^{4|4} (d_1 \mathcal{Z}_1 + d_2 \mathcal{Z}_2 + d_3 \mathcal{Z}_3 + d_4 \mathcal{Z}_4 + d_5 \mathcal{Z}_5) . \nonumber \\
\end{eqnarray}
The $GL(1)$ redundancy could be used to fix $d_5 = \la 1 \; 2 \; 3 \; 4\ra$, where the four-bracket is defined as
\begin{eqnarray}
\la i \; j \; k \; l\ra = \varepsilon_{A B C D} \mathcal{Z}_i^A \mathcal{Z}_j^B \mathcal{Z}_k^C \mathcal{Z}_l^D
\end{eqnarray}
The remaining four integration variables are then completely determined by $\delta$ - functions:
\begin{eqnarray}
d_1 = \la 2 \; 3 \; 4 \; 5\ra ,\quad d_2 = \la 3 \; 4 \; 5 \; 1\ra ,\quad
d_3 = \la 4 \; 5 \; 1 \; 2\ra ,\quad d_4 = \la 5 \; 1 \; 2 \; 3\ra
\end{eqnarray}
and $A^{*}_{3,3+1}$ is given by
\begin{eqnarray}
A^{*}_{3, 3+1} = A^{*}_{2, 3+1} \frac{[1 \; 2 \; 3 \; 4 \; 5]}{1+\frac{\la p 2\ra}{\la p 1\ra}\frac{\la 1 \; 2 \; 3 \; 4\ra}{\la 2 \; 3 \; 4 \; 5\ra}} , \label{A34momentumtwistors}
\end{eqnarray}
where the five-bracket is defined as
\begin{eqnarray}
[i \; j \; k \; l \; m] = \frac{\delta^4 (\la i \; j \; k \; l\ra\vle_m + \text{cyclic permutation})}{\la i \; j \; k \; l\ra\la j \; k \; l \; m\ra\la k \; l \; m \; i\ra\la l \; m \; i \; j\ra\la m \; i \; j \; k\ra} .
\end{eqnarray}
To compare (\ref{A34momentumtwistors}) with previously obtained formula (\ref{A34spinorhelicity}) it is convenient to use the following representation for the five-bracket
\begin{multline}
[n \; j-1 \; j \; k-1 \; k] = R_{n j k} = \\ \frac{\la j-1 \; j\ra^4\la k-1 \; k\ra^4\delta^4 (\Xi_{n j k})}{\la n \; j-1 \; j \; k-1\ra\la j-1 \; j \; k-1 \; k\ra\la j \; k-1 \; k \; n\ra\la k-1 \; k \; n \; j-1\ra\la k \; n \; j-1 \; j\ra} ,
\end{multline}
where
\begin{equation}
\Xi_{n j k} = \la n | y_{n k} y_{k j}|j\ra \vlet_j  + \la n|y_{n j} y_{j k} |k\ra \vlet_k - \la n | y_{n k} y_{k j} + y_{n j}y_{j k}|n\ra \vlet_n .
\end{equation}
and $y_{j k} = y_j - y_k = p_j + \ldots + p_{k-1}$. The four-brackets are easily calculated using the identity
\begin{eqnarray}
\la j-1 \; j \; k-1 \; k\ra = \la j-1 \; j\ra\la k-1 \; k\ra y_{j k}^2 .
\end{eqnarray}
This way we get
\begin{eqnarray}
[1 \; 2 \; 3 \; 4 \; 5] = \frac{\la 2 3\ra^4\la 4 5\ra^4 \delta^4 (\Xi_{135})}{\la 1 \; 2 \; 3 \; 4\ra\la 2 \; 3 \; 4 \; 5\ra\la 3 \; 4 \; 5 \; 1\ra\la 4 \; 5 \; 1 \; 2\ra\la 5 \; 1 \; 2 \; 3\ra},
\end{eqnarray}
where
\begin{eqnarray}
\Xi_{135} = \frac{\la 1 2\ra^2\kappa^{*}}{\la 2 p\ra}\Big\{
[p 2]\vlet_1 - [p 1]\vlet_2\Big\} ,
\end{eqnarray}
and
\begin{gather}
\la 1 \; 2 \; 3 \; 4\ra = \la 1 2\ra\la 3 p\ra\la 2 3\ra [3 2] , \quad
\la 2 \; 3 \; 4 \; 5\ra = -\la 1 2\ra\la 3 p\ra\la 2 3\ra [1 3] , \quad
\la 3 \; 4 \; 5 \; 1\ra = - \la 1 2\ra \la 3 p\ra k^2 , \nonumber \\
\la 4 \; 5 \; 1 \; 2\ra = \la 1 2\ra^2 \kappa^{*} [p 1] , \quad
\la 5 \; 1 \; 2 \; 3\ra = \la 1 2\ra^2\la 2 3\ra [1 2] . \nonumber \\
\end{gather}
Finally the expression for $A^{*}_{3,3+1}$ is given by
\begin{eqnarray}
A^{*}_{3, 3+1} = \frac{1}{\kappa}\frac{\delta^4 (p_1+p_2+p_3+k)}{[1 2][2 3][3 p][p 1]}
\frac{\delta^8 (\lambda_1 \vlet_1 + \lambda_2 \vlet_2 + \lambda_3 \vlet_3 + \lambda_p \vlet_p)\delta^4 ([p 2]\vlet_1 - [p 1]\vlet_2) }{\la 3 p\ra^4}.\nonumber\\
\end{eqnarray}

Now let us consider $\mbox{NMHV}_{4+1}$ off-shell amplitude. In this case the Grassmannian integral $\Omega$ is no longer localized on $\delta$ - functions and the result of integration depends on the choice of integration contour $\Gamma$:
\begin{equation}
\Omega^k_{4+1}[\Gamma]=\frac{\la \xi p\ra}{\kappa^{*}}
\int_{\Gamma}\frac{d^{3\times 6} C'}{\text{Vol}[GL (3)]}\frac{(612)}{(512)}
\frac{\delta^6 (C'\cdot \vltuu)
\delta^8 (C\cdot \vleuu)\delta^4 (C'^{\perp} \cdot \vlluu)}{(123)(234)(345)(456)(561)(612)}  ,
\end{equation}
Using integration technique described in appendix \ref{aB} (see also \cite{DualitySMatrix}) the Grassmannian integral may be reduced
to the integral over one complex parameter $\tau$. The minors $(ii+1i+2)$ in the denominator of integrand are linear functions of $\tau$ in this case and the integral over $\tau$ can be easily evaluated by taking residues. Next, we choose integration contour $\Gamma_{135}$ to encircle poles of inverse minors $1/(123)$, $1/(345)$
and $1/(561)$. The corresponding residues will be labeled as $\{1\}$, $\{3\}$ and $\{5\}$. In fact this particular integration contour is the same as the one in the case of $\mbox{NMHV}_6$ on-shell amplitude. The
ratios of minors $(612)/(512)$ evaluated at the mentioned residues are given by
\begin{eqnarray}
\frac{(612)}{(512)}\Big{|}_{\{1\}}=\frac{[45]}{[64]},~~
\frac{(612)}{(512)}\Big{|}_{\{5\}}=\frac{\langle16 \rangle}{\langle15 \rangle}.
\end{eqnarray}

Now, let us consider a particular Grassmann component proportional to $\tilde{\eta}^4_3\tilde{\eta}^4_4\tilde{\eta}^4_5$, which
corresponds to the helicities of on-shell particles $(1^+2^+3^-4^-)$, considered in the section \ref{offshellBCFW}.
Using decomposition (\ref{kT}) of off-shell momentum  $k$ in terms of a pair of on-shell momenta $p$ and $q=\xi\tilde{\xi}$ together with the definition (\ref{SpinorsInDeformadGrassmannian}) of spinors with numbers 5 and 6 (which are similar to $\mbox{NMHV}_4$ case considered before) the results of evaluation of residues could be written as
(hereafter we will drop total momentum conservation
$\delta$ - function)
\begin{eqnarray}
\{1\}&=&\frac{\la \xi p\ra}{\kappa^{*}}\frac{[45]}{[64]}\times\frac{\la 3|1+2|6]^3}{\la 1 2\ra\la 2 3\ra[4 5][5 6] p^2_{1,3} \la 1|2+3|4] }\nonumber\\
&=&\frac{\la \xi p\ra}{\kappa^{*}}
\frac{\langle \xi|k|4]}{\kappa^{*} [p4]}\times
\frac{\la 3|(1+2)k|p\rangle^3 \langle \xi p\rangle^{-3}}
{\la 1 2\ra\la 2 3\ra~\langle \xi|k|4]\langle \xi p\rangle^{-1}~k^2\langle \xi p\rangle^{-1} ~p^2_{1,3} \la 1|2+3|4] }\nonumber\\
&=&\frac{1}{\kappa}\frac{\la 3|1+2|p]^3}{\la 1 2\ra\la 2 3\ra ~[p4]~p^2_{1,3} \la 1|2+3|4]},
\end{eqnarray}
\begin{eqnarray}
\{5\}&=&\frac{\la \xi p\ra}{\kappa^{*}}\frac{\langle16 \rangle}{\langle15 \rangle}\times
\frac{\la 5| 3+4|2]^3}{\la 5 6\ra\la 6 1\ra [2 3] [3 4] p_{2,4}^2 \la 1|2+3|4]}\nonumber\\
&=&\frac{1}{\kappa^{*}}
 \frac{\la p| 3+4|2]^3}{\la p 1\ra [2 3] [3 4] p_{2,4}^2 \la 1|2+3|4]},
\end{eqnarray}
and $\{3\}=0$. So that, finally we get
\begin{eqnarray}
\Omega^3_{4+2}[\Gamma_{135}]\Big{|}_{\tilde{\eta}^4_3\tilde{\eta}^4_4\tilde{\eta}^4_5}
=\{1\}+\{3\}+\{5\}=A^{*}_{3,4+1}(1^+2^+3^-4^-5^*).
\end{eqnarray}
Other helicity configurations as well as supersymmetric (with respect to on-shell particles) result can be obtained in similar fashion.

At the end of this section let us reproduce the result for $A^{*}_{3,4+1}$ off-shell amplitude in a manifestly
supersymmetric way (with respect to on-shell particles) using momentum twistor representation. Considering Grassmannian integral
\begin{eqnarray}
\omega_{4+2}^3[\Gamma] = \int_{\Gamma}\frac{d^{1\times 6} D}{\text{Vol}[GL(1)]}\frac{1}{1+\frac{\la p \xi\ra}{\la p 1\ra}\frac{d_{6}}{d_1}}
\frac{1}{d_1 d_2 d_3 d_4 d_5 d_6}\delta^{4|4} \left(\sum_{i=1}^6 d_i \mathcal{Z}_i \right), \nonumber \\
\end{eqnarray}
and using integration method suggested in \cite{Grassmanians-N4SYM-ABJM} (see also appendix \ref{aB}) the result for integration contour $\Gamma_{246}$ encircling poles in $d_2$, $d_4$ and $d_6$ reads:
\begin{eqnarray}
\omega_{4+2}^3[\Gamma_{246}] = \frac{1}{1+\frac{\la p \xi\ra}{\la p 1\ra}\frac{\langle 1345\rangle}{\langle3456\rangle}}[13456]+\frac{1}{1+\frac{\la p \xi\ra}{\la p 1\ra}\frac{\langle1235\rangle}{\langle2356\rangle}}[12356]+[12345]=\frac{A^{*}_{3,4+1}}{A^{*}_{2,4+1}}.
\end{eqnarray}
We verified that this expression is free from spurious poles of the form $1/\langle abcd\rangle$\footnote{We are going to discuss this in more detail in a separate publication}. To simplify comparison with the results of BCFW recursion lets us rewrite answer for $A_{3,6}$  (\ref{NMHV6super}) in terms of momentum twistor variables:
\begin{eqnarray}
A_{3,6}/A_{2,6}=[12345]+[13456]+[12356].
\end{eqnarray}

In fact it is not difficult to consider a generalization of this result for an arbitrary number of on-shell legs $n$. Choosing integration contour $\Gamma_{3,n+2}$ similar to the case of $[12\rangle$
BCFW shift representation of $\mbox{NMHV}_{n+2}$ the amplitude (with an additional condition to avoid pole $\la p \xi\ra d_{n+2} = -\la p 1\ra d_1$) and following along the same lines as before
(see also discussion in appendix \ref{aB}) the Grassmannian integral:
\begin{equation}
\omega^3_{n+2}[\Gamma]= \int_{\Gamma}\frac{d^{1\times (n+2)} D}{\text{Vol}[GL(1)]}\frac{1}{1+\frac{\la p\xi\ra}{\la p 1\ra}\frac{d_{n+2}}{d_1}}
\frac{1}{d_1 d_2 \ldots d_{n+2}}\delta^{4|4} (D\cdot \mathcal{Z}),
\end{equation}
is evaluated to
\begin{eqnarray}
\omega_{n+2}^3[\Gamma_{3,n+2}] = \sum_{i<j}^{n+2}c_{j}[1\;i-1\;i\;j-1\;j].
\end{eqnarray}
with
\begin{eqnarray}
c_{n+2}= \frac{1}{1+\frac{\la p \xi\ra}{\la p 1\ra}\frac{\langle 1\;i-1\;i\; n+1\rangle}{\langle i-1\;i\; n+1\;n+2\rangle}},~
\mbox{and}~c_{j}=1~\mbox{if}~j<n+2.
\end{eqnarray}
It is natural to conjecture that
\begin{eqnarray}
\omega_{n+2}^3[\Gamma_{3,n+2}] = \frac{A^{*}_{3,n+1}}{A^{*}_{2,n+1}}.
\end{eqnarray}

\subsection{Regularization of soft limit and soft theorems for deformed
Grassmannian integral.}

As was noted in \cite{SoftTheoremsFormFactors,q2zeroFormFactors}, in a similar case of Grassmannian description of form factors of
operators from stress tensor operator supermultiplet the deformation of Grassmannian integral (the combination of non-consecutive minors in addition to the string of consecutive minors in the denominator of integrands in
(\ref{GrassmannianIntegralFinal}) and (\ref{GrassmannianMomentumTwistors})) can be considered as IR regulator of some sort. Namely, it regulates soft limit behavior with respect
to the momentum carried by operator $q\rightarrow0$.

Here we want to argue that similar behavior holds also in the case of deformations of the Grassmannian integrals considered here. Let us show, that the Grassmannian integral (\ref{GrassmannianIntegralFinal}), which we rewrite as
\begin{equation}
\Omega_{n+2}^k[\Gamma]=
\int_{\Gamma}\frac{d^{k\times (n+2)}C'}{\text{Vol}[GL (k)]}Reg.
\frac{\delta^{k\times 2} \left( C'
\cdot \vltuu \right)
\delta^{k\times 4} \left(C' \cdot \vleuu \right)
\delta^{(n+2-k)\times 2} \left(C'^{\perp} \cdot \vlluu \right)}{(1 \cdots k)\cdots (n+1 \cdots k-2) (n+2 \; 1\cdots k-1)},
\end{equation}
with
\begin{equation}\label{RegFunction}
Reg.=\frac{\la\xi p\ra}{\kappa^{*}}\frac{(n+2 \; 1\cdots k-1)}{(n+1 \; 1\cdots k-1)},
\end{equation}
is regular with respect to the holomorphic soft limit of the axillary  four-vector $q$. Here we are using spinorial decomposition of $q$ and $\vlluu_{n+2}$ from (\ref{GrassmannianIntegralFinal}):
$q\equiv\xi\tilde{\xi}$, $\vlluu_{n+2}\equiv\xi$. So, we are expecting to obtain finite limit $\epsilon\rightarrow 0$ for the following expression
\begin{eqnarray}
\Omega^{k}_{n+2}[\Gamma]\Big{|}_{\xi\mapsto \epsilon \xi}.
\end{eqnarray}

The behavior of $\Omega^{k}_{n+2}[\Gamma]$ with respect to holomorphic soft limit of $\vlluu_i$, $i\leq n$ is controlled by standard soft theorems and could be obtained for our Grassmannian integrals using the
method presented in \cite{N4SoftTheoremsGrassmannian} (here to simplify notation we used $\tilde{\lambda}\equiv\vltuuu$):
\begin{eqnarray}
\Omega^{k}_{n+2}[\Gamma_{k,n+2}]\Big{|}_{\lambda_i\mapsto \epsilon \lambda_i}=\left(\frac{\hat{S}_1}{\epsilon^2}+\frac{\hat{S}_2}{\epsilon}\right)
\Omega^{k}_{n+2}[\Gamma_{k,n+1}]+fin.+O(\epsilon),~\epsilon \rightarrow0,
\end{eqnarray}
with
\begin{eqnarray}
\hat{S}_1&=&\frac{\langle i-1i+1\rangle}{\langle ii+1\rangle\langle ii-1\rangle },
\nonumber\\
\hat{S}_2&=&
\frac{\tilde{\lambda}^{\dot{\alpha}}_i}{\langle ii-1\rangle}
\frac{\partial}{\partial\tilde{\lambda}^{\dot{\alpha}}_{i-1}}+
\frac{\tilde{\lambda}^{\dot{\alpha}}_i}{\langle ii+1 \rangle}
\frac{\partial}{\partial\tilde{\lambda}^{\dot{\alpha}}_{i+1}}+
\frac{\eta_{A,i}}{\langle ii-1 \rangle}\frac{\partial}{\partial \eta_{A,i-1}}+
\frac{\eta_{A,i}}{\langle ii+1 \rangle}\frac{\partial}{\partial \eta_{A,i+1}}.\nonumber\\
\end{eqnarray}

Now lets consider behavior of $\Omega^{k}_{n+2}[\Gamma]$ with respect to soft limit of $\xi$. We will consider $k=3$ as an example. It is convenient to fix $GL(3)$ gauge and parametrize $C'$ - matrix as in \cite{N4SoftTheoremsGrassmannian}  (the columns are numerated as $(1,2,\ldots,n,n+1,n+2)$):
\begin{eqnarray}
    C'=\left( \begin{array}{ccccccc}
        0 & c_{n2}  &\ldots& 1 & 0 & c_{nn+2} \\
        0 & c_{n+12}  &\ldots& 0 & 1 & c_{n+1n+2} \\
        1 & c_{12}  &\ldots& 0 & 0 & c_{1n+2}\end{array} \right).
\end{eqnarray}
In such parametrization the minors in $Reg.$ are given by
\begin{eqnarray}
    (n+212)=c_{n2}c_{n+1n+2}-c_{n+12}c_{nn+2},~(n+112)=c_{n2}.
\end{eqnarray}
Then $\Omega^{3}_{n+2}$ could be rewritten as (here as in \cite{N4SoftTheoremsGrassmannian} we suppress explicit dependence on Grassmann variables)
\begin{eqnarray}
   \Omega^{3}_{n+2}[\Gamma_{3,n+2}]\Big{|}_{\xi\mapsto \epsilon \xi}&=&
 \int d^3c_{In+2}\delta^2(\epsilon~\xi-\vlluu_Ic_{In+2})\frac{\epsilon Reg.(nn+11)'(n+112)'}{(nn+1n+2)(n+1n+21)(n+212)}
\nonumber\\
&\times&
\hat{\Omega}^{3}_{n+1}[\Gamma_{3,n+1}]\Big{|}_{\xi\mapsto \epsilon \xi},
\end{eqnarray}
where (index $I$ runs over $1,n,n+1$)
\begin{eqnarray}
   \int d^3c_{In+2}=\int dc_{1n+2}dc_{n+1n+2}dc_{nn+2},
\end{eqnarray}
and minors with primes like $(nn+11)'$ are evaluated in  $Gr(3,n+1)$ Grassmannian in contrast to other minors  evaluated in $Gr(3,n+2)$ Grassmannian,
$\Gamma_{3,n+2}$ contour contains the same poles as $\Gamma_{3,n+1}$ together with additional pole $(n+1\;n+2\;1)$. Here, we have also used the fact that
$\la\xi p\ra/\kappa^{*}\Big{|}_{\xi\mapsto \epsilon \xi}=\epsilon \la\xi p\ra/\kappa^{*}$. The hat in $\hat{\Omega}^{3}_{n+1}$ denotes the absence of $Reg.$ factor and that $\vltuuu_1$ and $\vltuuu_{n+1}$ spinors are shifted as
\begin{eqnarray}\label{SoftShift}
   \vltuuu_1&\mapsto& \vltuuu_1+c_{1n+2}\vltuuu_{n+2},\nonumber\\
   \vltuuu_{n+1}&\mapsto&\vltuuu_{n+1}+c_{n+1n+2}\vltuuu_{n+2}.
\end{eqnarray}
The integral $\int d^3c_{In}$ can be evaluated by taking residue at pole $1/(n+1n+21)$, which
fixes the values of $c_{nn+2}$ and $c_{n+1n+2}$, $c_{1n+2}$ coefficients to be (we
use spinor definitions from (\ref{SpinorsInDeformadGrassmannian}))
\begin{eqnarray}\label{CcoefficientsInSoftLimit}
   c_{nn+2}=0,~c_{n+1n+2}=\frac{\langle 1\xi \rangle \epsilon}{\langle 1p \rangle},
   ~c_{1n+2}=\frac{\langle p\xi \rangle \epsilon}{\langle 1p \rangle}.
\end{eqnarray}
Then the result of integration could be written as
\begin{equation}
\int d^3c_{In+2}\delta^2(\epsilon~\xi-\vlluu_Ic_{In+2})\frac{\epsilon Reg.(nn+11)'(n+112)'}{(nn+1n+2)(n+1n+21)(n+212)}=
\frac{\langle 1p\rangle}{\epsilon^2\langle 1\xi \rangle \langle \xi p\rangle}
~\epsilon Reg.,\nonumber\\
\end{equation}
with $Reg.$ function being evaluated at $(n+1n+21)$ residue, which is given by
\begin{eqnarray}
\epsilon Reg.\Big{|}_{(n+1n+21)} =\epsilon^2 \frac{\la\xi p\ra}{\kappa^{*}}\frac{\langle 1\xi \rangle}{\langle 1p \rangle}.
\end{eqnarray}
So, taking $\epsilon \rightarrow 0$
limit and accounting for (\ref{SoftShift}) and (\ref{CcoefficientsInSoftLimit}) we get
\begin{eqnarray}
\Omega^{3}_{n+2}[\Gamma_{3,n+2}]\Big{|}_{\xi\mapsto \epsilon \xi}=\frac{1}{\kappa^{*}}\hat{\Omega}^{3}_{n+1}[\Gamma_{3,n+1}]\Big{|}_{\epsilon=0}+O(\epsilon).
\end{eqnarray}
Thus, we see that presented here deformation of Grassmannian integral could be also considered as a regularization of soft limit behavior. The case of general $k$ is  more involved, but we should get similar result, that is
\begin{eqnarray}
\Omega^{k}_{n+2}[\Gamma_{k,n+2}]\Big{|}_{\xi\mapsto \epsilon \xi}=fin.+O(\epsilon).
\end{eqnarray}
The soft limit considered could also serve as a prescription of how to obtain corresponding on-shell amplitudes from our off-shell expressions. Indeed, $\hat{\Omega}^{3}_{n+1}[\Gamma_{3,n+1}]\Big{|}_{\epsilon=0}$ in the above expressions is nothing else but the  Grassmannian representation of $\mbox{NMHV}_{n+1}$ amplitude with $n+1$ on-shell particles, where momentum of $i$'th particle $p_{n+1}$ is given by
\begin{eqnarray}
p_{n+1}=\lambda_p\frac{k|\xi\rangle}{\langle \xi p \rangle}.
\end{eqnarray}
This observation is in agreement with the on-shell limit prescription of \cite{vanHamerenBCFW}. In addition we would like to note, that double soft limit with respect to $\vlluu_{n+2}$ and $\vlluu_{n+1}$ will be singular and controlled by soft theorems.

\section{Amplitudes with one leg off-shell and integrability}

Yangian symmetry, which combines invariance under superconformal and dual superconformal transformations \cite{DualSuperConformalSymmetry}, for on-shell tree-level amplitudes was proven in \cite{YangianSymmetryTreeAmplitudes}. It was further claimed \cite{Drummond_Grassmannians_Tduality,Drummond_Yangian_origin_Grassmannian_integral}, that the Grassmannian integral representation for amplitudes (\ref{GrassmannianIntegralAmplitudes}) is the most general form of rational Yangian invariant making all symmetries of the theory manifest. The existence of Yangian symmetry allows us to reformulate the problem of finding expressions for the scattering amplitudes in the language of integrable systems, in particular in the language of Quantum Inverse Scattering Method (QISM).

The study of tree-level scattering amplitudes within the context of QISM was started in \cite{AmplitudesSpectralParameter1,AmplitudesSpectralParameter2}, where the notion of spectral parameter was introduced. The latter was interpreted as a deformed particle helicity. Later the authors of \cite{Frassek_BetheAnsatzYangianInvariants,Chicherin_YangBaxterScatteringAmplitudes} proposed to study certain auxiliary spin chain monodromies build from local Lax operators. The introduced monodromies depended on an auxiliary spectral parameter, while the spectral parameters of \cite{AmplitudesSpectralParameter1,AmplitudesSpectralParameter2} played the role of inhomogeneities of Lax operators.  Yangian invariants and thus on-shell amplitudes are then found as the eigenstates of these monodromies.
Further, in \cite{Kanning_ShortcutAmplitudesIntegrability,Broedel_DictionaryRoperatorsOnshellGraphsYangianAlgebras} a systematic classification of Yangian invariants obtained within QISM was provided. Yangian invariance can be defined in a very compact form as a system of eigenvalue equations for the elements of a suitable monodromy matrix $M(u)$ \cite{Frassek_BetheAnsatzYangianInvariants,Chicherin_YangBaxterScatteringAmplitudes,Kanning_ShortcutAmplitudesIntegrability}:
\begin{eqnarray}
M_{ab}(u)|\Psi\rangle = {C}_{ab}|\Psi\rangle, \label{monodromyEq}
\end{eqnarray}
where $u$ is the auxiliary spectral parameter mentioned above, $C_{ab}$ are monodromy eigenvalues and monodromy eigenvectors $|\Psi\rangle$ are elements of the Hilbert space $V = V_1\otimes\ldots\otimes V_n$ with $V_i$ being a representation space of a particular $\mathfrak{gl}(N|M)$ representation. To describe tree-level scattering amplitudes one will need to specialize to the case of $N|M = 4|4$ and its non-compact representations  build using a single family of Jordan-Schwinger harmonic superoscillators $\overline{\bf w}^\mathcal{A}, {\bf w}^\mathcal{B}$, $\mathcal{A},\mathcal{B} = 1\ldots 8$. The latter could be conveniently realized in terms of Heisenberg pairs
\begin{eqnarray}
J^{\mathcal{AB}} = \overline{\bf w}^\mathcal{A} {\bf w}^\mathcal{B} = x^\mathcal{A} p^\mathcal{B},\quad x^\mathcal{A} = \left(\vll^{\alpha},-\frac{\partial}{\partial\vlt^{\dotalpha}}, \frac{\partial}{\partial\vlet^A} \right) , \quad p^\mathcal{A} = \left(\frac{\partial}{\partial\vll^{\alpha}}, \vlt^{\dotalpha} , \vlet^A\right),
\end{eqnarray}
such that $[x^{\mathcal{A}}, p^{\mathcal{B}}\} = (-1)^{|\mathcal{A}|}\delta^{\mathcal{AB}}$. Here $[\cdot , \cdot\}$ denotes graded commutator and $|\cdot|$ - grading. A vacuum state for the Hilbert space $V$ used to construct Yangian invariants $|\Psi\rangle_{n,k}$ corresponding to the on-shell N$^{k-2}$MHV $n$-point tree-level amplitudes $A_{n,k}$ is given by
\begin{eqnarray}
|{\bf 0}\rangle_{k,n} = \delta_1^+ \cdots \delta_{n-k}^+\delta_{n-k+1}^-\cdots \delta_n^- ,
\end{eqnarray}
where $\delta_i^+ \equiv \delta^2 (\vll_i)$ is the vacuum for positive helicity state at position $i$ and $\delta_i^- \equiv \delta^2 (\vlt_i)\delta^4 (\vlet_i)$ is the corresponding vacuum for negative helicity state at the same position. The monodromy matrix of the auxiliary spin chain reads
\begin{eqnarray}
M(u,\{v_i\}) = \mathcal{L}_1 (u,v_1)\ldots\mathcal{L}_k (u,v_k)
\mathcal{L}_{k+1} (u,v_{k+1})\ldots\mathcal{L}_n (u,v_n), \label{monodromy-matrix}
\end{eqnarray}
where $u$ is again the auxiliary spectral parameter, $v_i$ are spin chain inhomogeneities and Lax operators $\mathcal{L}_i (u,v)$ are given by
\begin{eqnarray}
\mathcal{L} (u,v) = u - v + \sum_{a,b} e_{ab} J_{ba}\, , \label{LaxOperator}
\end{eqnarray}
where matrix $e_{ab}$ acting in the auxiliary space is given by $(e_{ab})_{cd} = \delta_{ac}\delta_{bd}$. It is easy to see, that the action of Lax operators on particles vacuums is given by
\begin{eqnarray}
\mathcal{L}_i (u)\;\delta_i^+ = (u-1)\; \mathbb{I}\;\delta_i^+, \qquad
\mathcal{L}_i (u)\;\delta_i^- = u\;\mathbb{I}\;\delta_i^- . \label{Lax-vacua}
\end{eqnarray}
The solution of the eigenvalue equation\footnote{It should be noted that this eigenvalue equation is different from the usual Bethe ansatz equations, which diagonalize only the trace of monodromy matrix.} for monodromy matrix (\ref{monodromyEq}) leads to the expressions for Yangian invariants labeled by the permutations $\sigma$ with minimal\footnote{The decomposition is minimal in a sense, that there exists no other decomposition of $\sigma$ into a smaller number of transpositions.} decomposition $\sigma = (i_1,j_1)\ldots (i_P,j_P)$ \cite{Chicherin_YangBaxterScatteringAmplitudes,Kanning_ShortcutAmplitudesIntegrability,Broedel_DictionaryRoperatorsOnshellGraphsYangianAlgebras}:
\begin{eqnarray}
|\Psi\rangle = \mathcal{R}_{i_1,j_1}(\bar{u}_1)\ldots\mathcal{R}_{i_Pj_P}(\bar u_P)|{\bf 0}\rangle_{k,n}
\end{eqnarray}
with \cite{Chicherin_YangBaxterScatteringAmplitudes} (see also \cite{Frassek_BetheAnsatzYangianInvariants})
\begin{eqnarray}
\mathcal{R}_{ij} (u) = \Gamma (-u)(x_j\cdot p_i)^u = \int_0^\infty \frac{d\alpha}{\alpha^{1+u}}e^{-\alpha (x_j\cdot p_i)} , \label{RoperatorDef}
\end{eqnarray}
where $\Gamma$ is the Euler gamma function and
\begin{eqnarray}
\bar u_p = v_{\tau_p (i_p)} - v_{\tau_p  (j_p)},\qquad  \tau_p=\tau_{p-1}\circ(i_p,j_p)=(i_1,j_1)\cdots(i_p,j_p).
\end{eqnarray}
As we already mentioned, amplitudes with one leg off-shell can be considered as form factors of Wilson line operator (\ref{WilsonLineOperDef}) corresponding to the off-shell leg. It turns out, that QISM machinery could be also used in the case of form factors \cite{FormFactorsGrassmanians}, see also \cite{SoftTheoremsFormFactors}. The only new ingredient needed is the spin chain vacuum state corresponding to the minimal form factor. The latter in the case of form factors of stress-tensor operator supermultiplet is given by \cite{FormFactorsGrassmanians}:
\begin{eqnarray}
\mathcal{F}_{2,2} (1,2) = \delta^2 (\vltu_1)\delta^4 (\vleu_1)\delta^2 (\vltu_2) \delta^4 (\vleu_2) ,
\end{eqnarray}
where\footnote{See appendix A of \cite{SoftTheoremsFormFactors} for the notation used in harmonic superspace description of form factors of operators from stress-tensor operator supermultiplet.}
\begin{eqnarray}
\begin{aligned}
\underline{\tilde{\lambda}}_1 = \tilde{\lambda}_1 - \frac{\la 2| q}{\la 2 | 1\ra} \quad
\underline{\eta}_1^{-} = \eta_1^{-} - \frac{\la 2|\gamma^{-}}{\la 2 |1\ra}\quad \underline{\eta}_1^{+} = \eta_1^{+} \\
 \underline{\tilde{\lambda}}_2 = \tilde{\lambda}_2 - \frac{\la 1| q}{\la 1 | 2\ra} \quad
\underline{\eta}_2^{-} = \eta_2^{-} - \frac{\la 1|\gamma^{-}}{\la 1 |2\ra}\quad \underline{\eta}_2^{+} = \eta_2^{+}
\end{aligned}
\end{eqnarray}
The vacuum state for the deformed minimal amplitude (vertex) with one leg off-shell could be easily obtained by performing integrations in (\ref{deformed-offshell-vertex}). This way we get:
\begin{eqnarray}
A^*_{2,2+1}(1^*,2,3) = \frac{1}{\kappa^*}\frac{\la 2 3\ra}{\la p 2\ra\la p 3\ra}\left(\frac{\la p 3\ra}{\la p 2\ra}\right)^{u_{32}}
\left(\frac{\la 2 3\ra}{\la p 3\ra}\right)^{u_{31}}
\delta^2 (\vltu_2)\delta^2 (\vltu_3)\delta^4 (\vleu_2)\delta^4 (\vleu_3) , \nonumber \\
\end{eqnarray}
where ($k$ is the off-shell gluon momentum and $p$ is its direction as before)
\begin{eqnarray}
\vltu_2 = \vlt_2 + \frac{\la 3|k}{\la 3 2\ra},
\qquad \vltu_3 = \vlt_3 + \frac{\la 2|k}{\la 2 3\ra}
 \qquad \vleu_2 = \vlet_2 + \frac{\la p 3\ra}{\la 2 3\ra}\vlet_p ,  \qquad \vleu_3 = \vlet_3 + \frac{\la p 2\ra}{\la 3 2\ra}\vlet_p . \nonumber \\ \label{lambda-eta-offshell-vertex}
\end{eqnarray}
Here we will however restrict ourselves to the case of undeformed minimal off-shell amplitude given by
\begin{eqnarray}
A^*_{2,2+1}(2,3) = A^*_{2,2+1}(1^*,2,3) = \frac{1}{\kappa^*}\frac{\la 2 3\ra}{\la p 2\ra\la p 3\ra}
\delta^2 (\vltu_2)\delta^2 (\vltu_3)\delta^4 (\vleu_2)\delta^4 (\vleu_3).
\end{eqnarray}
Next, similar to the case of form factors of operators from stress-tensor operator supermultiplet \cite{FormFactorsGrassmanians} let us consider off-shell amplitudes defined as\footnote{Actually it is just one of BCFW contributions and the expression for amplitude is obtained as linear combination of such terms.}
\begin{eqnarray}
\hat{\mathcal{A}} = \mathcal{R}_{i_1 j_1}(\bar u_1)\cdots\mathcal{R}_{i_P j_P}\; A^{*,\delta}_{2,2+1} (m-1, m) \label{amplitudesvacua}
\end{eqnarray}
with
\begin{eqnarray}
A^{*,\delta}_{2,2+1}(m-1,m) = \delta_1^+\cdots\delta_{m-2}^+\;
A^*_{2,2+1}(m-1,m)\;\delta_{m+1}^-\cdots\delta_n^- .
\end{eqnarray}
As was shown in \cite{Chicherin_YangBaxterScatteringAmplitudes,Kanning_ShortcutAmplitudesIntegrability,Broedel_DictionaryRoperatorsOnshellGraphsYangianAlgebras,FormFactorsGrassmanians}, the monodromy matrix (\ref{monodromy-matrix}) (as a consequence of Yang-Baxter equation) satisfies the following relation
\begin{eqnarray}
M (u, \{v_i\}) \; \mathcal{R}_{i_1 j_1} (\bar u_1)\cdots
\mathcal{R}_{i_P j_P} (\bar u_P) = \mathcal{R}_{i_1 j_1} (\bar u_1)\cdots \mathcal{R}_{i_P j_P}(\bar u_P) \; M (u, \{v_{\sigma (i)}\}) ,
\end{eqnarray}
where $M (u, \{v_{\sigma (i)}\})$ is the monodromy matrix with inhomogeneities $v_i$ replaced with $v_{\sigma (i)}$. Now, commuting monodromy matrix through $\mathcal{R}$-chain in (\ref{amplitudesvacua}) we get
\begin{eqnarray}
M_n (u, \{v_i\})\;\hat{\mathcal{A}} =  \mathcal{R}_{i_1 j_1}(\bar u_1)\cdots \mathcal{R}_{i_P j_P} (\bar u_P) \; M_n (u, \{v_{\sigma (i)}\})\; A^{*,\delta}_{2, 2+1}(m-1,m) .
\end{eqnarray}
Taking into account, that Lax operators act diagonally on vacua (\ref{Lax-vacua}) we can write:
\begin{multline}
M_n (u, \{v_{\sigma (i)}\})\; A^{*,\delta}_{2, 2+1}(m-1,m) =
\prod_{i=1}^{m-2} (u - v_{\sigma (i)} - 1)
\prod_{j=m+1}^n (u-v_{\sigma (j)}) \\
\times \delta_1^+\cdots \delta_{m-2}^+
\Bigg[
M_2 (u, \{v_{\sigma (i)}\}) A^*_{2, 2+1}(m-1,m)
\Bigg] \delta_{m+1}^-\cdots\delta_n^- ,  \label{monodromy-commute}
\end{multline}
where length 2 monodromy matrix is given by
\begin{eqnarray}
M_2 (u, \{v_{\sigma (i)}\}) = \mathcal{L}_{m-1} (u-v_{\sigma (m-1)})
\mathcal{L}_m (u - v_{\sigma (m)}) .
\end{eqnarray}
The minimal off-shell vertex $A^*_{2, 2+1}(m-1,m)$ is not an eigenstate of monodromy matrix $M_2 (u, \{v_{\sigma (i)}\})$ and Yangian invariance for off-shell amplitudes is broken similar to the case of form factors with $q^2\neq 0$ \cite{FormFactorsGrassmanians}. This conclusion easily follows from the fact, that the momentum-like generators do not contain off-shell momentum and the action of monodromy matrix on minimal off-shell vertex produces \cite{FormFactorsGrassmanians}
\begin{eqnarray}
(\vll_{m-1}\vlt_{m-1}+\vll_m\vlt_m)\delta^4 (\vll_{m-1}\vlt_{m-1}+\vll_m\vlt_m + k) ,
\end{eqnarray}
which does not vanish. On the other hand, again similar to the case of form factors with $q^2\neq 0$ the amplitudes with one leg off-shell turn out to be eigenvectors of transfer matrix. The later is defined as the super trace of monodromy matrix over auxiliary space:
\begin{eqnarray}
\mathcal{T}_n (u, \{v_i\}) = \text{str} M_n (u, \{v_i\}) .
\end{eqnarray}
One can use Yang-Baxter equation to show, that this transfer matrix is $\mathfrak{gl}(4|4)$ invariant, that is
\begin{eqnarray}
[\mathcal{T}_n (u, \{v_i\}), \sum_{i=1}^n x_i^{\mathcal{A}}p_i^{\mathcal{B}}] = 0 .
\end{eqnarray}
To show, that amplitudes with one leg off-shell are annihilated by transfer matrix, let us consider first the minimal off-shell vertex and length 2 transfer matrix with equal inhomogeneities
\begin{eqnarray}
\mathcal{T}_2 (u-v) = \text{str} \mathcal{L}_m (u - v)
\mathcal{L}_{m-1} (u - v).
\end{eqnarray}
Next, we can exploit $\mathfrak{gl}(4|4)$ invariance of transfer matrix and consider the particular component of the minimal off-shell amplitude $A^*_{2, 2+1}(m-1,m)$, for example:
\begin{eqnarray}
\frac{1}{\kappa^*}\frac{\la p \; m-1\ra\la p \; m\ra}{\la m-1\; m\ra}\;
\vlet_{m-1}^2\vlet_m^2\vlet_p^4 \;\delta^4 (\vll_{m-1}\vlt_{m-1}+\vll_m\vlt_m+k) ,
\end{eqnarray}
that is scalar - scalar - off-shell gluon vertex. Here, superscripts next to Grassmann variables are the numbers of copies of corresponding Grassmann variables and not $SU (4)_R$ indexes. Similar to \cite{FormFactorsGrassmanians} we have
\begin{eqnarray}
\mathcal{T}_2 (u-v)\; \delta^4 (\vll_{m-1}\vlt_{m-1} + \vll_m\vlt_m + k) = 0 ,
\end{eqnarray}
\begin{multline}
\mathcal{T}_2 (u-v) \;\vlet_{m-1}^2\vlet_m^2\vlet_p^4 = \Bigg(
(u-v-1) x_{m-1}^{\mathcal{A}}p_{m-1}^{\mathcal{A}} +
(-1)^{\mathcal{A}} p_{m-1}^{\mathcal{A}}x_{m-1}^{\mathcal{B}}
p_m^{\mathcal{B}}x_m^{\mathcal{A}}
\Bigg) \;\vlet_{m-1}^2\vlet_m^2\vlet_p^4 = 0
\end{multline}
and we have also checked, that
\begin{eqnarray}
\mathcal{T}_2 (u-v) \; \frac{\la p \; m-1\ra\la p \; m\ra}{\la m-1\; m\ra} = 0
\end{eqnarray}
This shows, that the minimal off-shell amplitude $A^*_{2, 2+1}(m-1,m)$ is an eigenstate of the transfer matrix, that is
\begin{eqnarray}
\mathcal{T}_2 (u-v) A^*_{2, 2+1}(m-1,m)  =  0.
\end{eqnarray}
Moreover, as the transfer matrix satisfies a relation similar to  (\ref{monodromy-commute}) the same statement also applies to any planar on-shell diagram with a minimal off-shell vertex insertion:
\begin{eqnarray}
\mathcal{T}_n (u, \{v_i\})\hat{\mathcal{A}} = 0.
\end{eqnarray}
with $v_i = v$. This property is in fact a consequence of multiplicative renormalizability of minimal off-shell vertex, similar to multiplicative renormalizability of stress-tensor operator supermultiplet  in \cite{FormFactorsGrassmanians}.

At the end of this section let us show on a simple example of $A^*_{2,3+1}$ off-shell amplitude how one can use QISM technique to get explicit expressions for  spectral deformations of off-shell amplitudes and Yangian invariants they are build from. First we perform the minimal decomposition of corresponding permutation $\sigma = (3,1,2) = (2,3)(1,2)$. Then $A^*_{2,3+1}(\bar u_1,\bar u_2)$ is given by
\begin{eqnarray}
A^*_{2,3+1}(\bar u_1,\bar u_2) = \mathcal{R}_{23} (\bar u_1)\mathcal{R}_{12} (\bar u_2)
\delta^2 (\vll_1)\frac{1}{\kappa^*}\frac{\la 2 3\ra}{\la p 2\ra\la p  3\ra}\delta^2 (\vltu_2)\delta^2 (\vltu_3)\delta^4 (\vleu_2) \delta^4 (\vleu_3) ,
\end{eqnarray}
where $\vllu_i$, $\vleu_i$ are defined in (\ref{lambda-eta-offshell-vertex}) and $\bar u_1 = v_{32} = v_3 - v_2$, $\bar u_2 = v_{31} = v_3 - v_1$. Using definition of $\mathcal{R}$ operators (\ref{RoperatorDef}) we get:
\begin{multline}
	A_{2,3+1}^*(v_1,v_2,v_3) =
	\frac{1}{\kappa^{*}}\int\frac{d\alpha_2}{\alpha_2}
	\int\frac{d\alpha_1}{\alpha_1}\frac{\la 2 3\ra}{\la p 2\ra\la p 3\ra \left(1-\alpha_2\frac{\la p 3\ra}{\la p 2\ra}\right)}\times  \\  \times \delta^2 (\vll_1 - \alpha_1\vll_2 + \alpha_1\alpha_2\vll_3)
	\delta^2 (\vltu_2 + \alpha_1\vlt_1) \delta^4 (\vleu_2 + \alpha_1 \vlet_1)
	\delta^2 (\vltu_3 + \alpha_2\vltu_2)\delta^4 (\vleu_3 + \alpha_2\vleu_2)  \\
	=  \frac{1}{\kappa^*}\frac{1}{\la p 1\ra\la 1 2\ra\la 2 3\ra\la 3 p\ra}\left(\frac{\la 2 3\ra}{\la 1 3\ra}\right)^{v_{31}}
	\left(\frac{\la 1 3\ra}{\la 1 2\ra}\right)^{v_{32}}
	\delta^4 (\sum_{i=1}^3\vll_i\vlt_i + k)
	\delta^8 (\sum_{i=1}^3\vll_i\vlet_i + \vll_p\vlet_p) . \\
\end{multline}
The off-shell amplitude $A^*_{2,3+1}$ is then recovered by setting to zero deformation parameters $v_i = 0$.

\section{Conclusion}
In this article we considered Grassmannian integral representation
for tree level gauge invariant $\mathcal{N}=4$ SYM   off-shell amplitudes with one leg off-shell (Wilson line form factors).
We presented a conjecture for Grassmannian integral representation
for amplitudes with all on-shell particles treated in a manifestly supersymmetric manner and the only off-shell gluon (Wilson line insertion) treated in non-supersymmetric way. We have considered spinor helicity, twistor and momentum twistor versions of Grassmannian integral representation and successfully checked them by reproducing BCFW results for $\mbox{MHV}_{n+1}$, $\mbox{NMHV}_{3+1}$ and
$\mbox{NMHV}_{4+1}$ gauge invariant off-shell amplitudes. In addition from our Grassmannian integral representation we reproduced appropriate soft (on-shell) limit. Using Grassmannian integral representation we have also derived a conjecture for $\mbox{NMHV}_{n+1}$ gauge invariant off-shell amplitudes. We have investigated integrability properties of gauge invariant off-shell amplitudes and showed that similar to the case of form factors of local gauge invariant operators
gauge invariant off-shell amplitudes are no longer eigenvectors of monodromy matrix of the auxiliary $\mathfrak{gl}(4|4)$ spin chain. The latter, however, turn out to be eigenvectors of corresponding transfer matrix.

There are several possible generalizations of the construction and ideas presented in this article. First, it would be interesting to investigate loop corrections to the gauge invariant off-shell amplitudes. Off-shellness of the gluon should play the role of
natural IR regularization\footnote{This brings up the question of existence of IR regularization preserving all symmetries of tree level amplitudes in $\mathcal{N}=4$ SYM.}. This will help to clarify the relation between Grassmannian integrals considered here and leading singularities of off-shell loop amplitudes. The rigorous supersymmetric consideration of the off-shell gluon (Wilson line) is another direction to follow. Finally, it would be extremely interesting to consider generalization of the Grassmannian integral representation and spin chain construction considered here to the case of gauge invariant off-shell amplitudes with an arbitrary number of off-shell gluons (Wilson line insertions).

At the end we want to mention an important  conceptual question, which arises in this and similar studies \cite{HarmonyofFF_Brandhuber,BORK_NMHV_FF,BORK_POLY,FormFactorsGrassmanians}. We have seen, that structures
such as Grassmannian integrals and spin chains considered here, which (at least naively) could be considered as a purely on-shell objects, also appear for partially off-shell objects, such as form factors and off-shell amplitudes. Next, the natural question is how far on-shell techniques could be extended in the case of off-shell kinematics.

\section*{Acknowledgements}

The authors would like to thank D.I. Kazakov and S.E. Derkachov both for drawing our attention to this problem as well  as for interesting and stimulating discussions. This work was supported by RFBR grants \# 14-02-000494 , \# 16-02-00943 and contract \# 02.A03.21.0003 from 27.08.2013 with Russian Ministry of Science and Education.

\appendix

\section{$\mathcal{N}=4$ SYM and spinor helicity formalism}\label{aA}

$\mathcal{N}=4$ supersymmetric gauge theory is a maximally supersymmetric gauge theory in four spacetime dimensions. The field content of $\mathcal{N}=4$ SYM consists from six scalars $\phi^{AB}$ (antisymmetric in the $SU(4)_R$ indices $A,B = 1\ldots 4$), four Weyl fermion fields $\psi_{\alpha}^A$ and gauge field strength tensor $F^{\mu\nu}$, all transforming in the adjoint representation of the $SU(N_c)$ gauge group. The lagrangian of $\mathcal{N}=4$ SYM is given by \cite{N4SYM_1,N4SYM_2}:
\begin{eqnarray}
\mathcal{L}_{\mathcal{N}=4~\text{SYM}} &=& \mbox{tr}\Big\{
-\frac{1}{4}F_{\mu\nu}F^{\mu\nu} + \frac{1}{4}(D_{\mu}\phi^{AB})(D^{\mu}\bar{\phi}_{AB})
+ \frac{1}{32}g^2[\phi^{AB},\phi^{CD}][\bar{\phi}_{AB},\bar{\phi}_{CD}]
 \nonumber \\
&&  +  i\bar{\psi}_{\dotalpha A}\sigma_{\mu}^{\dotalpha\beta}D^{\mu}\psi_{\beta}^A
- \frac{1}{2}g\psi^{\alpha A} [\bar{\phi}_{AB}, \psi_{\alpha}^B]
+ \frac{1}{2}g\bar{\psi}_{\dotalpha A}[\phi^{AB}, \bar{\psi}_B^{\dotalpha}]
\Big\} , \label{N4SYMlagrangian}
\end{eqnarray}
where the field strength is $F_{\mu\nu} = \partial_{\mu}A_{\nu} - \partial_{\nu}A_{\mu} - \frac{i g}{\sqrt{2}}[A_{\mu},A_{\nu}]$ and the covariant derivative is $D_{\mu} = \partial_{\mu} - \frac{i g}{\sqrt{2}} [A_{\mu}, \cdot ]$. All fields are matrix valued in $SU(N_c)$ gauge group, i.e. $\Phi\equiv \Phi^a t^a$  with generators normalized as $\text{tr}~t^at^b = \delta^{ab}$. $\bar{\psi}_{\dotalpha A} = (\psi_{\alpha}^A)^{*}$ and
\begin{eqnarray}
\bar{\phi}_{AB} = (\phi^{AB})^{*} = \frac{1}{2}\epsilon_{ABCD}\phi^{CD},\quad \epsilon_{ABCD} = \epsilon^{ABCD}
\end{eqnarray}
The lagrangian of $\mathcal{N}=4$ is invariant under the following supersymmetry transformations
\begin{eqnarray}
\delta A^{\mu} &=& -i \xi^{\alpha A} \bar{\sigma}^{\mu}_{\alpha\dotbeta}\bar{\psi}_A^{\dotbeta} - i \bar{\xi}_{\dotalpha A}\sigma^{\mu~\dotalpha\beta}\psi_{\beta}^A , \nonumber \\
\delta\phi^{AB} &=& -i\sqrt{2}\left\{
\xi^{\alpha A}\psi_{\alpha}^B - \xi^{\alpha B}\psi_{\alpha}^A
-\epsilon^{ABCD}\bar{\xi}_{\dotalpha C}\bar{\psi}_D^{\dotalpha}
\right\} , \nonumber \\
\delta\psi_{\alpha}^A &=& \frac{i}{2}F_{\mu\nu}{\sigma^{\mu\nu}}_{\alpha}^{~\beta}\xi_{\beta}^A
-\sqrt{2}(D_{\mu}\phi^{AB})\bar{\sigma}^{\mu}_{\alpha\dotbeta}\bar{\xi}_B^{\dotbeta}
+ \frac{i g}{\sqrt{2}} [\phi^{AB},\bar{\phi}_{BC}]\xi_{\alpha}^C ,
\nonumber \\
\delta\bar{\psi}_A^{\dotalpha} &=& \frac{i}{2}F_{\mu\nu}{\bar{\sigma}^{\mu\nu\dotalpha}}_{~~~~\dotbeta}\bar{\xi}^{\dotbeta}_A
+\sqrt{2}(D_{\mu}\bar{\phi}_{AB})\sigma^{\mu\dotalpha\beta}\xi^B_{\beta}
+ \frac{i g}{\sqrt{2}} [\bar{\phi}_{AB},\phi^{BC}]\bar{\xi}^{\dotalpha}_C ,
\end{eqnarray}
where
\begin{eqnarray}
\sigma^{\mu ~\dotalpha\beta} = \epsilon^{\beta\gamma}\bar{\sigma}^{\mu}_{\gamma\dotdelta}\epsilon^{\dotdelta\dotalpha}, \quad
\bar{\sigma}^{\mu}_{~\alpha\dotbeta} = \epsilon_{\dotbeta\dotgamma}\sigma^{\mu ~\dotgamma\delta}
\epsilon_{\delta\alpha} ,
\end{eqnarray}
and
\begin{eqnarray}
{\sigma^{\mu\nu}}_{\alpha}^{~\beta}\equiv \frac{i}{2}
\left[{\bar{\sigma}^{\mu}}_{~~\alpha\dotgamma}\sigma^{\nu ~\dotgamma\beta}
- \bar{\sigma}^{\nu}_{~\alpha\dotgamma}\sigma^{\mu ~\dotgamma\beta}
\right], \quad \bar{\sigma}^{\mu\nu ~\dotalpha}_{~~~~~\dotgamma}\equiv
\frac{i}{2}\left[
\sigma^{\mu ~\dotalpha\gamma}\bar{\sigma}^{\nu}_{~~\gamma\dotbeta}
- \sigma^{\nu ~\dotalpha\gamma}\bar{\sigma}^{\mu}_{~~\gamma\dotbeta}
\right]
\end{eqnarray}
To rewrite scattering amplitudes in color ordered form one uses a representation of color factors for adjoint $SU(N_c)$ states $f^{abc}$  in terms of color factors $t^a$ associated with smaller fundamental  $SU(N_c)$ representation (\ref{adjointTofundamental}). The idea behind the spinor helicity formalism is similar, that is one consider trading Lorentz vector $p_i^{\mu}$ for kinematical quantities, that transform under a smaller representation of Lorentz group. The latter are given by two-dimensional (Weyl) spinors. So, the four momentum $p_i^{\mu}$ is exchanged with a pair of spinors:
\begin{eqnarray}
p_i^{\mu} \; \to \; u_+ (p_i) = \frac{1}{2}(1+\gamma_5)u(p_i)\equiv |i \ra\equiv\lambda_{i, \alpha}, \quad
u_{-}(p_i)=\frac{1}{2}(1-\gamma_5)u(p_i)\equiv |i]\equiv\vlt_i^{\dotalpha} \nonumber \\
\end{eqnarray}
These spinors satisfy the massless Dirac equation
\begin{eqnarray}
\hat{p}_iu_{\pm}(p_i) = \hat{p}_i|i\ra = \hat{p}_i|i] = 0.
\end{eqnarray}
There are also negative-energy solutions $v_{\pm}$, but for $p_i^2=0$ they are not distinct from $u_{\mp}(p_i)$. The undotted and dotted spinor indices correspond to two different spinor representations of Lorentz group $L_+^\uparrow = SO(3,1) = SO(4,C)_{\downarrow R}\approx (SL(2,C)\otimes SL(2,C))_{\downarrow R}$, which are labeled by a pair $(j_1,j_2)$ of eigenvalues $j_i(j_i+1)$ of the $SL(2)$ Casimir operators ${\bf J}_i^2$: $\lambda_{\alpha}\sim (\frac{1}{2},0)$ and $\vlt^{\dotalpha}\sim (0,\frac{1}{2})$. The raising and lowering of Weyl spinor indices is done with help of antisymmetric tensors $\varepsilon^{\alpha\beta}$ and $\varepsilon^{\dotalpha\dotbeta}$:
\begin{eqnarray}
\lambda^{\alpha} = \varepsilon^{\alpha\beta}\lambda_{\beta},\quad \vlt_{\dotalpha} = \varepsilon_{\dotalpha\dotbeta}\vlt^{\dotbeta},\quad \lambda_{\alpha} = \lambda^{\beta}\varepsilon_{\beta\alpha},\quad \vlt^{\dotalpha} = \vlt_{\dotbeta}\varepsilon^{\dotbeta\dotalpha} ,
\end{eqnarray}
where $\varepsilon^{\alpha\beta} = -\varepsilon_{\dotalpha\dotbeta}$ and $\varepsilon^{12} = \varepsilon_{12} = -\varepsilon_{\dot{1}\dot{2}} = -\varepsilon^{\dot{1}\dot{2}} = 1$. Within spinor helicity formalism one defines the spinor products as:
\begin{eqnarray}
\begin{aligned}
\la ij\ra\equiv \varepsilon^{\alpha\beta}\lambda_{i,\alpha}\lambda_{j,\beta} = \bar{u}_{-}(p_i)u_{+}(p_j), \\
[i j]\equiv \varepsilon^{\dotalpha\dotbeta}\vlt_{i,\dotalpha}\vlt_{j,\dotbeta} =
\bar{u}_{+}(p_i)u_{-}(p_j) .
\end{aligned}
\end{eqnarray}
Rewriting the massless positive energy projector
\begin{eqnarray}
u_{+}(p_i)\bar{u}_{+}(p_i) = |i\ra [ i| = \frac{1}{2}(1+\gamma_5)\hat{p}_i\frac{1}{2}(1-\gamma_5)
\end{eqnarray}
in two-component notation we get
\begin{eqnarray}
\lambda_{i,\alpha}\vlt_{i,\dotalpha} = p_{i, \mu}\sigma^{\mu}_{\alpha\dotalpha} = \hat{p}_{i, \alpha\dotalpha} = \begin{pmatrix}
p_i^t+p_i^z & p_i^x-ip_i^y \\ p_i^x+ip_i^y & p_i^t-p_i^z
\end{pmatrix} \label{hatmomentum}
\end{eqnarray}
It should be noted, that the determinant of this $2\times 2$ matrix vanishes, $\text{det} (\hat{p}_i) = p_i^2 = 0$, which is consistent with its factorization into a column vector $\lambda_{i, \alpha}$ times a row vector $\vlt_{i, \dotalpha}$. Contracting  (\ref{hatmomentum}) with $\sigma^{\nu ~\dotalpha\alpha}$ it is possible to reconstruct the four-momenta $p_i^{\mu}$ from the spinors:
\begin{eqnarray}
2p_i^{\mu} = [i|\gamma^{\mu}|i\ra\equiv \vlt_{i,\dotalpha}\sigma^{\mu ~\dotalpha\alpha}\lambda_{i,\alpha} .
\end{eqnarray}
There are also the following useful in calculations properties of the spinor products:
\begin{eqnarray}
\begin{aligned}
&\text{anti-symmetry:}\quad \la ij\ra = - \la ji\ra,\quad [i j] = -[j i],\quad \la ii\ra = [ii] = 0 ,\\
&\text{squaring:}\quad \la ij\ra [j i] = s_{ij} , \\
&\text{momentum conservation:}\quad \sum_{j=1}^{n} \la ij\ra [j k] = 0 ,\\
&\text{Schouten identity:}\quad \la i j\ra\la k l\ra - \la i k\ra\la j l\ra = \la i l\ra\la k j\ra .
\end{aligned}
\end{eqnarray}
where $s_{ij}=(p_i+p_j)^2$. One could also use spinor helicity formalism to write down the {\it physical} polarization vectors for massless vector particles with definite helicity in terms of a pair of massless spinors:
\begin{alignat}{2}
&\varepsilon^{+}_{i,\mu} = \varepsilon^{+}_{\mu} (p_i,r_i) = \frac{\la r_i|\gamma_{\mu}|i]}{\sqrt{2}\la r_i i\ra},&\qquad
&\varepsilon^{-}_{i,\mu} = \varepsilon^{-}_{\mu} (p_i,r_i) = -\frac{[r_i|\gamma_{\mu}|i\ra}{\sqrt{2}[r_i i]}, \label{polarizationvectors} \\
&\hat{\varepsilon}^{+}_{\alpha\dotalpha} = \hat{\varepsilon}^{+}_{\alpha\dotalpha} (p_i,r_i)
= \sqrt{2}\frac{\lambda_{r_i,\alpha}\vlt_{i,\dotalpha}}{\la r_i i\ra},&\qquad &\hat{\varepsilon}^{-}_{i,\alpha\dotalpha} = \hat{\varepsilon}^{-}_{\alpha\dotalpha} (p_i,r_i) = -\sqrt{2}\frac{\vlt_{r_i,\dotalpha}\lambda_{i,\alpha}}{[r_i i]} ,
\end{alignat}
where we have also written $2\times 2$ matrix version by contracting polarization vectors with $\sigma$ matrix and using Fierz identity for Pauli matrices $\sigma^{\mu}_{\alpha\dotalpha}\sigma_{\mu}^{\dotbeta\beta} = 2\delta_{\alpha}^{\beta}\delta_{\dotalpha}^{\dotbeta}$. Here $p_i^{\mu}$ is  particle momentum and $r_i^{\mu}$ is the null reference momentum accounting for gauge dependence of polarization vectors (local gauge invariance allows an independent choice of reference momenta for different particles). $\lambda_{r_i}$ and $\vlt_{r_i}$ are two-component left and right-handed spinors
associated to reference momentum. It is easy to see, that polarization vectors (\ref{polarizationvectors}) obey the required transversality conditions with respect to particle momentum:
\begin{eqnarray}
\varepsilon_i^{\pm}\cdot p_i = 0.
\end{eqnarray}
They are also transverse with respect to reference momentum $r_i$: $\varepsilon_i^{\pm}\cdot r_i = 0$.
Besides the physical polarization vectors defined above, there are two other {\it unphysical} polarizations vectors:
\begin{eqnarray}
\varepsilon_{i,\mu}^{L} = p_i^{\mu}, \quad
\varepsilon_{i,\mu}^{T} = \frac{\la r_i|\gamma_{\mu}|r_i ]}{2p_i\cdot r_i}
= \frac{r_i^{\mu}}{p_i\cdot r_i}.
\end{eqnarray}
It is easy to check, that polarization vectors satisfy the following orthogonality
\begin{eqnarray}
0 = \varepsilon^+\cdot\varepsilon^+	 = \varepsilon^+\cdot\varepsilon^L = \varepsilon^+\cdot\varepsilon^T = \varepsilon^-\cdot\varepsilon^- =
\varepsilon^-\cdot\varepsilon^L = \varepsilon^-\cdot\varepsilon^T =
\varepsilon^T\cdot\varepsilon^T = \varepsilon^L\cdot\varepsilon^L \nonumber
\end{eqnarray}	
and normalization
\begin{eqnarray}
1 = \varepsilon^+\cdot\varepsilon^- = \varepsilon^L\cdot\varepsilon^T .\nonumber
\end{eqnarray}
conditions.

\section{Grassmannian integral evaluation}\label{aB}
For the evaluation of Grassmannian integrals we used the strategy of \cite{DualitySMatrix}. First, consider bosonic $\delta$ - unctions in Grassmannian integral
\begin{equation}
\delta^{k\times 2} \left( C
\cdot \tilde{\lambda} \right)
\delta^{(n+2-k)\times 2} \left(C^{\perp} \cdot \lambda \right).
\end{equation}
Fixing $GL(k)$ gauge so that first $k$ columns of $C_{al}$ form unity matrix we get
\begin{equation}
\delta^{k\times 2} \left( C
\cdot \tilde{\lambda} \right)
\delta^{(n+2-k)\times 2} \left(C^{\perp} \cdot \lambda \right)=\prod_{a=1}^k
 \delta^{2}\left(\tilde{\lambda}_a+\sum_{i=k+1}^nc_{ai}\tilde{\lambda}_i\right)
 \prod_{i=k+1}^n
 \delta^{2}\left(\lambda_i+\sum_{a=1}^k c_{ai}\lambda_a \right).
\end{equation}
Next, $\delta$ - function constraints lead to the following  underdetermined system of linear equations
\begin{eqnarray}\label{underdeterminedSystemOfEquations}
	c_{ai}\lambda_a&=&-\lambda_i,\nonumber\\
	c_{ai}\tilde{\lambda}_i&=&-\tilde{\lambda}_a,
\end{eqnarray}
where $a=1\ldots k,~i=k+1\ldots n$. For other $GL(k)$ gauges the structure of the above equations will be identical, the only difference are the values taken by indexes $a$ and $i$. The general solution of this system of equations could be  parametrized by $(k-2)(n-k-2)$ complex parameters $\tau_A$:
\begin{eqnarray}\label{GeneralSolutionOfUSoLE1}
	c_{ai}(\tau)=c_{ai}^*+d_{aiA}\tau_A,
\end{eqnarray}
where $d_{aiA}$ are some rational functions of $\lambda,\tilde{\lambda}$'s and $c_{ai}^*$ is some particular solution of (\ref{underdeterminedSystemOfEquations}).
Using these solutions the bosonic $\delta$ - functions could be rewritten as
\begin{eqnarray}\label{FromCtoTau1}
	&&\prod_{a=1}^k
 \delta^{2}\left(\tilde{\lambda}_a+\sum_{i=k+1}^nc_{ai}\tilde{\lambda}_i\right)
 \prod_{i=k+1}^n
 \delta^{2}\left(\lambda_i+\sum_{a=1}^k c_{ai}\lambda_a \right)=\nonumber\\
 &=&\delta^4\left(\sum_{j=1}^n\lambda_j\tilde{\lambda}_j\right)~J(\lambda,\tilde{\lambda})~
\int d^{(k-2)(n-k-2)}\tau_A ~\prod_{a=1}^k\prod_{i=k+1}^n\delta\left(c_{ai}-c_{ai}(\tau)\right),
\end{eqnarray}
where $J(\lambda,\tilde{\lambda})$ is Jacobian of transformation. Note that the number of $\delta$ - functions in LHS and RHS of the above equation is the same. In LHS we have $2n$, while in RHS -
$k(n-k)+4-(k-2)(n-k-2)$ of them. Now, the integration $\int \frac{d^{n\times k}C_{al}}{Vol[GL(k)]}$ could be removed using $\delta$ - functions and the only integrations remained will be with respect to  $\tau_A$ variables ($A = 1, (k-2)(n-k-2)$). Expressing minors of $C_{al}$ - matrix and the Grassmann $\delta$ - functions in terms of $\tau_A$ using (\ref{GeneralSolutionOfUSoLE1}) the integrand of the original Grassmannian integral becomes a rational function
of $\tau_A$ variables and the corresponding integral over $\tau_A$ could be evaluated using (multidimensional) residue theorem.

In the case of $Gr(3,6)$ Grassmannian integral considered in the paper  it is convenient to choose $GL(3)$ gauge as
\begin{eqnarray}
    C=\left( \begin{array}{cccccc}
        1 & c_{12} & 0 & c_{14} & 0 & c_{16} \\
        0 & c_{32} & 1 & c_{34} & 0 & c_{36} \\
        0 & c_{52} & 0 & c_{54} & 1 & c_{56}\end{array} \right).
\end{eqnarray}
Then (\ref{FromCtoTau1}) reduces to
\begin{eqnarray}\label{FromCtoTau}
	&&\prod_{i'=1,3,5}
 \delta^{2}\left(\tilde{\lambda}_{i'}+\sum_{j=2,4,6}c_{i'j}\tilde{\lambda}_j\right)
 \prod_{j=2,4,6}
 \delta^{2}\left(\lambda_j+\sum_{i'=1,3,5} c_{i'j}\lambda_{i'} \right)=\nonumber\\
 &=&\delta^4\left(\sum_{i=1}^6\lambda_i\tilde{\lambda}_i\right)
\int d\tau~\prod_{i'=1,3,5}\prod_{j=2,4,6}\delta\left(c_{i'j}-c_{i'j}(\tau)\right),
\end{eqnarray}
with
\begin{eqnarray}\label{GeneralSolutionOfUSoLE}
	c_{i'j}(\tau)=c_{ij'}^*+\epsilon_{i'k'p'}\epsilon_{jlm}\langle k'p'\rangle [lm]~\tau.
\end{eqnarray}

The minors $M_{1}=(123),\ldots,M_6=(612)$ of $C_{al}$ matrix are linear functions of $\tau$ and the corresponding integral over
$\tau$ could be evaluated using residues (we assume that we are considering
particular component of $\delta^{k\times 2} \left( C
\cdot \tilde{\eta} \right)$ expansion in Grassmann variables such that the overall behavior of the integrand is no worse than $1/\tau^2$ at infinity. At the end we may supersymmetrize the result if necessary).
We were interested in residues at poles $1/M_1$, $1/M_3$ and $1/M_5$. In the gauge chosen the corresponding minors are given by
$M_1=c_{52}(\tau)$, $M_3=c_{14}(\tau)$ and $M_5=c_{36}(\tau)$. To simplify the evaluation of residues even further one can notice that for each of the residues the particular solution $c^*_{i'j}$ could be chosen independently, so that $c_{52}^*=0$ for $M_1$, $c_{14}^*=0$ for $M_3$, and $c_{36}^*=0$ for $M_5$. Then the residue theorem fixes  $\tau=0$ at each residue and all other $c_{i'j}(\tau=0)$ matrix elements are easily evaluated. This way the coefficients of $C_{al}$ matrix for residues $\{1\},\{3\},\{5\}$ at poles $1/M_{1,3,5}$ are given by
\begin{eqnarray}
    C\big{|}_{\{1\}}=\left( \begin{array}{cccccc}
        1 & c_{12}=\dfrac{\langle23\rangle}{\langle13\rangle} & 0 &
        c_{14}=\dfrac{\langle3|1+2|6]}{\langle13\rangle[46]} & 0 & c_{16}=\dfrac{\langle3|1+2|4]}{\langle13\rangle[46]} \\
        0 & c_{32}=\dfrac{\langle12\rangle}{\langle13\rangle} & 1 &
        c_{34}=\dfrac{\langle1|2+3|6]}{\langle13\rangle[46]} & 0 & c_{36}=\dfrac{\langle1|2+3|4]}{\langle13\rangle[46]} \\
        0 & c_{52}=0 & 0 & c_{54}=\dfrac{[56]}{[46]} & 1 & c_{56}=\dfrac{[54]}{[46]}\end{array} \right),
\end{eqnarray}
\begin{eqnarray}
    C\big{|}_{\{3\}}=\left( \begin{array}{cccccc}
        1 & c_{12}=\dfrac{[16]}{[16]} & 0 & c_{14}=0 & 0 & c_{16}=\dfrac{[12]}{[62]} \\
        0 & c_{32}=\dfrac{\langle5|3+4|6]}{\langle35\rangle[26]} & 1 & c_{34}=\dfrac{\langle45\rangle}{\langle35\rangle} & 0 & c_{36}=\dfrac{\langle5|3+4|2]}{\langle 35\rangle[26]} \\
        0 & c_{52}=\dfrac{\langle3|4+5|6]}{\langle53\rangle[26]} & 0 & c_{54}=\dfrac{\langle 43\rangle}{\langle 53 \rangle} & 1 & c_{56}=\dfrac{\langle 3|4+5|2]}{\langle53\rangle [62]}\end{array} \right),
\end{eqnarray}
\begin{eqnarray}
    C\big{|}_{\{5\}}=\left( \begin{array}{cccccc}
        1 & c_{12}=\dfrac{\langle5|1+6|4]}{\langle15\rangle[24]} & 0 & c_{14}=\dfrac{\langle5|1+6|2]}{\langle15\rangle[24]} & 0 & c_{16}=\dfrac{\langle 56\rangle}{\langle15\rangle} \\
        0 & c_{32}=\dfrac{[34]}{[24]} & 1 & c_{34}=\dfrac{[32]}{[42]} & 0 & c_{36}=0 \\
        0 & c_{52}=\dfrac{\langle1|5+6|4]}{\langle15\rangle[24]} & 0 & c_{54}=\dfrac{\langle1|5+6|2]}{\langle15\rangle[24]} & 1 & c_{56}=\dfrac{\langle 61\rangle}{\langle 51 \rangle}\end{array} \right).
\end{eqnarray}

The general case of $Gr(k,n)$ Grassmannian is more involved since one have to consider integral over multiple complex parameters and multidimensional generalization of residue theorems. In the $\mbox{NMHV}_n$ case, which we were discussing in the main body of the paper, the situation is simplified for $n>6$ if we are considering Grassmannian integral in momentum twistor representation (here we  follow \cite{Grassmanians-N4SYM-ABJM}):
\begin{eqnarray}
\omega_{n+2}^3[\Gamma] &=& \int_{\Gamma}\frac{d^{1\times (n+2)} D}{\text{Vol}[GL(1)]}
\frac{Reg.}{d_1 d_2 d_3\ldots d_{n+2}}\delta^{4|4} \left(\sum_{i=1}^{n+2} d_i \mathcal{Z}_i \right),
\nonumber \\
Reg&=&\frac{1}{1+\frac{\la p \xi\ra}{\la p 1\ra}\frac{d_{n+2}}{d_1}},
\end{eqnarray}
which is integral over $Gr(1,n+2)$ Grassmannian. Here,
 $d^{1\times (n+2)} D$ is $n+2$-dimensional volume form in $\mathbb{C}^{n+2}$
$d^{1\times (n+2)} D=\bigwedge_{i=1}^{n+2} \text{d}d_i$.
One of the variables may be set to a prescribed value by fixing $GL(1)$ gauge, while four other variables may be fixed by solving bosonic part of $\delta$ - function constraints $\delta^{4|0} \left(\sum_{i=1}^{n+2} d_i Z_i \right)$. So, we are left with $(n+2)-5$ dimensional integral. Now, let us rearrange integrations in $\omega_{n+2}^3$ in the following way. First, we fix $GL(1)$ gauge by imposing $\delta$ - function constraint for some $d_a$, from $d_1,\ldots,d_{n+2}$:
 \begin{eqnarray}
\int_{\Gamma}\frac{d^{1\times n+2} D}{\text{Vol}[GL(1)]}=
\int_{\Gamma}d^{1\times n+2} D~d^{(0)}_a~\delta\left(d^{(0)}_a-d_a\right),
\end{eqnarray}
were $d^{(0)}_a$ is some number, for example we can choose $d^{(0)}_a=1$. Next, we can use the bosonic part of $\delta$ - functions $\delta^{4|0} \left(\sum_{i=1}^{n+2} d_i Z_i \right)$ to solve these constraints for four arbitrary (but different from
$d_a$) $d_b,d_c,d_d,d_e$ variables and rewrite the above $\delta$ - functions as:
 \begin{eqnarray}
\delta^{4|0} \left(\sum_{i=1}^{n+2} d_i Z_i \right)=
\frac{1}{\langle bcde\rangle}\prod_{i=b,c,d,e}\delta\left(d^{(0)}_i-d_i\right),
\end{eqnarray}
with
 \begin{eqnarray}
d_b^{(0)}&=&\sum_{j\neq b,c,d,e}\frac{\langle cdej \rangle}{\langle bcde \rangle}d_j,~
d_c^{(0)}=\sum_{j\neq b,c,d,e}\frac{\langle bejb \rangle}{\langle bcde \rangle}d_j,~\nonumber\\
d_d^{(0)}&=&\sum_{j\neq b,c,d,e}\frac{\langle ejbc \rangle}{\langle bcde \rangle}d_j,~
d_e^{(0)}=\sum_{j\neq b,c,d,e}\frac{\langle jbcd \rangle}{\langle bcde \rangle}d_j.
\end{eqnarray}
Finally we rewrite the expression for $\omega_{n+2}^3$ as
\begin{equation}
\omega_{n+2}^3[\Gamma] = \frac{d^{(0)}_a}{\langle bcde\rangle}
\int_{\Gamma}d^{1\times n+2} D
\prod_{i=a,b,c,d,e}\delta\left(d^{(0)}_i-d_i\right)\frac{Reg.}{d_1\ldots d_{n+2}}\delta^{0|4} \left(\sum_{i=1}^{n+2} d_i \chi_i \right).
\end{equation}
Now we can chose contour $\Gamma=\Gamma_{abcde}$ in $\mathbb{C}^{n+2}$ to encircle points $d^{(0)}_i$ for $i=a,b,c,d,e$ and $d_i=0$ for all other $i$. This allows us to replace first four $\delta$ - functions left with $1/(d^{(0)}_i-d_i)$ and rewrite the above integral as
\begin{equation}
\omega_{n+2}^3[\Gamma_{abcde}] = \frac{d^{(0)}_a}{\langle bcde\rangle}
\int_{\Gamma_{abcde}}d^{1\times n+2} D
\prod_{i=a,b,c,d,e}\frac{1}{d^{(0)}_i-d_i}\prod_{j\neq a,b,c,d,e}\frac{1}{d_j}
\frac{Reg.\delta^{0|4} \left(\sum_{i=1}^{n+2} d_i \chi_i \right)}{d_a d_b d_c d_d d_e}.
\end{equation}
Using multidimensional generalization of residue theorem we get
\begin{equation}
\omega_{n+2}^3[\Gamma_{abcde}] = Reg.[abcde],
\end{equation}
where
\begin{eqnarray}
Reg.= \frac{1}{1+\frac{\la p \xi\ra}{\la p 1\ra}\frac{\langle a\;b\;c\;d\rangle}{\langle b\;c\; d\;e\rangle}},~\mbox{if one of}~a,b,c,d,e~\mbox{equals to}~n+2,
\end{eqnarray}
and
\begin{eqnarray}
Reg.=1~\mbox{in all other cases}.
\end{eqnarray}
Choosing combinations of $\Gamma_{abcde}$ contours we will get linear combinations of the above terms $Reg.[abcde]$. In the main text we choose contour similar to the case of $[12\rangle$-shift BCFW representation of $\mbox{NMHV}_{n+1}$ on-shell amplitude. In particular, in the case of
$n+2=6$ this choice provides us with the local expression (free from spurious poles) and we hope that similar pattern will hold for $n+2>6$. Cancellation of spurious poles is a little tricky
question and we will consider it in more detail in separate publication.

\end{document}